\documentclass[twocolumn]{aastex62} 

\usepackage{xspace}
\usepackage{color}
\usepackage{rotating}
\usepackage{amsmath}
\usepackage{subfigure}
\usepackage[ampersand]{easylist}
\usepackage[version=4]{mhchem} 
\usepackage[normalem]{ulem}    

\PassOptionsToPackage{hyphens}{url}\usepackage{hyperref}

\begin{document}

\bibliographystyle{apj}


\title{The Detectability and Characterization of the TRAPPIST-1 Exoplanet Atmospheres with JWST}

\shorttitle{Characterizing the TRAPPIST-1 System with JWST}
\shortauthors{Lustig-Yaeger, Meadows, \& Lincowski}

\correspondingauthor{Jacob Lustig-Yaeger}
\email{jlustigy@uw.edu}

\author[0000-0002-0746-1980]{Jacob Lustig-Yaeger}
\affiliation{Department of Astronomy and Astrobiology Program, University of Washington, Box 351580, Seattle, Washington 98195, USA}
\affiliation{NASA NExSS Virtual Planetary Laboratory, Box 351580, University of Washington, Seattle, Washington 98195, USA}

\author[0000-0002-1386-1710]{Victoria S. Meadows}
\affiliation{Department of Astronomy and Astrobiology Program, University of Washington, Box 351580, Seattle, Washington 98195, USA}
\affiliation{NASA NExSS Virtual Planetary Laboratory, Box 351580, University of Washington, Seattle, Washington 98195, USA}

\author[0000-0003-0429-9487]{Andrew P. Lincowski}
\affiliation{Department of Astronomy and Astrobiology Program, University of Washington, Box 351580, Seattle, Washington 98195, USA}
\affiliation{NASA NExSS Virtual Planetary Laboratory, Box 351580, University of Washington, Seattle, Washington 98195, USA}


\begin{abstract}

The James Webb Space Telescope (JWST) will offer the first opportunity to characterize terrestrial exoplanets with sufficient precision to identify high mean molecular weight atmospheres, and TRAPPIST-1's seven known transiting Earth-sized planets are particularly favorable targets. To assist community preparations for JWST observations, we use simulations of plausible post-ocean-loss and habitable environments for the TRAPPIST-1 exoplanets, and test simulations of all bright object time series spectroscopy modes and all MIRI photometry filters to determine optimal observing strategies for atmospheric detection and characterization using both transmission and emission observations. 
We find that transmission spectroscopy with NIRSpec Prism is optimal for detecting terrestrial, CO$_2$ containing atmospheres, potentially in fewer than 10 transits for all seven TRAPPIST-1 planets, if they lack high altitude aerosols. If the TRAPPIST-1 planets possess Venus-like H$_2$SO$_4$ aerosols, up to 12 times more transits may be required to detect an atmosphere. We present optimal instruments and observing modes for the detection of individual molecular species in a given terrestrial atmosphere and an observational strategy for discriminating between evolutionary states. We find that water may be prohibitively difficult to detect in both Venus-like and habitable atmospheres due to its presence lower in the atmosphere where transmission spectra are less sensitive. Although the presence of biogenic O$_2$ and O$_3$ will be extremely challenging to detect, abiotically produced oxygen from past ocean loss may be detectable for all seven TRAPPIST-1 planets via O$_2$-O$_2$ collisionally-induced absorption at 1.06 and 1.27 $\mu$m, or via NIR O$_3$ features for the outer three planets. Our results constitute a suite of hypotheses on the nature and detectability of highly-evolved terrestrial exoplanet atmospheres that may be tested with JWST.

\end{abstract}


\section{Introduction\label{sec:intro}}

The discovery of Earth-sized planets in temperate orbits around nearby, low-mass stars opens a new door into the era of terrestrial exoplanet atmospheric characterization \citep{Berta-Thompson2015, Anglada-Escude2016, Gillon2016, Gillon2017, Luger2017, Dittmann2017}. Transmission and emission (secondary eclipse) spectroscopy of transiting rocky worlds with the upcoming James Webb Space Telescope (JWST) may offer a first glimpse into the atmospheres of terrestrial exoplanets \citep{Morley2017, Kalirai2018} and a first opportunity to search for signs of habitability \citep{Lincowski2018} and biosignatures beyond the Solar System \citep{Cowan2015}.

The TRAPPIST-1 system of seven transiting Earth-sized exoplanets \citep{Gillon2016, Gillon2017, Luger2017} is observationally favorable for the atmospheric characterization of small exoplanets. TRAPPIST-1 is a late M dwarf \citep[M8V;][]{Liebert2006} with a small radius \citep[0.121R$_{\odot};$][]{VanGrootel2018}, which increases planetary transit and eclipse depths; it has a low effective temperature \citep[2511 K;][]{VanGrootel2018}, which increases the eclipse depth; and it is nearby to Earth \citep[12.2 pc;][]{Gillon2016}. These system properties increase sensitivity to atmospheric spectral features in transmission and emission spectroscopy, particularly for small, temperate planets. 

Observations of the TRAPPIST-1 planets with the \textit{Kepler}, \textit{Hubble} (HST), and \textit{Spitzer} space telescopes suggest that the innermost 6 planets do not have primordial, low mean molecular weight atmospheres, but whether they have high molecular weight atmospheres or no atmospheres at all requires observations with future facilities.  
HST Wide Field Camera 3 (WFC3) transmission spectroscopy has ruled out \ce{H2}-dominated atmospheres for most of the TRAPPIST-1 planets \citep{deWit2016, deWit2018}, and instead they may have secondary outgassed atmospheres composed of relatively high mean molecular weight gases \citep{Moran2018}.
Secondary atmospheres are more difficult to detect---or rule out---than primordial atmospheres, and they may span a large range in temperature, pressure, and composition. Consequently a broad variety of potential atmospheres are consistent with the current modest observational constraints \citep{Delrez2018,Lincowski2018}.

Alternatively, the planets could have no atmospheres, although outgassing from the possibly high-volatile-content interiors of the TRAPPIST-1 planets may make that outcome less likely.  Simulations suggest that the TRAPPIST-1 planets could have had their atmospheres completely stripped \citep[e.g.][]{Dong2018, Airapetian2017, Roettenbacher2017}, although some models of M dwarf planets suggest that atmospheric loss rates may be less than the replenishment rate via outgassing from a planetary interior \citep{Garcia-Sage2017, Bolmont2017}, such that atmospheres may be retained. The TRAPPIST-1 planets may also have larger volatile reservoirs than Solar System terrestrials. Masses derived from transit timing variation (TTV) analyses of \textit{Kepler} \citep{Grimm2018} and \textit{Spitzer} observations \citep{Delrez2018} suggest that the planets have densities between 0.6 to 1.0 $\rho_{\oplus}$, consistent with a rocky composition with a significant fraction of ices \citep{Grimm2018}.  This conclusion is supported by the resonant chain structure of the TRAPPIST-1 system, which suggests inward migration from more volatile-rich formation orbits \citep{Luger2017}. Determining whether the interplay of planetary processes over time will allow M dwarf terrestrial planets to maintain high-molecular weight atmospheres and support habitability will be key science questions for JWST.  

Previous simulations have demonstrated the plausibility of detecting terrestrial atmospheres with JWST for M dwarf planetary systems in general \citep[e.g.][]{Belu2011, Barstow2016}, and for the TRAPPIST-1 planets in particular \citep{Barstow2016b, Morley2017, Batalha2018, Krissansen-Totton2018}. 
\citet{Morley2017} considered Venus-like, Earth-like, and Titan-like atmospheres in thermochemical equilibrium and found that CO$_2$-dominated atmospheres could be detected for 6 of the 7 TRAPPIST-1 planets in fewer than 20 transits with JWST NIRSpec/G235M, and secondary eclipse photometry with JWST/MIRI could readily detect thermal emission from TRAPPIST-1 b and c, and possibly d, e, and f.  \citet{Batalha2018} investigated optimal strategies for JWST observations of the TRAPPIST-1 system and found that allowing the NIRSpec Prism to slightly saturate at the peak of the stellar spectral energy distribution (SED) could allow the dominant absorber to be detected in 10 transits of H$_2$O, CO$_2$, and N$_2$ dominated atmospheres. 
\citet{Krissansen-Totton2018} investigated the potential detectability of anoxic biosignatures in the atmosphere of TRAPPIST-1 e and concluded that 10 transits observed with JWST NIRSpec Prism may be sufficient to detect CO$_2$ and constrain the CH$_4$ abundance enough to rule out non-biological CH$_4$ production. 
Recently, \citet{Wunderlich2019} considered the detectability of photochemically self-consistent Earth-like planets in the habitable zone of various M dwarfs spectral types, including TRAPPIST-1, and found that \ce{H2O}, \ce{CH4}, and \ce{CO2} may be detectable in ${\sim}10$ transits with JWST. However, this study did not consider non Earth-like planetary compositions, or the effect of clouds and hazes.  

These previous simulations of TRAPPIST-1 planetary atmospheres did not include photochemical forcing by the late M dwarf SED on multiple plausible planetary environments, both habitable and uninhabitable.  Photochemistry can have significant impacts on terrestrial atmospheric composition \citep{Segura2005, Rugheimer2015} and haze formation \citep{Arney2017, Arney2018}, which can in turn modify the resultant temperature structure \citep{Lincowski2018}, impacting the predicted spectrum in both transmission and emission.  Models that include photochemistry and haze formation for terrestrial atmospheres will better predict the atmospheric composition, and inform the preparation and interpretation of upcoming observations with JWST \citep{Lincowski2018}.  

In this paper we explore the potential for JWST to detect and characterize the TRAPPIST-1 planetary atmospheres and distinguish between model predictions for evolutionary outcomes and different atmospheric states. As input, we use the climatically and photochemically self-consistent atmospheric and spectral simulations of \citet{Lincowski2018}, who used a rigorous line-by-line 1D radiative-convective-equilibrium climate model coupled with a 1D photochemical model to simulate different habitable and post-ocean-loss environments for the TRAPPIST-1 planets.  The atmospheric bulk compositions considered by \citet{Lincowski2018} are motivated by the early high luminosity of late M dwarf stars \citep{Baraffe2015} like TRAPPIST-1, which may drive early ocean loss and the generation of tens to thousands of bars of O$_2$ \citep{Luger2015, Bolmont2017, Meadows2018, Lincowski2018, Wordsworth2018}. The final inventory of \ce{O2} may be severely reduced by atmospheric and surface loss processes such that only a few bars of \ce{O2} remain \citep[][]{Schaefer2016, Wordsworth2018}.  Depending on the initial water inventory, this process may have exhausted the entire planet's water supply, leaving it desiccated, or the planet may have formed with enough water to endure such vigorous loss.  For the ocean loss planets with efficient O$_2$ sinks and ongoing outgassing of volatiles, large quantities of CO$_2$ may build up, forming a Venus-like atmosphere \citep{Meadows2018}.  \citet{Lincowski2018} also considered a habitable ocean world for TRAPPIST-1 e, for the scenario where it formed with an appreciable H$_2$ envelope, which was subsequently stripped to reveal a habitable core \citep{Luger2015b}.

To understand the nature and possibly habitability of terrestrial exoplanets, we will need a systematic approach to environmental assessment that starts with the most scientifically significant and least expensive observation, and builds from there. For terrestrial planets orbiting M dwarfs, the first property to be determined is whether or not they have an atmosphere.  If an atmosphere can be confirmed, then subsequent studies will focus on determining the nature of that atmosphere, and whether there are atmospheric characteristics that could discriminate between evolutionary outcomes. 
Finally, a deeper dive to search for signs of habitability, including the presence of an ocean \citep[see][]{Robinson2018, Lustig-Yaeger2018} and biosignatures \citep[see][]{Schwieterman2018} may be warranted for planets whose initial characterization does not preclude habitable conditions.

Here we determine the feasibility of atmospheric characterization for the seven known TRAPPIST-1 exoplanets by first identifying  optimal instrument selection and experiments to test whether or not the TRAPPIST-1 planets have atmospheres. 
We then determine how to best detect specific molecules as a second step of atmospheric characterization, and discriminate between different plausible climate and photochemically self-consistent atmospheres (e.g. \ce{O2}-dominated, \ce{CO2}-dominated; \citealp{Lincowski2018}) using JWST transmission and emission photometry and spectroscopy. 

The different post-runaway, evolved planet atmospheres considered here are by no means a comprehensive set of evolutionary outcomes, but rather a representative subset of potential physically- and chemically-motivated atmospheres for which we can predict spectra. 
By understanding the detectability of spectral discriminants for the TRAPPIST-1 planets we develop informed hypotheses on the nature of these planets that may later be tested with JWST observations.  

The structure of the paper is as follows: In \S\ref{sec:methods} we present the models and methods used to simulate JWST data and find the optimal JWST modes for detecting and characterizing plausible planet compositions. We describe our results in \S\ref{sec:results}, offer a discussion of the significance of those results in \S\ref{sec:discussion}, and conclude in \S\ref{sec:conclusion}.

\section{Methods}
\label{sec:methods}

In the following subsections we present our methods for assessing the detectability of different self-consistent atmospheric compositions for the seven TRAPPIST-1 planets with different instruments and observational modes available to JWST. 
We first describe the JWST noise models used in this work (\S\ref{sec:methods:JWST}), which include a MIRI photometry component and a spectroscopy component using \texttt{PandExo} \citep{Batalha2017b}. 
We then detail our model inputs (\S\ref{sec:methods:inputs}), and outline a series of experiments that can be used to successively characterize terrestrial planet environments and determine optimal observing modes (\S\ref{sec:results:experiments}).  

\subsection{JWST Noise Modeling}
\label{sec:methods:JWST}

We simulate synthetic exoplanet time-series spectroscopy and photometry with JWST to consider observations during transit (transmission) and secondary eclipse (emission). In the following two subsections we detail our modeling of JWST/MIRI photometry (\S \ref{sec:methods:jwst_phot}) and JWST spectroscopy (\S \ref{sec:methods:jwst_spec}). 

\subsubsection{JWST/MIRI Photometry}
\label{sec:methods:jwst_phot}

Filter photometry with JWST's Mid-Infrared Instrument (MIRI) imager has been suggested to offer an efficient means of performing an initial characterization of Earth-sized planets around low-mass stars \citep{Morley2017}. To assess MIRI photometry, we develop a basic MIRI imaging noise model for exoplanet transit and secondary eclipse observations. 

The number of photons from the planet incident upon the detector is
\begin{equation}
\label{eqn:N_photons}
    N_{p} = T_{\text{exp}} \frac{F_p \mathcal{T} A \lambda \Delta \lambda}{hc}
\end{equation}
where $T_{\text{exp}}$ is the exposure time, $F_p$ is the spectral flux density (e.g. W/m$^2$/um) from the planet, $\mathcal{T}$ is the filter throughput, $A$ is the telescope collecting area (25 m$^2$), $\lambda$ is wavelength, $\Delta \lambda$ is the width of the wavelength bin, $h$ is Planck's constant, and $c$ is the speed of light.
Note that the exposure time used in our photometry model is discretized in terms of the simulated planet's transit duration, but does not explicitly depend on the MIRI integration times or number of groups per integration, as our time-series spectroscopy noise modeling with \texttt{PandExo} does.
Photon conversion efficiency curves for the MIRI imager were acquired online\footnote{\url{http://ircamera.as.arizona.edu/MIRI/pces.htm}} \citep{Glasse2015}. Photons from the star are calculated analogously using Equation \ref{eqn:N_photons} by replacing the planet flux with the stellar flux. 

The signal-to-noise (SNR) ratio on the transit depth is given by 
\begin{align}
\label{eqn:transit_SNR}
    \text{SNR}_T = \frac{N_{s}(R_p / R_s)^2}{\sqrt{(N_{s} + N_{bg})/n_{\text{out}} + N_{bg} + N_s[1-(R_p/R_s)^2]}}. 
\end{align}
where $N_{s}$ is the number of photons from the star, $N_{bg}$ is the number of photons from background sources, and $n_{\text{out}}$ is the number of out-of-transit transit durations observed. Background photon noise is calculated using the seven component grey-body model of \citet{Glasse2015}, which includes telescope thermal and scattered zodiacal noise contributions. 
The SNR on the photons detected from a planet observed in secondary eclipse is given by
\begin{equation}
\label{eqn:eclipse_SNR}
    \text{SNR}_E = \frac{N_{p}}{\sqrt{(N_p + N_s + N_{bg})/n_{\text{out}} + N_s + N_{bg}}}.
\end{equation}
where $N_{p}$ is the number of photons from the planet. Derivations of Equations \ref{eqn:transit_SNR} and  \ref{eqn:eclipse_SNR} are provided in Appendix \ref{sec:appendix:snr}. 

Saturation must be considered when planning long exposures necessary to characterize small exoplanets around nearby stars. Using the bright source limits of \citet{Glasse2015} we find that TRAPPIST-1 will saturate MIRI in the two shortest wavelength filters, F560W and F770W, for the shortest exposures allowed in the standard imaging mode. Although these shorter wavelength filters may saturate, we nonetheless consider MIRI photometry in all nine filters (F560W, F770W, F1000W, F1130W, F1280W, F1500W, F1800W, F2100W, and F2550W) to assess the atmospheric information contained in each. 

\subsubsection{JWST Spectroscopy}
\label{sec:methods:jwst_spec}

We use the JWST time-series spectroscopy simulator \texttt{PandExo}\footnote{\url{https://natashabatalha.github.io/PandExo/}} \citep[version 1.1.2;][]{Batalha2017b, Pandexo2018} to model different observing modes and their associated noise sources for transmission and emission spectroscopy. \texttt{PandExo} leverages the core of the Space Telescope Science Institute's Exposure Time Calculator, \texttt{Pandeia}\footnote{\url{https://jwst.etc.stsci.edu/}} \citep[version 1.2.2;][]{Pontoppidan2016}, to calculate 3-D data cubes for realistic PSF modeling. We refer the reader to \citet{Batalha2017b} for a thorough description of the model and its bench-marking. 

We consider a broad variety of JWST instruments and modes that are capable of exoplanet transmission and emission spectroscopy and available using \texttt{PandExo}. We include the Near-Infrared Camera \citep[NIRCam;][]{Greene2007, Greene2017} using the grism time-series mode; 
the Near-Infrared Spectrograph \citep[NIRSpec;][]{Bagnasco2007, Ferruit2014}; 
the Near Infrared Imager and Slitless Spectrograph \citep[NIRISS;][]{Doyon2012} using the single object slitless spectroscopy (SOSS) mode; 
and the Mid-Infrared Instrument \citep[MIRI;][]{Bouchet2015} low resolution spectrometer \citep[LRS;][]{Kendrew2015}. 

\begin{deluxetable*}{llllllll||lll}
\tablewidth{\linewidth}
\tablecaption{\label{tab:instruments} JWST instruments used in this study and their observability of TRAPPIST-1}
\tablehead{ \multicolumn{8}{c}{} & \multicolumn{3}{c}{TRAPPIST-1} \\
\cline{9-11}
\colhead{Instrument} & \colhead{Mode}  & \colhead{Disperser} & \colhead{Filter} & \colhead{Subarray} & \colhead{Read Mode} & \colhead{$\lambda$ [$\mu$m]} & \colhead{$R$ [$\lambda / \Delta \lambda$]} & \colhead{$N_{groups}$} & \colhead{Efficiency} & \colhead{$N_{sat}$}  }
\startdata
NIRCam	& ssgrism	 &  grism R	& f322w2	    & subgrism64	    & rapid	    & $2.42 - 4.15$	& ${\sim}$1600	& 303	& 99.34\%	& 0 \\
NIRCam	& ssgrism	 &  grism R	& f444w	        & subgrism64	    & rapid	    & $3.70 - 5.00$	& ${\sim}$1600	& 342	& 99.42\%	& 0 \\
NIRISS	& SOSS	     &  gr700xd	& None	        & substrip96	    & nisrapid	& $0.6 - 2.8$	& ${\sim}$700	& 52	& 96.23\%	& 0 \\
NIRISS	& SOSS	     &  gr700xd	& None	        & substrip256	    & nisrapid	& $0.6 - 2.8$   & ${\sim}$700	& 21	& 90.91\%	& 0 \\
NIRSpec	& fixed slit &	g140h	& f100lp	    &     sub2048	    & nrsrapid	& $0.97 - 1.82$	& ${\sim}$2700	& 39	& 95.00\%	& 0 \\
NIRSpec	& fixed slit &	g140m	& f100lp	    &     sub2048	    & nrsrapid	& $0.97 - 1.84$	& ${\sim}$1000	& 14	& 86.67\%	& 0 \\
NIRSpec	& fixed slit &	g235h	& f170lp	    &     sub2048	    & nrsrapid	& $1.66 - 3.05$	& ${\sim}$2700	& 41	& 95.24\%	& 0 \\
NIRSpec	& fixed slit &	g235m	& f170lp	    &     sub2048	    & nrsrapid	& $1.66 - 3.07$	& ${\sim}$1000	& 14	& 86.67\%	& 0 \\
NIRSpec	& fixed slit &	g395h	& f290lp	    &     sub2048	    & nrsrapid	& $2.87 - 5.14$	& ${\sim}$2700	& 82	& 97.59\%	& 0 \\
NIRSpec	& fixed slit &	g395m	& f290lp	    &     sub2048	    & nrsrapid	& $2.87 - 5.10$	& ${\sim}$1000	& 29	& 93.33\%	& 0 \\
NIRSpec	& fixed slit &	prism 	& clear	        & sub512	        & nrsrapid	& $0.6 - 5.3$	& ${\sim}$100	& 2	    & 33.33\%	& 0 \\
NIRSpec	& fixed slit &	prism 	& clear	        & sub512s	        & nrsrapid	& $0.6 - 5.4$	& ${\sim}$100	& 3	    & 50.00\%   & 0 \\
NIRSpec	& fixed slit &	prism 	& clear	        & sub512	        & nrsrapid	& $0.6 - 5.5$	& ${\sim}$100	& 6	    & 71.43\%	& 47 \\
NIRSpec	& fixed slit &	prism 	& clear	        & sub512s	        & nrsrapid	& $0.6 - 5.6$	& ${\sim}$100	& 6	    & 71.43\%	& 19 \\
MIRI	& LRS	     &    p750l	& None	        & slitlessprism	    & fast	    & $0.5 - 12.0$	& ${\sim}$100	& 139	& 98.57\%	& 0 
\enddata
\tablecomments{$N_{groups}$ is the number of groups per integration and $N_{sat}$ is the number of saturated pixels at the end of the ramp. The right three columns are outputs from the \texttt{PandExo} JWST noise model specifically for observations of the TRAPPIST-1 system.}
\end{deluxetable*}

Table \ref{tab:instruments} summarizes the different JWST instruments and modes used to simulate transmission and emission spectroscopy of the TRAPPIST-1 system. Specifically, Table \ref{tab:instruments} lists the instrument, mode, disperser, filter, subarray, read mode, wavelength range, and nominal spectral resolving power used in our \texttt{PandExo} calculations. Table \ref{tab:instruments} also presents the number of groups per integration, the observing efficiency, and the number of saturated pixels at the end of the ramp for our simulated observations of the TRAPPIST-1 system with each instrument.
We use the NIRSpec Prism in three configurations. First, we use the SUB512 subarray (frame time: 0.226 s) with 2 groups per integration set by \texttt{PandExo} to avoid pixel saturation. Second, we use the SUB512s subarray (frame time: 0.144 s) with 3 groups per integration, again set by \texttt{PandExo}. Finally, we simulate a partial saturation strategy by using the SUB512 and SUB512s subarrays with 6 groups per integration to allow for slight pixel saturation near the peak of the SED \citep{Batalha2018}. This modification improves the duty cycle from 33.3\% (50\%) to 71.4\% for the SUB512 (SUB512s) subarray, as shown in Table \ref{tab:instruments} (see ``Efficiency'' column). We refer to this partially saturated NIRSpec Prism mode as ``NIRSpec Prism*'' hereafter.

\subsection{Noise Model Inputs}
\label{sec:methods:inputs}

For the stellar input to the \texttt{PandExo} noise model, we approximate the TRAPPIST-1 stellar spectrum, which has yet to be observed, using a PHOENIX stellar model \citep{Husser2013} with an effective temperature of T = 2511 K, metallicity of [Fe/H] = 0.04, and surface gravity $\log g = 5.23$ \citep{Delrez2018}, normalized to the K band magnitude of TRAPPIST-1 \citep[$K = 10.30$;][]{Grimm2018}. 
However, we ignore the effects of stellar opacity in the MIR, stellar variability due to rotation and flaring during periods of observation \citep{Vida2017, Morris2018}, and heterogeneous stellar photospheres \citep{Rackham2018, Zhang2018}.

We use the modeled transmission and emission spectra of \citet{Lincowski2018} as inputs into the noise models to assess the detectability of photochemically and climatically self-consistent TRAPPIST-1 planet's atmospheres. 
The climate model developed by \citet{Robinson2018b} and \citet{Lincowski2018} uses line-by-line radiative transfer computed by the Spectral Mapping and Atmospheric Radiative Transfer (SMART) code \citep[][developed by D. Crisp]{Meadows1996}, and can generate top-of-atmosphere planetary radiances and transmission spectra \citep{Robinson2017a} of its equilibrium climate and photochemical states. 
See \citet{Lincowski2018} for a thorough description of the climate and photochemical modeling and subsequent climate results. 
The stellar spectrum used in the optical through the MIR in \citet{Lincowski2018} is identical to the stellar spectrum used here for the JWST noise model input.  

For target exposure times per transit we use the median transit durations for the TRAPPIST-1 planets from \citet{Grimm2018}, and for photometry and spectroscopy noise calculations we assume that an equal amount of time is spent observing in transit/eclipse versus out of transit/eclipse. For \texttt{PandExo}, we assume saturation to be an exposure level 80\% of the full well and we impose no strict noise floor for JWST spectroscopy \citep[c.f.][]{Greene2016}. We compute noise calculations across a grid in the number of transits/eclipses ([1, 100]), and then use these results to derive and report the number of transits/eclipses needed to meet specific atmospheric detection and characterization metrics, as described in the following section. 

\subsection{Observing Experiments}
\label{sec:results:experiments}

We aim to identify the optimal observing approaches for JWST to (1) detect the presence of the TRAPPIST-1 planet atmospheres, and (2) characterize the composition of the atmospheres, assuming the TRAPPIST-1 planets possess atmospheres similar in nature to the evolved atmospheres modeled in \citet{Lincowski2018}. 
We consider an atmosphere to be detected when sufficient SNR is achieved on \textit{any} spectral feature in a transmission or emission spectrum. 
For atmospheric characterization, we consider a specific molecule in the atmosphere to be detected when sufficient SNR is attained on the contribution to the spectrum from that molecule, which may include multiple bands from a given molecule.
Next, we detail our SNR approach for ruling out a fiducial spectrum, and then describe how the method is used to quantify the detectability of atmospheres and specific molecules within them. 

\subsubsection{SNR Approach}

We now define a signal-to-noise ratio approach to determine the confidence with which we can rule out that our data match a fiducial transmission or emission spectrum---which is a featureless spectrum for the case of detecting atmospheres. 

For each atmospheric model and JWST instrument considered, we employ the following procedure for both transit and secondary eclipse geometries. 
First, we run the \texttt{PandExo} JWST noise model across a grid in number of transits/eclipses ($n_{occ}$) from 1-100, which is sufficient to establish a simple SNR scaling relationship. Second, we determine the signal-to-noise on the difference between the model spectrum and the fiducial spectrum, and calculate the total expected signal-to-noise $\left < \mathrm{SNR} \right >$ (defined below) by summing this difference over wavelength. Finally, we solve for number of transits/eclipses, $n_{occ}$, such that a given $\left < \mathrm{SNR} \right >$ is achieved.

We define the total expected signal-to-noise using a $\Delta \chi^2$ test formalism, which is common for model selection applications. 
For many random drawings of synthetic data with Gaussian noise, the expected value for $\Delta \chi^2$ between two models ($m_{1}$ and $m_{2}$) is simply
\begin{equation}
    \label{eqn:delta_chi2_1}
    \left < \Delta \chi^2 \right > = \sum_{i=1}^{N_{\lambda}} \left ( \frac{m_{1,i} - m_{2,i}}{\sigma_i} \right )^2 
\end{equation}
assuming that the observations with uncertainties $\sigma$ are truly sampled from one of the models ($m_{1}$ in this case), and where the sum is over all $N_{\lambda}$ spectral elements for a particular instrument. 
Gaussian noise is not added to the synthetic spectra both to speed up calculation of the mean result, and to avoid any single random data realization from biasing our results \citep[e.g.][]{Feng2018}.
The numerator in the sum in Equation \ref{eqn:delta_chi2_1} is the ``signal'' used to discriminate between the two models, while the denominator is the ``noise'' on the observations. For convenience, we define 
\begin{equation}
    \label{eqn:snr_i}
    \mathrm{SNR}_i = \frac{m_{1,i} - m_{2,i}}{\sigma_i} ,
\end{equation}
which is the individual signal-to-noise contribution from each spectral element to model $m_{2}$ being ruled out in favor of model $m_{1}$.   
Equation \ref{eqn:delta_chi2_1} may then be rewritten in terms of $\mathrm{SNR}_i$ and what we refer to as the total expected signal-to-noise 
\begin{equation}
    \label{eqn:SNR_delta_chi2}
    \left < \mathrm{SNR} \right > = \sqrt{\left < \Delta \chi^2 \right >} = \sqrt{\sum_{i=1}^{N_{\lambda}} \mathrm{SNR}_i^2}.     
\end{equation}
Equation \ref{eqn:SNR_delta_chi2} is particularly useful because the quadrature sum over wavelength allows for the comparison between multiple different JWST instruments that may have different spectral resolutions and wavelength ranges, and for a comparison between transmission and emission spectroscopy for the same hypothesis (e.g. \textit{the planet does not have an atmosphere}). 

We caution that Equation \ref{eqn:SNR_delta_chi2} is equal to the confidence in the detection of model $m_{1}$ in units of standard deviations (number of ``sigma'' $n_{\sigma}$), only in the case that each model has one degree of freedom. To avoid assumptions on the degrees of freedom and degeneracies associated with spectral models, we simply report $\left < \mathrm{SNR} \right >$ with the understanding that these are upper limits on the  confidence in the detection. 
Furthermore, when we report the number of transits/eclipses to rule out the fiducial model to a given $\left < \mathrm{SNR} \right >$, these are lower limits; additional sources of uncertainty and/or retrieval model complexity may require that more transits/eclipses be observed.  

We now detail the fiducial spectral models that are used to detect the presence of an atmosphere and the individual molecules with them. 

\subsubsection{Detecting the presence of an atmosphere}
\label{sec:results:detect_spec}

The spectra of planets with atmospheres can be discriminated from the featureless spectra of airless worlds by the presence of spectral absorption features.  This approach works best for atmospheres with strong absorption features, but will be challenging for transmission spectroscopy if molecular absorption is suppressed by a high mean molecular weight atmosphere \citep{Miller-Ricci2009}, the presence of clouds and hazes \citep{Berta2012, Ehrenreich2014, Knutson2014, Kreidberg2014, Nikolov2015}, or atmospheric refraction \citep{Betremieux2014, Misra2014a}. In these cases, higher SNR observations will be required to detect the presence of an atmosphere. Here we assess the detectability of realistic terrestrial atmospheres with clouds and hazes in transmission by comparing their spectra to a baseline featureless spectrum that is modeled as the best-fitting constant planet radius with wavelength. For the corresponding emission spectrum test, we model the featureless spectrum as the ratio of two blackbodies at the stellar effective temperature and best-fitting planetary equilibrium temperature (set by a variable bond albedo).  

While there are several ``false negative'' processes that could suppress the signal from an atmosphere as described above, one ``false positive'' process is worth considering: the possibility that surface mineralogy could produce wavelength dependent features in emission spectra of airless worlds. 
However, these features are unlikely to be as prominent as atmospheric features, and this is especially true in realistic cases where multiple reflections occur within the mineral surface.
Specifically, by Kirchoff's law of thermal radiation, the emissivity $\epsilon$ is related to the reflectance $R$ by \citep{Nicodemus1965}, 
\begin{equation}
    \epsilon = 1 - R.
\end{equation}
The emissivity then regulates the efficiency of thermally radiated flux $F_{\lambda}$ relative to a perfect blackbody $B_{\lambda}(T)$ via 
\begin{equation}
\label{eqn:greybody} 
    F_{\lambda} = \epsilon_e B_{\lambda}(T), 
\end{equation}
for an object with an equilibrium temperature $T$. However, Equation \ref{eqn:greybody} uses the \textit{effective} emissivity, 
\begin{equation}
\label{eqn:epsilon_e}
    \epsilon_e = 1 - (1 - \epsilon)^{(n+1)}, 
\end{equation}
which accounts for the number of reflections $n$ within the material, and ultimately decreases the contrast of mineralogical features in the thermal emission spectrum \citep{Kirkland2003, Hu2012}.  

\begin{figure*}[t!]
\centering
\includegraphics[width=0.97\textwidth]{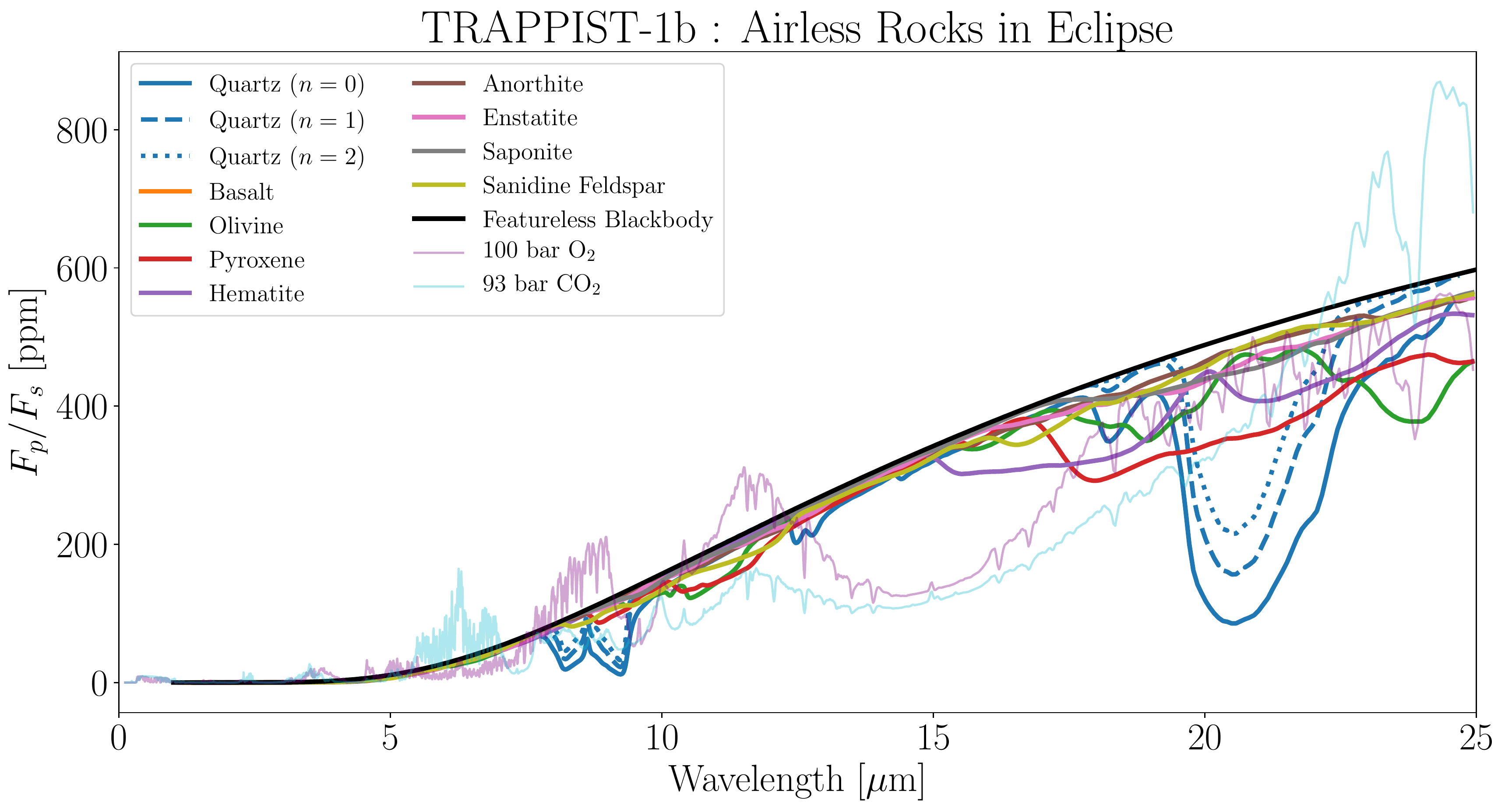}
\caption{Secondary eclipse spectrum models of TRAPPIST-1b assuming different end-member planet mineralogical compositions. The black line shows a featureless blackbody curve, corresponding to the zero bond albedo planet equilibrium temperature, from which all of the thermal emission curves from rock forming minerals (thick color lines) deviate due to non-unity, wavelength-dependent emissivities. The three line styles for the quartz curves demonstrate the reduction in effective emissivity due to $n$ reflections within the rock (see equation \ref{eqn:epsilon_e}). The thin purple and teal lines show the expected emission spectrum for TRAPPIST-1b if it were to possess a climatically and photochemically self-consistent thick O$_2$- or a CO$_2$-dominated atmosphere, respectively \citep{Lincowski2018}.}
\label{fig:airless_rocks}
\end{figure*} 

Figure \ref{fig:airless_rocks} compares secondary eclipse spectrum models for different assumed end-member planet mineralogical compositions. Emission spectra are shown for planets at the zero bond albedo equilibrium temperature composed solely of quartz, basalt, olivine, pyroxene, hematite, anthorite, enstatite, saponite, and feldspar---a selection of common rocks and minerals found in terrestrial solar system bodies. For comparison, also shown are emission spectra with strong atmospheric absorption features from the climatically and photochemically self-consistent 100 bar O$_2$-dominated and 92 bar CO$_2$-dominated atmospheres from \citet{Lincowski2018}. 

In some cases, mineralogical surface emission features compare in signal contrast to model atmospheric thermal emission features, but under most plausible physical scenarios an airless rock would likely have significantly lower spectral variation than atmospheric features. The single-reflection quartz silicate features between ${\sim}8 - 10$ $\mu$m and ${\sim}19 - 23$ $\mu$m are strong and rival the strength of the atmospheric features caused by O$_3$, CO$_2$, and H$_2$O over this wavelength range. 
However, single reflections are unlikely, and both multiple reflections and blends of different minerals would decrease the contrasts on features from any one of the representative end-member cases, producing a relatively featureless thermal emission spectrum.

Throughout the rest of the paper, unless otherwise stated, we adopt the assumption that detecting deviations from blackbody emission in a secondary eclipse spectrum are evidence of an atmosphere. However, in \S\ref{sec:results} we further explore the detectability of quartz silicate emissivity features in the emission spectrum of TRAPPIST-1b observed with MIRI LRS as an optimistic limiting case on the potential signal from an airless body. 

\begin{figure}[t]
\centering
\includegraphics[width=0.49\textwidth]{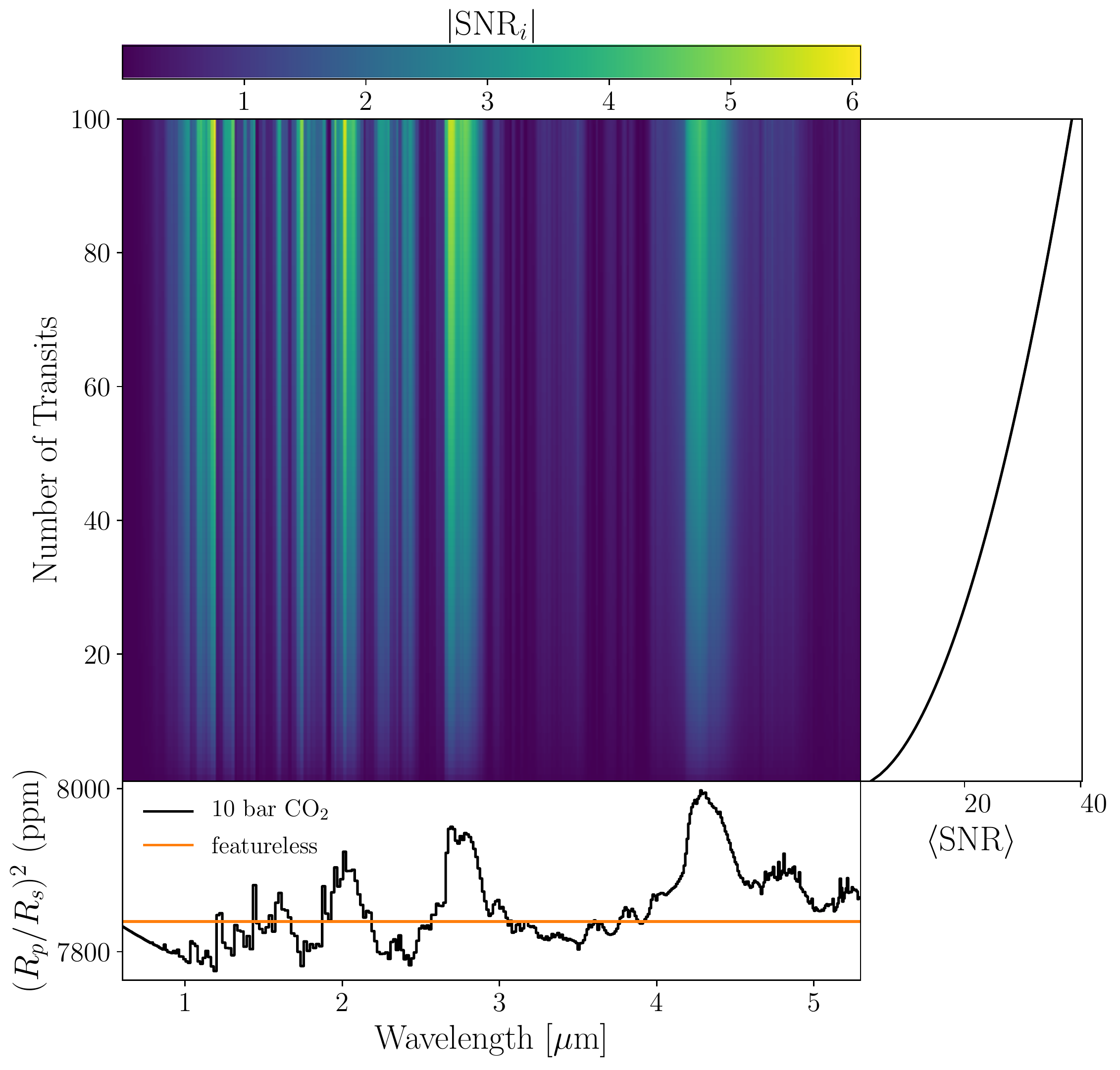}
\caption{Signal-to-noise contours on the simulated 10 bar high CO$_2$ transmission spectrum, observed with JWST/NIRSpec Prism with 6 groups per integration,
relative to a featureless spectrum as a function of the number of transits observed and wavelength of the instrument. The bottom panel shows the noiseless spectrum (black) and the best-fitting featureless spectrum (orange), both convolved to the instrument resolution. The right panel shows how the total expected SNR from Equation \ref{eqn:SNR_delta_chi2} increases with more transits observed.}
\label{fig:featureless_transit}
\end{figure}

\begin{figure}[t]
\centering
\includegraphics[width=0.49\textwidth]{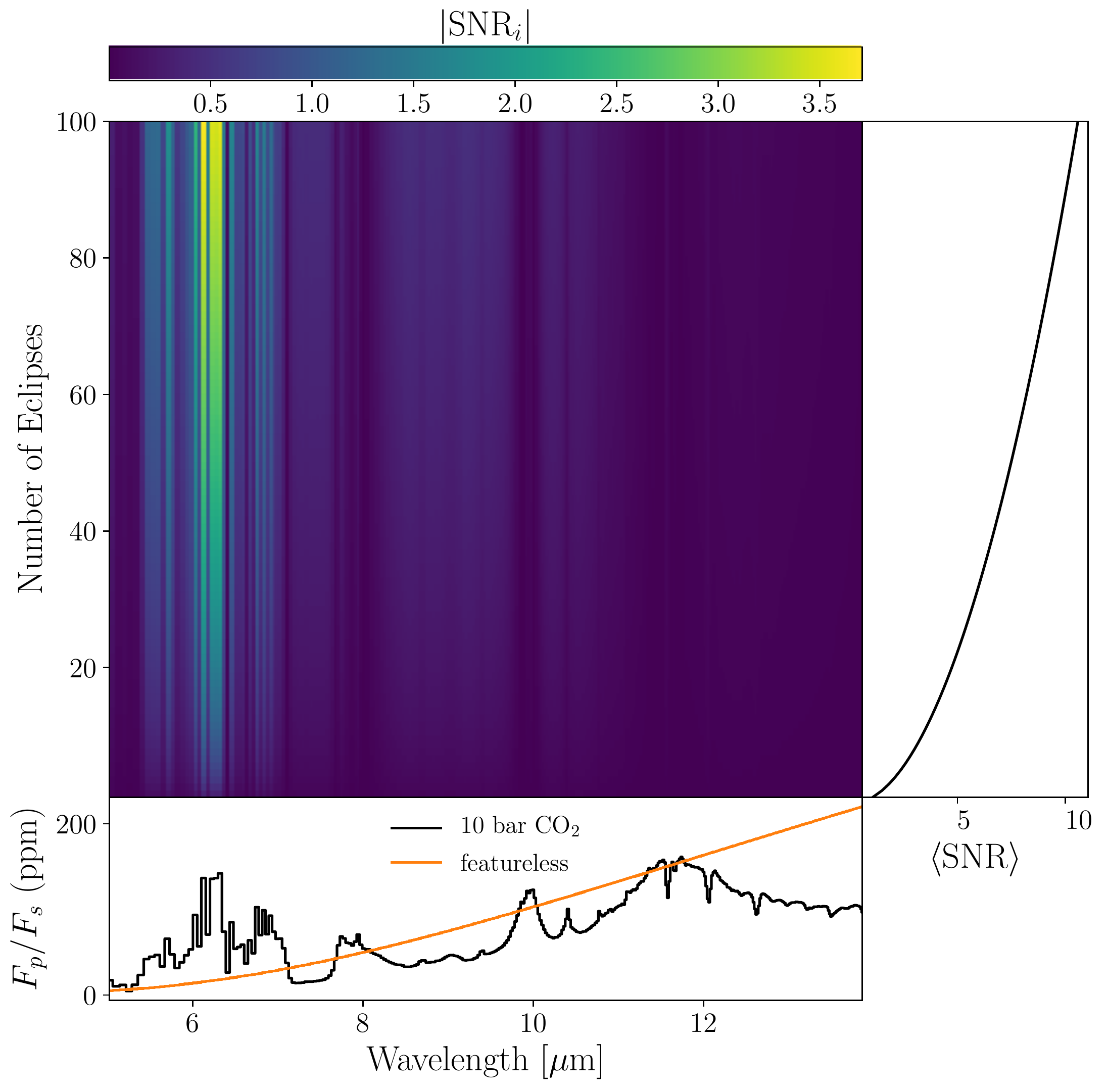}
\caption{Signal-to-noise contours on the simulated 10 bar high CO$_2$ emission spectrum observed with JWST/MIRI/LRS relative to a featureless spectrum as a function of the number of occultations observed and wavelength of the instrument. The bottom panel shows the noiseless spectrum (black) and the best-fitting featureless spectrum (orange), both convolved to the instrument resolution. The right panel shows how the total expected SNR from Equation \ref{eqn:SNR_delta_chi2} increases with more eclipses observed.}
\label{fig:featureless_emission}
\end{figure}

Figures \ref{fig:featureless_transit} and \ref{fig:featureless_emission} provide a specific example of a test for the presence of a 10-bar \ce{CO2} atmosphere, using both transmission (with NIRSPEC Prism*) and emission (with MIRI LRS) spectroscopy, respectively. 
The bottom panel of each figure shows a direct comparison between the template spectrum and the featureless model that best fits the template spectrum. The color-contours show the magnitude of difference between the two spectra in the bottom panel divided by the noise ($| \mathrm{SNR}_i |$) as a function of wavelength and number of occultations. The right panels show the total expected SNR, $\left < \mathrm{SNR} \right >$, as a function of number of occultations, which is the total signal of the atmosphere over the wavelength range of the instrument. The quadrature sum over wavelength not only allows for a comparison between different instruments that naturally accounts for the native resolution of and noise incurred by the instrument, but also a comparison between transmission and emission spectroscopy. Comparing Figures \ref{fig:featureless_transit} and \ref{fig:featureless_emission} we see that if TRAPPIST-1b possesses a 10 bar high CO$_2$ atmosphere, detecting that atmosphere by molecular features in the spectrum will require fewer transits with NIRSpec Prism* than secondary eclipses with MIRI LRS. This example demonstrates how we identify optimal observing modes for atmosphere detections, and enables a comprehensive study to determine the exposure times needed detect the presence of an atmosphere as a function of observing mode and atmosphere type. 

Unless otherwise stated, throughout the rest of the paper we adopt the convention that an atmosphere is detected if  $\left < \mathrm{SNR} \right > \ge 5$ is achieved on absorption features in the spectrum. We report our results for the detectability of atmospheres at this threshold, but encourage readers to scale our results to their own desired detection thresholds.

\subsubsection{Detecting Specific Molecules}

To detect individual molecules in the spectrum, we apply the methods described in the previous sections to spectra with and without the absorption features from a given molecular species.
To perform these tests we generate additional transmission and emission spectra by running our radiative transfer model for a given atmosphere with each spectrally active molecule removed one at a time. We then use the spectra that are missing contributions from individual molecules as $m_2$ in Equation \ref{eqn:delta_chi2_1} and \ref{eqn:SNR_delta_chi2} to calculate the $\left < \mathrm{SNR} \right >$ on the contribution from each molecule to the spectrum. 
In this case, rather than a ``flat line'' test, where we attempt to rule out a featureless spectrum and thereby detect the presence of an atmosphere, here we attempt to rule out a spectrum that does not have a particular gas---for instance \ce{H2O}---and thereby detect the presence of \ce{H2O}.
This procedure enables the identification of which JWST instruments and modes are sensitive to detecting individual molecules in an observed spectrum and how much time must be spent on any given target to reduce the noise enough to measure the spectral contributions from each molecule. 

\begin{figure}[tb]
\centering
\includegraphics[width=0.49\textwidth]{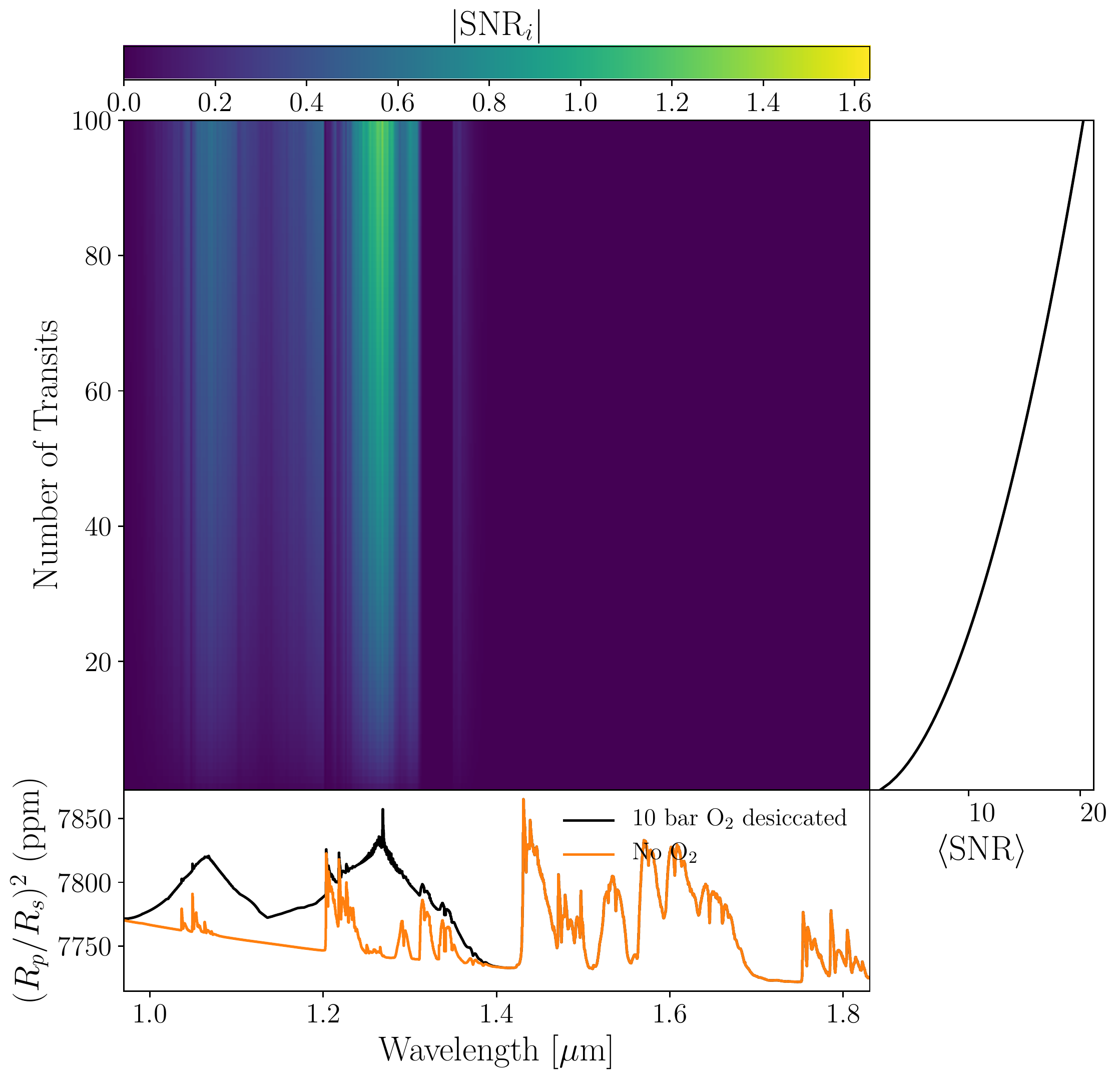}
\caption{Signal-to-noise contours on the O$_2$ contribution to the transmission spectrum of TRAPPIST-1b if it possesses a 10 bar desiccated high O$_2$ atmosphere observed with NIRSpec G140H as a function of the number of occultations observed and wavelength of the instrument. The bottom panel shows the full model spectrum (black) and the model spectrum with O$_2$ removed (orange), both convolved to the resolution of NIRSpec G140H. The right panel shows the total expected SNR ($\left < \mathrm{SNR} \right >$) from Equation \ref{eqn:SNR_delta_chi2}. The O$_2$ features at 1.06 and 1.27 $\mu$m are due to O$_2$-O$_2$ collisionally-induced absorption (CIA), and could lead to $\left < \mathrm{SNR} \right > = 5$ detection of O$_2$ in ${\sim}7$ transits.}
\label{fig:T-1b_O2sensitivity_nirspecG140}
\end{figure}

Figure \ref{fig:T-1b_O2sensitivity_nirspecG140} shows the detectability of O$_2$ in the atmosphere of TRAPPIST-1b with NIRSpec G140H if the planet possesses a desiccated 10 bar O$_2$ atmosphere. The strong O$_2$-O$_2$ collisionally-induced absorption (CIA) features at 1.06 and 1.27 $\mu$m lead to a $\left < \mathrm{SNR} \right > = 5$ detection of O$_2$ in 7 transits. Therefore, an oxygen dominated atmosphere for TRAPPIST-1b could be \textit{ruled out} by not detecting these features in 7 transits. 

Unless otherwise stated, throughout the rest of the paper we adopt the convention that molecules in the  atmosphere are weakly detected if $\left < \mathrm{SNR} \right > \ge 3$ is achieved on that molecule's contribution to the spectrum, and we report our results for the characterization of atmospheres at this weak detection threshold. Keep in mind that we use a weaker threshold $\left < \mathrm{SNR} \right >$ to report detecting individual molecules than for simply detecting the atmosphere, but we encourage readers to scale our molecular detection results to their own desired thresholds.

\section{Results} \label{sec:results}

Here we present the full results of our simulations on the detectability and characterization of the TRAPPIST-1 exoplanet atmospheres using JWST. 
First, we assess the JWST observations needed to detect the presence of an atmosphere for the TRAPPIST-1 planets ({\S}\ref{sec:results:atmospheres}). We then address the detectability of individual molecules within the TRAPPIST-1 planet spectra that may be used to distinguish between different atmospheric states and evolutionary scenarios ({\S}\ref{sec:results:detect_molecules}). 

\subsection{Detecting Atmospheres}
\label{sec:results:atmospheres}

We simulate the detectability of the TRAPPIST-1 planetary atmospheres with MIRI photometry ({\S}\ref{sec:results:photometry}), transmission spectroscopy ({\S}\ref{sec:results:ntran_featureless}), and emission spectroscopy ({\S}\ref{sec:results:neclip_featureless}). 

\subsubsection{JWST/MIRI Photometry}
\label{sec:results:photometry}

MIRI photometry may be advantageous for initial assessments prior to the potentially long time commitment necessary to observe the spectrum of the Earth-sized TRAPPIST-1 planets with JWST. We investigate both transit and eclipse photometry with the nine MIRI photometric filters spanning wavelengths from about 5 to 27 $\mu$m \citep{Bouchet2015}. We first present results that may help to constrain the presence of an atmosphere from secondary eclipse observations in a \textit{single} MIRI photometric band using brightness temperature arguments. We then present results for transit and eclipse observations in multiple filters, using the tests described in {\S\ref{sec:results:detect_spec}}, to assess the observational requirements for ruling out a featureless spectrum. 

\paragraph{Single-Band Constraints}

\begin{figure*}[t!]
\centering
\includegraphics[width=0.97\textwidth]{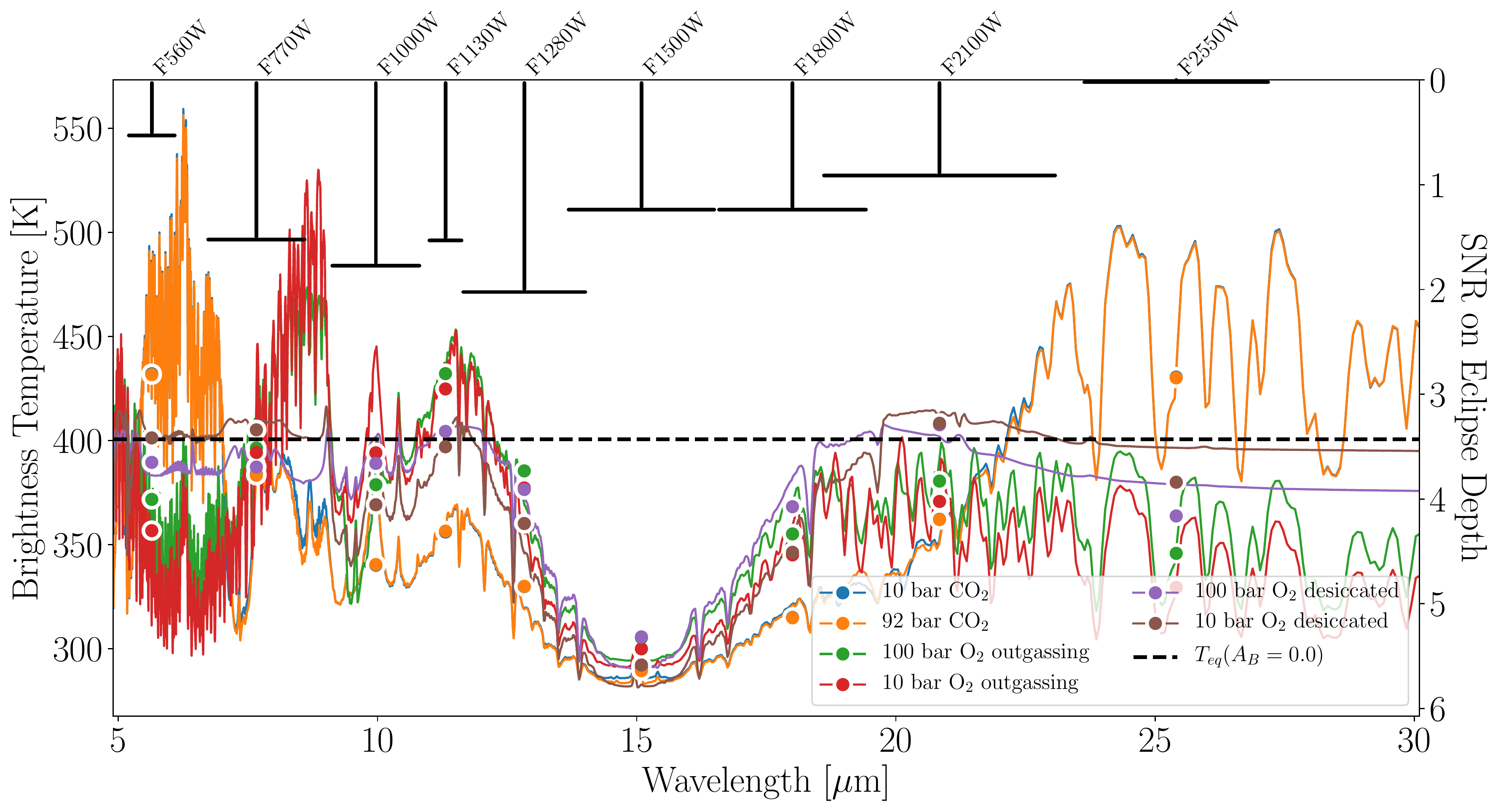}
\caption{Brightness temperatures for different TRAPPIST-1b atmospheric models across the JWST/MIRI imaging wavelength range. The average brightness temperature integrated across each MIRI filter are shown as color points.  The zero bond albedo equilibrium temperature of TRAPPIST-1b (black horizontal dashed line) is shown to compare against the model brightness temperatures, which vary as a function of wavelength due to the interplay between atmospheric temperature structure and gaseous opacities. The black lines from the top of the plot correspond to the (atmospheric model averaged) SNR on a single secondary eclipse, shown increasing from the top of the right y-axis to the bottom. These SNR indicators are shown capped with the effective width of each MIRI filter, and help to identify which filters may best offer secondary eclipse detections.}
\label{fig:T-1b_miri_brightness_temps}
\end{figure*}

We calculate brightness temperatures for each of our self-consistent atmosphere models \citep{Lincowski2018} to help plan and interpret photometric assessments of the TRAPPIST-1 planets in secondary eclipse. 
Figure \ref{fig:T-1b_miri_brightness_temps} shows brightness temperature as a function of wavelength for TRAPPIST-1b for different assumed atmospheres. The wavelength-dependent fluxes were also convolved with the nine MIRI filters to calculate the brightness temperature of each atmosphere model, shown as color points in Fig. \ref{fig:T-1b_miri_brightness_temps}. The horizontal dashed line shows the zero bond albedo equilibrium temperature of TRAPPIST-1b---a limit which a planet without additional internal geothermal or atmospheric greenhouse heating would not be expected to exceed. Figure \ref{fig:T-1b_miri_brightness_temps} also provides the SNR on the depth of a single observed secondary eclipse, averaged over the atmospheric models and displayed increasing from top to bottom on the right y-axis. 

A few of the possible TRAPPIST-1b atmospheres, in particular the Venus-like and outgassing \ce{O2} atmospheres, have brightness temperatures that exceed the zero bond albedo equilibrium temperature in a handful of the MIRI photometric bands. Like Venus, the 10 and 92 bar CO$_2$ atmospheres have 6 $\mu$m windows that provide a glimpse into their hotter, greenhouse heated, lower atmospheres. The F560W MIRI filter could potentially detect this emission. The 10 and 100 bar O$_2$ outgassing atmospheres also have strong emission windows near 11.5 $\mu$m, between the 9.6 $\mu$m O$_3$ band and the 15 $\mu$m CO$_2$ band, which could be detected with the F1130W MIRI filter. The 10 and 92 bar CO$_2$ atmospheres have sufficiently strong CO$_2$ absorption to saturate the wings of the 15 $\mu$m band and cause a significantly lower brightness temperature at 10-12 $\mu$m compared to atmospheres not dominated by CO$_2$. The 10 and 92 bar CO$_2$ atmospheres also exceed the maximum equilibrium brightness temperature beyond about 22 $\mu$m, however the F2550W MIRI filter may lack the SNR to provide constraining information. 

Despite the few cases with potentially detectable high brightness temperatures, the overwhelming majority of atmospheric models viewed through MIRI filters have brightness temperatures that are consistent with a plausible planetary bond albedos (between 0 and 1), including in the 12-18 $\mu$m wavelength range where outgoing thermal radiation is strongly absorbed by the broad 15 $\mu$m \ce{CO2} band. 
In the atmospheres that we considered, no \textit{single} photometric band stands out above the rest as providing a definitive detection of an atmosphere via an emission window.
However, using secondary eclipse observations in multiple MIRI filters in and out of the strong 15 $\mu$m CO$_2$ feature may be used to detect the presence of an atmosphere. 
We discuss this point later in this section.  

Although Figure \ref{fig:T-1b_miri_brightness_temps} only shows brightness temperatures for TRAPPIST-1b, Table~\ref{tab:bright_temps} in Appendix \ref{app:bright_temp} lists the brightness temperature in each MIRI imaging filter for each TRAPPIST-1 planet atmosphere considered here. Table~\ref{tab:bright_temps} also contains calculations for the Warm Spitzer photometric filters. We note, however, that the SNR on secondary eclipses decreases rapidly with planet equilibrium temperature, making precise eclipse photometry beyond TRAPPIST-1c largely infeasible with JWST. 

\paragraph{Multi-Band Constraints} 

We now turn to constraints that may be placed on the existence of an atmosphere using a combination of any of the nine MIRI photometric bands to observe either transits or eclipses. 
For each type of atmosphere, we determine the number and set of MIRI filters that can detect the atmosphere using transit and eclipse photometry in the minimum number of total occultations, assuming an equal number of occultations are observed in each filter. In all cases, 2-3 MIRI filters is optimal for detecting deviations from a featureless spectrum and additional filters are costly given their marginal increase in atmospheric detectability. The F1500W filter is always optimal to include due to the presence of \ce{CO2} in these atmospheres. For transit photometry, the F1500W filter is best combined with F560W for \ce{CO2}-dominated atmospheres and F770W for \ce{O2}-dominated atmospheres, and typically just 2 filters is optimal. For eclipse photometry, the F1500W filter is best combined with F560W, F770W, and/or F1130W, and typically 3 filters is optimal.

\begin{figure*}[t!]
\centering
\includegraphics[width=0.47\textwidth]{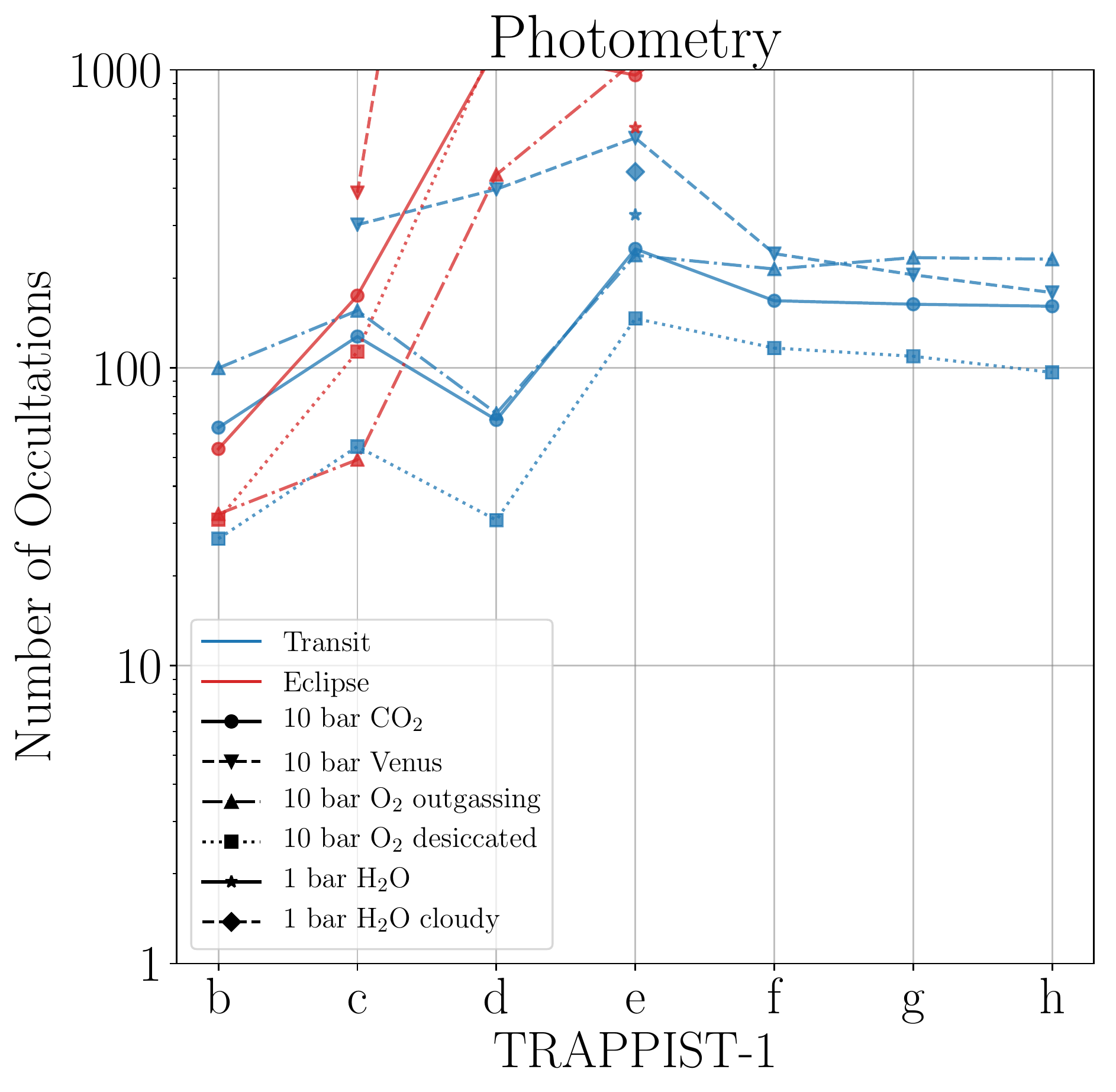}
\includegraphics[width=0.47\textwidth]{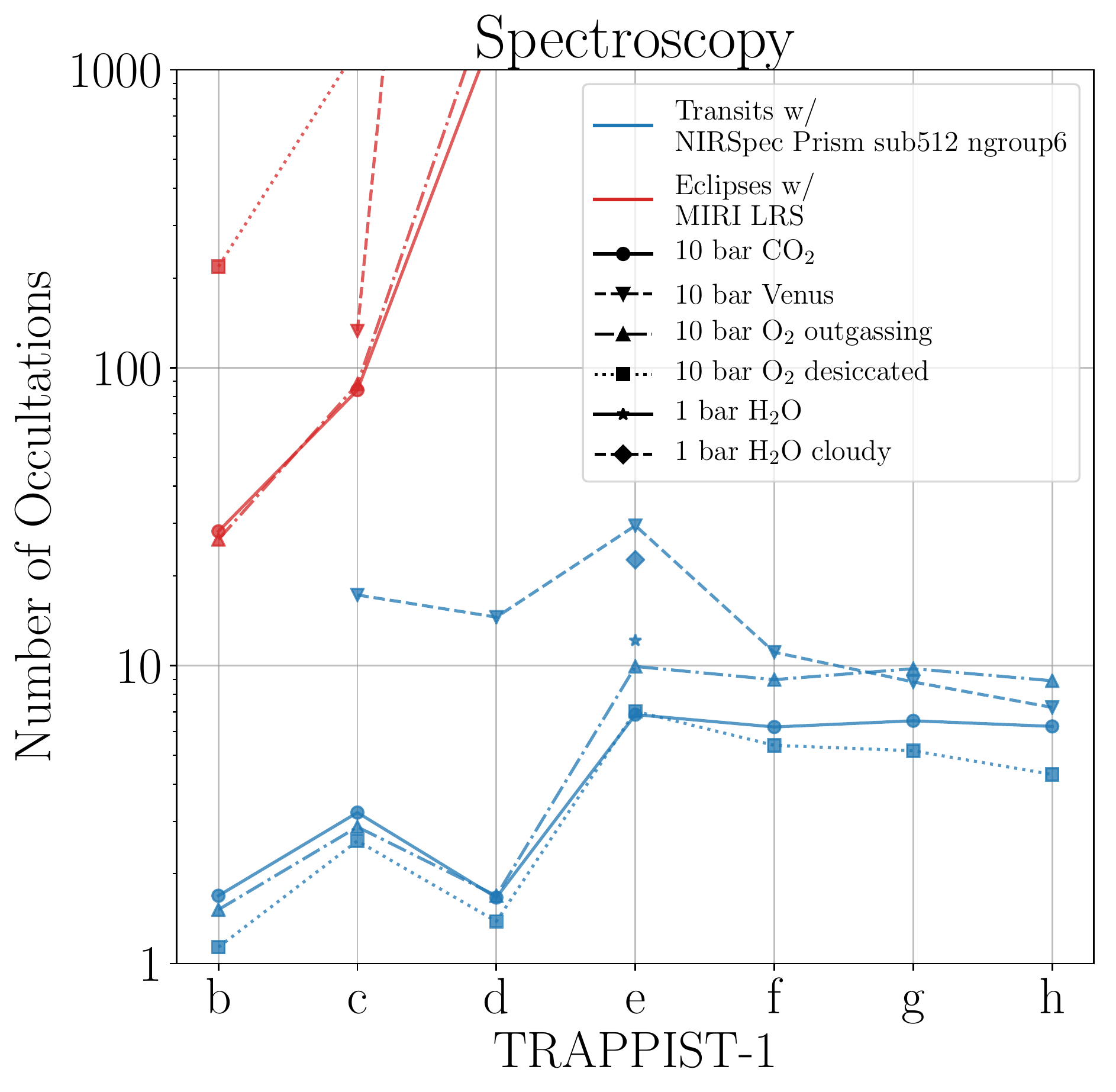}
\caption{Number of transits (blue) or eclipses (red) needed to detect an atmosphere for all seven known TRAPPIST-1 planets with MIRI photometry (left) and JWST spectroscopy (right). Line and marker styles designate the type of atmosphere assumed. 
MIRI imaging assumes that the number of occultations are split evenly between the 2-3 photometric filters that are optimal to detect each atmosphere (see \S\ref{sec:results:photometry}).
Transmission spectroscopy is shown using NIRSpec Prism* (see \S\ref{sec:results:ntran_featureless}) and emission spectroscopy is shown using MIRI LRS (see \S\ref{sec:results:neclip_featureless}). Detecting the atmospheres of the TRAPPIST-1 planets is optimal using transmission spectroscopy with NIRSpec Prism across a range of terrestrial atmospheric compositions.}
\label{fig:miri_imaging_nocc}
\end{figure*}

The left panel of Fig. \ref{fig:miri_imaging_nocc} shows the total number of transits (blue lines) and eclipses (red lines) needed to detect different atmospheric compositions for each of the seven known TRAPPIST-1 planets using the optimal 2 or 3 MIRI photometric filters. Plotting the number of occultations (transits or eclipses) as a function of the TRAPPIST-1 planets, ordered by semi-major axis, reveals a general trend according to the observation method: emission photometry is comparable with transmission photometry at detecting atmospheres for the innermost/hottest planets (e.g. TRAPPIST-1b and c), but becomes increasingly less efficient as the planets decrease in equilibrium temperature; whereas transmission photometry increases in observational time much more gradually with equilibrium temperature. This strong scaling with temperature occurs because the planet emission, at wavelengths contributing most to the detection of molecular features, is not in the Rayleigh-Jeans limit. The left panel of Fig. \ref{fig:miri_imaging_nocc} implies that determining whether or not the potentially habitable TRAPPIST-1 planets (e, f, and g) have atmospheres will be much more efficient with transit photometry than eclipse photometry. 

The high CO$_2$ atmospheres may be surprisingly difficult to distinguish with eclipse photometry because the wings of the 15 $\mu$m CO$_2$ band saturate and extend many microns on either side. 
This strong absorption effectively mutes the planet's emitted flux, creating a nearly featureless spectrum.
Consequently, MIRI eclipse photometry at 12.8 $\mu$m and 18 $\mu$m (F1280W and F1800W filters, respectively) may not be sufficiently separated from 15 $\mu$m (F1500W) to avoid substantial contamination from the wings of strong \ce{CO2} absorption (see the orange curve in Fig. \ref{fig:T-1b_miri_brightness_temps}), making these filters ineffective for continuum measurement. Instead, the F1130W filter may be better at probing deeper into the atmosphere.

\subsubsection{JWST Transmission Spectroscopy}
\label{sec:results:ntran_featureless}

We now present results on the detectability of the TRAPPIST-1 planet atmospheres using transmission spectroscopy with JWST. 
We begin with a few specific examples of detectable TRAPPIST-1 atmospheres to demonstrate the size of molecular absorption features relative to the expected JWST noise, and then present our full ensemble of results for each TRAPPIST-1 planet, atmosphere, and JWST instrument.  

\paragraph{Specific Examples of Atmospheric Detection}

\begin{figure}[t!]
\centering
\includegraphics[width=0.47\textwidth]{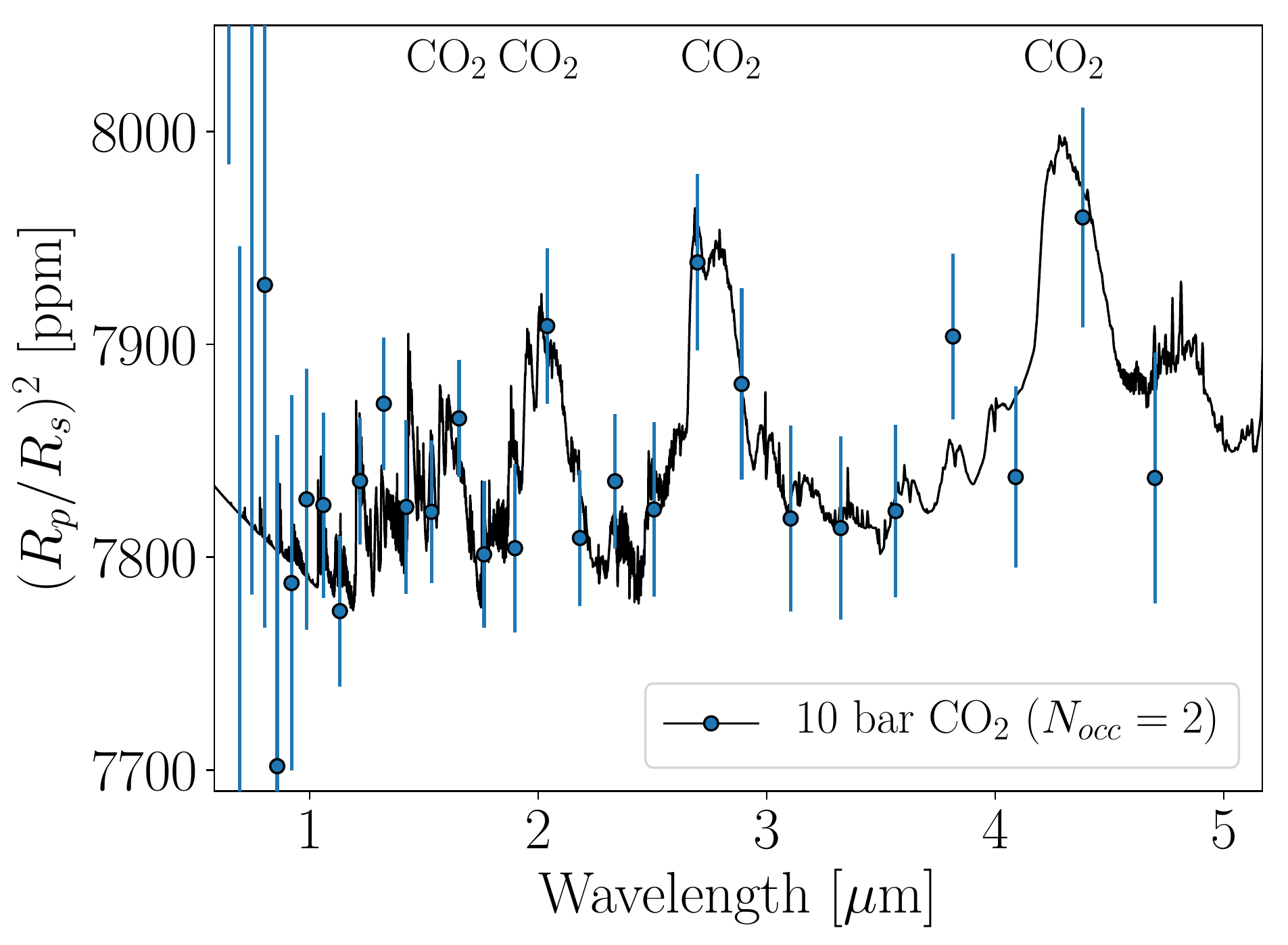}
\caption{Model transmission spectrum of TRAPPIST-1b with a 10 bar CO$_2$ atmosphere. Synthetic data are simulated for two transits observed with NIRSpec Prism* and binned to a resolution of $R=8$.}
\label{fig:T1b_SNR5_example}
\end{figure}

Figure \ref{fig:T1b_SNR5_example} shows an example detection of absorption features in the transmission spectrum of TRAPPIST-1 b assuming it possesses a clear 10 bar \ce{CO2} atmosphere. The transmission spectrum is shown with synthetic data simulated for two transits observed with NIRSpec Prism*, which we find to be sufficient to rule out a featureless spectrum with $\left < \mathrm{SNR} \right > = 5$ (our fiducial detection limit). The synthetic data are shown binned to a resolution of $R=8$, however the featureless spectrum was ruled out at the native resolution of the NIRSpec Prism ($R \approx 100$). CO$_2$ absorption features at 1.6, 2.0, 2.8, 4.3 $\mu$m drive the detectability of this atmosphere and are apparent in the synthetic data.  

\begin{figure*}[t!]
\centering
\includegraphics[width=0.97\textwidth]{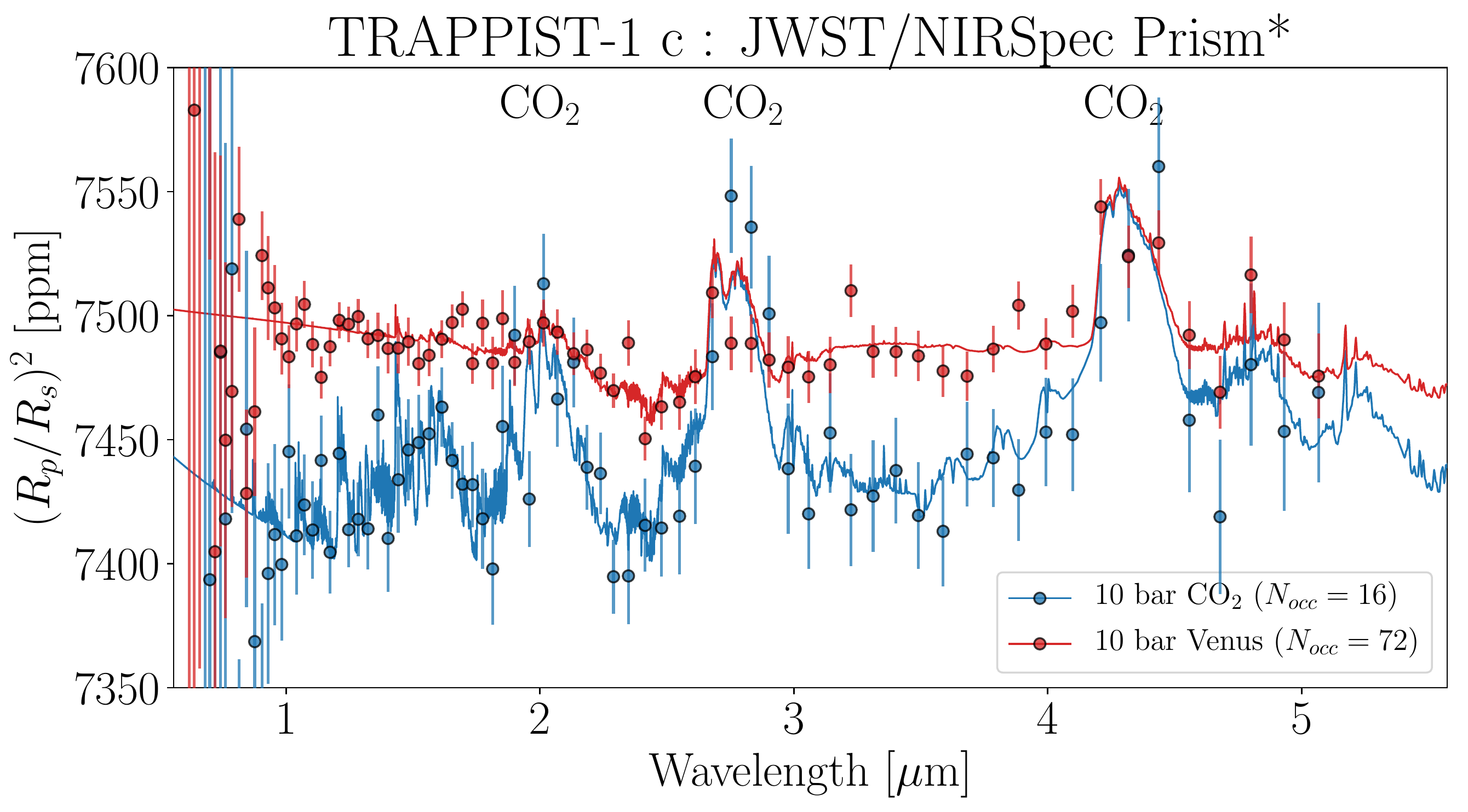}
\caption{Simulated JWST/NIRSpec Prism transmission spectra of TRAPPIST-1c for different CO$_2$-dominated atmospheric states. The blue model shows a clear sky spectrum with large CO$_2$ absorption features and the red model includes a self-consistent H$_2$SO$_4$ cloud. The number of coadded transits simulated for each NIRSpec observation are set so that the atmosphere is strongly detected ($\left < \mathrm{SNR} \right > = 10$). That is, to obtain approximately equal constraints on the presence and composition of TRAPPIST-1c's atmosphere will require 16 transits for the clear sky CO$_2$ case and 72 transits for the cloudy case.}
\label{fig:T1c_nirspec_clouds}
\end{figure*}

Figure \ref{fig:T1c_nirspec_clouds} demonstrates possible transmission spectra of TRAPPIST-1 c for two different aerosol conditions. Both cases are CO$_2$-dominated spectral models from \citet{Lincowski2018} with simulated noise for observations with NIRSpec Prism*. The number of transits observed for each model is calculated such that the atmosphere is strongly detected with $\left < \mathrm{SNR} \right > = 10$. The blue model shows a clear sky spectrum with large CO$_2$ absorption features and data uncertainties calculated for 16 observed transits. The red model includes H$_2$SO$_4$ clouds at altitudes consistent with H$_2$SO$_4$ condensation, with data uncertainties calculated for 72 observed transits. 

The aerosol-free atmosphere has strong CO$_2$ features that can be detected in considerably fewer transits than the case with Venus-like H$_2$SO$_4$ aerosols. The self-consistent clouds are located high enough in the atmosphere that over 70 transits are required to detect the CO$_2$ absorption features with the same confidence as the clear sky case. If TRAPPIST-1 c possess Venus-like aerosols, then about 5$\times$ more transits must be observed to achieve comparable constraints on the presence and composition of TRAPPIST-1 c's atmosphere, than our estimates for the clear-sky case.  

\begin{figure*}[t!]
\centering
\includegraphics[width=0.97\textwidth]{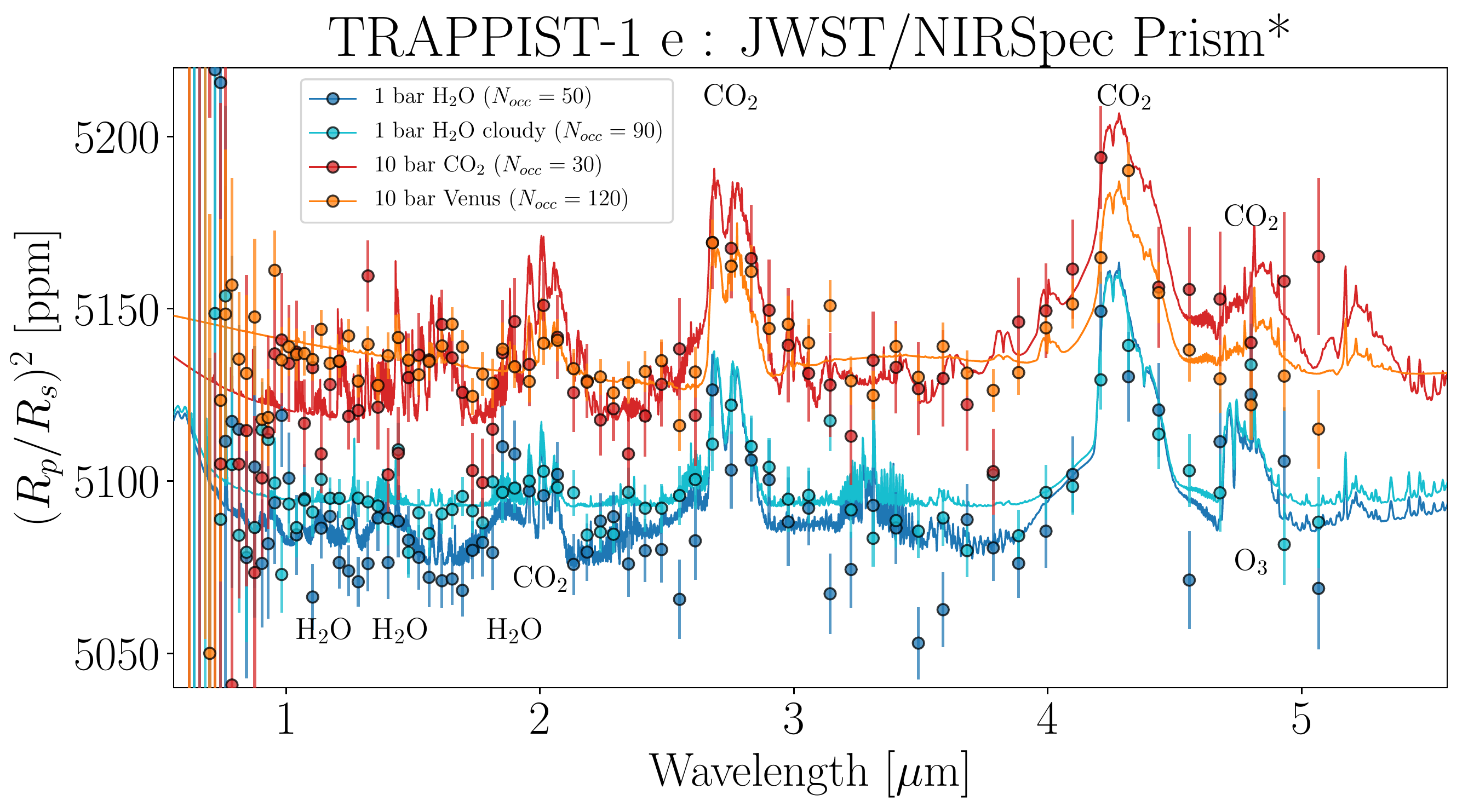}
\caption{Simulated JWST/NIRSpec Prism* transmission spectra of TRAPPIST-1e for different possible atmospheric compositions. The teal and blue models show spectra for water-dominated atmospheres, with and without a water cloud, respectively. The orange and red spectra are for CO$_2$-dominated atmospheres, with and without H$_2$SO$_4$ aerosols, respectively. The number of coadded transits simulated for each NIRSpec observation are set so that the atmosphere is strongly detected ($\left < \mathrm{SNR} \right > = 10$). That is, to obtain approximately equal constraints on the presence of TRAPPIST-1e's atmosphere will require ${\sim}$30 transits for a clear sky CO$_2$ atmosphere, ${\sim}$50 transits for a clear sky H$_2$O atmosphere, ${\sim}$90 transits for a cloudy H$_2$O atmosphere, and ${\sim}$120 transits for a CO$_2$-dominated Venus-like atmosphere.}
\label{fig:T1e_nirspec}
\end{figure*} 

We find a factor of ${\sim}4$ variation in the number of transits needed to detect the atmosphere of TRAPPIST-1e depending on the type of terrestrial atmosphere it possesses. Figure \ref{fig:T1e_nirspec} explores the detectability of molecular features in the transmission spectrum of TRAPPIST-1 e. Like Fig. \ref{fig:T1c_nirspec_clouds}, Fig. \ref{fig:T1e_nirspec} shows possible TRAPPIST-1 e model spectra with simulated NIRSpec Prism* observations assuming the number of transits needed to strongly detect features in each atmosphere with $\left < \mathrm{SNR} \right > \approx 10$. Spectra for water-covered environments with and without a water cloud are shown in teal and blue, respectively, and CO$_2$-dominated atmospheres, with and without H$_2$SO$_4$ aerosols are shown in orange and red, respectively.
To detect features in each spectrum with $\left < \mathrm{SNR} \right > \approx 10$ ($ \left < \mathrm{SNR} \right > \approx 5$), and thereby obtain approximately equal constraints on the presence of TRAPPIST-1e's atmosphere, will require ${\sim}$30 (${\sim}$7) transits for a clear sky CO$_2$ atmosphere, ${\sim}$50 (${\sim}$13) transits for a clear sky aqua planet atmosphere, ${\sim}$90 (${\sim}$22) transits for a cloudy aqua planet atmosphere, and ${\sim}$120 (${\sim}$30) transits for a CO$_2$-dominated atmosphere with \ce{H2SO4} clouds. 

\begin{figure}[t!]
\centering
\includegraphics[width=0.47\textwidth]{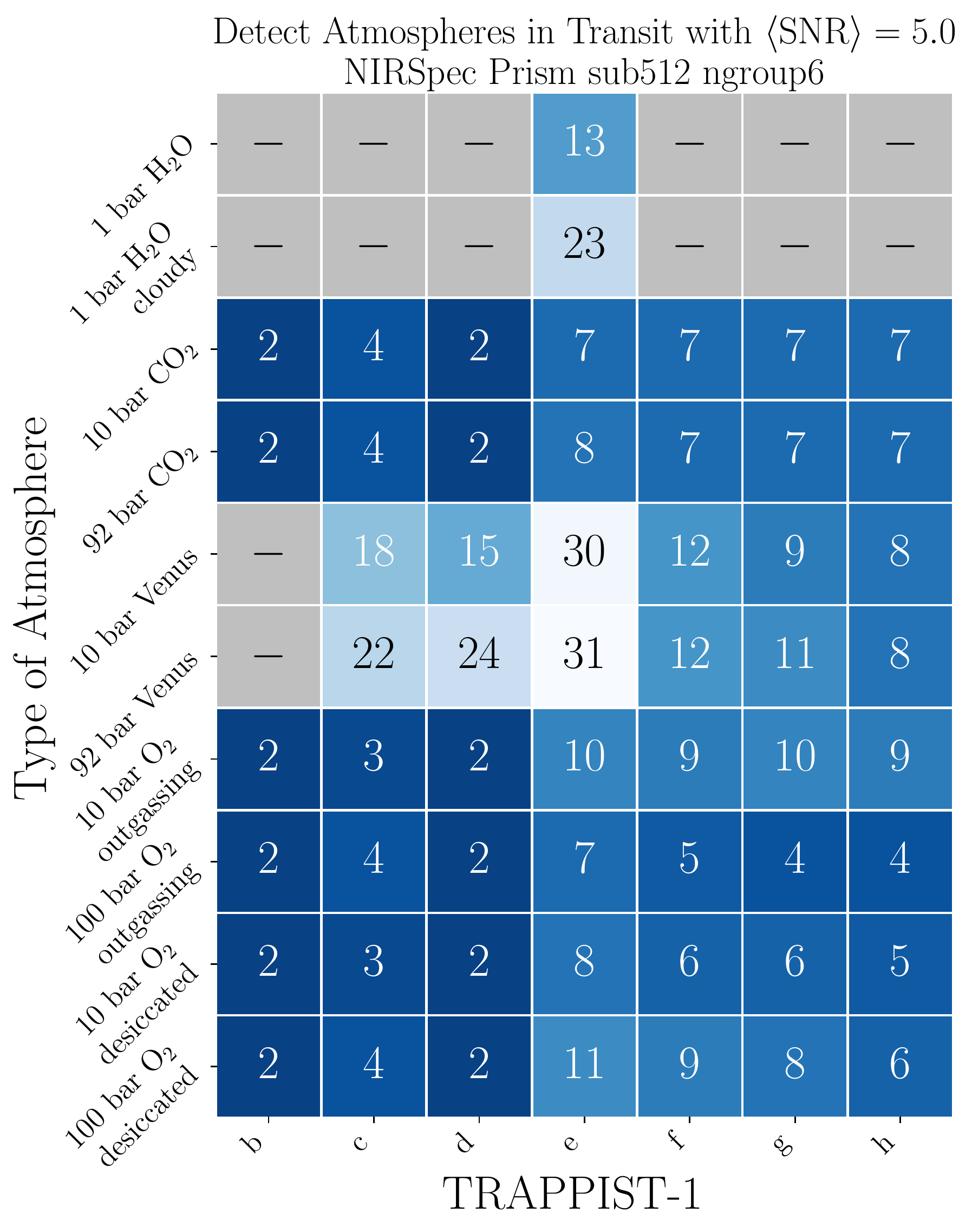}
\caption{Number of transits for each TRAPPIST-1 planet necessary to rule out a featureless spectrum with $\left < \mathrm{SNR} \right > = 5$ for different self-consistent atmospheric compositions using JWST NIRSpec Prism with the optimized readout mode of \citet{Batalha2018}.}
\label{fig:colortable_featureless_transit_summary}
\end{figure} 

\paragraph{Optimal JWST Observing Modes for Atmospheric Detection}

Thus far we have only discussed specific cases for the detectability of the TRAPPIST-1 planet atmospheres with JWST transmission spectroscopy; we now report results from our comprehensive study to determine the exposure times needed detect the presence of atmospheres as a function of JWST observing mode and atmosphere type, for each TRAPPIST-1 planet.  
The number of transits necessary to detect spectral features in the transmission spectrum with $\left < \mathrm{SNR} \right > = 5$ for TRAPPIST-1 b, c, d, e, f, g, and h are shown in Appendix \ref{app:atmos_detect} in Figures \ref{fig:colortable_featureless_T-1b_transit},  \ref{fig:colortable_featureless_T-1c_transit},  \ref{fig:colortable_featureless_T-1d_transit},  \ref{fig:colortable_featureless_T-1e_transit},  \ref{fig:colortable_featureless_T-1f_transit},  \ref{fig:colortable_featureless_T-1g_transit}, and  \ref{fig:colortable_featureless_T-1h_transit}, respectively, as a function of both atmospheric compositions and JWST instrument/mode. 
Color is used in these figures to guide the eye to small (more blue) and large (more white) values for the number of transits. 

These results suggest that under most circumstances the JWST/NIRSpec Prism is the optimal instrument for detecting the presence of an atmosphere using transmission spectroscopy of the TRAPPIST-1 planets. 
If a partial saturation strategy is used with the NIRSpec Prism \citep{Batalha2018}, then it will be more capable of detecting atmospheric features than any other JWST instrument or mode. 
However, if it turns out that the systematics introduced via partial saturation are not beneficial, the SUB512S subarray will offer improved performance over the SUB512 subarray due to its shorter readout time.   Although we note that SUB512 should be considered the subarray of choice because SUB512s has limited access to important background pixels and it may be difficult to keep the curved trace inside only 16 pixels in the cross-dispersion direction. 
The NIRSpec G395M/H disperser offers comparable results with the standard (no partial saturation) NIRSpec Prism.

Most of the atmospheres that we consider may require fewer than 12 transits to detect for all of the TRAPPIST-1 planets using the optimized NIRSpec Prism mode.
Figure \ref{fig:colortable_featureless_transit_summary} displays the number of transits to detect the TRAPPIST-1 planet atmospheres using only the NIRSpec Prism* instrument for each TRAPPIST-1 planet and for each atmosphere considered in this work.  

Our results show that the outer TRAPPIST-1 planets require only 2-7 more transits than TRAPPIST-1 b to detect clear-sky atmospheres.
The right panel of Fig. \ref{fig:miri_imaging_nocc} summarizes the best-case-scenario results for detecting the atmospheres of the TRAPPIST-1 planets with JWST transit spectroscopy (blue lines). The number of transits required to detect the atmosphere at $\left < \mathrm{SNR} \right > = 5$ for each of the atmospheric models is shown for observations with NIRSpec Prism*. 
Like our transit photometry results, the number of transits only weakly increases with semi-major axis. 
However, significantly fewer transits are required for transmission spectroscopy atmosphere detections with NIRSpec Prism* than with MIRI filter photometry. The emission spectroscopy results in Fig. \ref{fig:miri_imaging_nocc} (red lines) are discussed in \S\ref{sec:results:neclip_featureless}. 

The atmospheric composition of the planets has a relatively minimal effect on the number of transits needed to rule out a featureless spectrum, with Venus-like clouds being a significant exception. All atmospheric cases considered for TRAPPIST-1 c could be detected in 3-4 transits with NIRSpec Prism*, except for the 10 and 92 bar Venus atmospheres (with \ce{H2SO4} aerosols), which would require $4.5 {\times}$ and $5.5 {\times}$ the number of transits to detect, respectively, compared to the 10 and 92 bar clear-sky \ce{CO2} counterparts. In general, more transits tend to be required to detect the atmospheres of the cooler worlds, except for TRAPPIST-1 d, which is more comparable to TRAPPIST-1 b in the detectability of its atmosphere. While the atmosphere of TRAPPIST-1 d may be relatively easy to detect if it is without clouds, with clouds the 10 and 92 bar \ce{CO2} atmospheres require $7.5 {\times}$ and $12 {\times}$ the number of transits to detect, respectively, the largest increase due to clouds seen in the sample. Beyond TRAPPIST-1 e the effect of Venus-like clouds has a diminished impact on the atmospheric detectability, with fewer than $2 {\times}$ the number of transits required to detect the atmospheres for TRAPPIST-1 f, g, and h if they have clouds. 

Note that the numerical values in Figure \ref{fig:colortable_featureless_transit_summary} and Figures  \ref{fig:colortable_featureless_T-1b_transit} - \ref{fig:colortable_featureless_T-1h_transit} can easily be scaled to higher or lower $\left < \mathrm{SNR} \right >$ thresholds that more or less confidently rule out a featureless spectrum. Since the SNR on an observation (and $\left < \mathrm{SNR} \right >$) scales with the square-root of the exposure time, and the number of occultations is a proxy for exposure time, we can obtain a new value for the number of occultations:   
\begin{equation}
    N_{\mathrm{occ}}' = N_{\mathrm{occ}} \left ( \frac{\left < \mathrm{SNR} \right >'}{\left < \mathrm{SNR} \right >} \right )^2
\end{equation}
where $N_{\mathrm{occ}}$ is the number of occultations necessary to distinguish features in the spectrum with $\left < \mathrm{SNR} \right >$ (5 in Figures \ref{fig:colortable_featureless_transit_summary},  \ref{fig:colortable_featureless_T-1b_transit} - \ref{fig:colortable_featureless_T-1h_transit}) and $\left < \mathrm{SNR} \right >'$ is the new signal-to-noise threshold. 

The increased transit duration with semi-major axis make our results for the outer TRAPPIST-1 planets appear more optimistic relative to our results for the inner planets. Note that for JWST planning, our reported number of transits/eclipses leads to different telescope time for each TRAPPIST-1 planet since each has a different transit duration. For each observed transit (eclipse) we assumed one transit duration worth of out-of-transit (out-of-eclipse) observing time. Considering the following median transit durations for the TRAPPIST-1 planets from \citet{Grimm2018}: 36.40 mins for b, 42.37 mins for c, 49.13 mins for d, 57.21 mins for e, 62.60 mins for f, 68.40 mins for g, and 76.7 mins for h, one can calculate expected on-target time per transit/eclipse (before overheads) by multiplying the transit duration by two plus any overheads. 

JWST is not always able to point at TRAPPIST-1 due to the star's proximity to the ecliptic plane. As a result, the star will only be observable to JWST for ${\sim} 100$ days per year\footnote{\url{https://jwst-docs.stsci.edu/display/JTI/JWST+Target+Viewing+Constraints}}. Over the course of JWST's nominal 5 year mission the maximum number of observable transits/eclipses is approximately: 331 for b, 206 for c, 123 for d, 81 for e, 54 for f, 40 for g, and 26 for h. Of course, an extended mission lifetime would allow considerably more observations of the TRAPPIST-1 system.  

\subsubsection{JWST Emission Spectroscopy}
\label{sec:results:neclip_featureless}

Unlike transmission spectroscopy, which can make use of many JWST instruments that span a broad wavelength range (e.g. NIRCam, NIRSpec, NIRISS, and MIRI LRS), secondary eclipse spectroscopy of the TRAPPIST-1 planets will only be viable at the longer wavelengths accessible to JWST. This is primarily driven by the increase in eclipse depths with wavelength as the planet-star contrast ratio becomes more favorable.

We find that only MIRI LRS observations of TRAPPIST-1 b and c may potentially rule out a featureless emission spectrum with $\left < \mathrm{SNR} \right > = 5$ in fewer than 100 observed secondary eclipses. 
Using an analogous approach to that shown in Figures \ref{fig:colortable_featureless_T-1b_transit} - \ref{fig:colortable_featureless_T-1h_transit}, MIRI LRS observations of TRAPPIST-1 b may rule out a featureless emission spectrum in 27-47 secondary eclipses if the planet possess a (10 or 92 bar) clear-sky CO$_2$ atmosphere or a (10 or 100 bar) outgassing \ce{O2} atmosphere, with the 10 bar outgassing \ce{O2} atmosphere the most readily detectable. 
Over 100 secondary eclipses are required if the planet possesses a desiccated O$_2$ atmosphere (10 or 100 bars), as these emission spectra appear remarkably featureless between 5-9 $\mu$m. 
For TRAPPIST-1 c we find that MIRI LRS observations could rule out a featureless emission spectrum in 85-100 secondary eclipses if the planet possesses a (10 or 92) bar clear \ce{CO2} atmosphere or a 10 bar O$_2$ atmosphere with outgassing.
Over 100 secondary eclipses are required for all other atmospheric compositions considered for TRAPPIST-1 c, including the \ce{CO2} atmospheres with \ce{H2SO4} clouds, which have emission spectra similar to the clear \ce{CO2} atmospheres, but with reduced temperature contrasts in the absorbing and emitting spectral regions that effectively mute the features and drive the spectrum towards a featureless blackbody. All of the exterior TRAPPIST-1 planets have emission spectra that will appear indistinguishable from cool blackbodies due to insufficient SNR. 

The right panel of Fig. \ref{fig:miri_imaging_nocc} summarizes the best-case-scenario results for detecting the atmospheres of the TRAPPIST-1 planets with JWST emission spectroscopy (red lines) to compare against our filter photometry and transmission spectroscopy results. The number of eclipses required to detect the atmosphere at $\left < \mathrm{SNR} \right > = 5$ for each of the atmospheric models is shown for observations with MIRI LRS. 
Compared with transmission spectroscopy, emission spectroscopy is an inefficient method for \textit{detecting} the TRAPPIST-1 planetary atmospheres, particularly for the cooler planets.  

\begin{figure}[tb]
\centering
\includegraphics[width=0.49\textwidth]{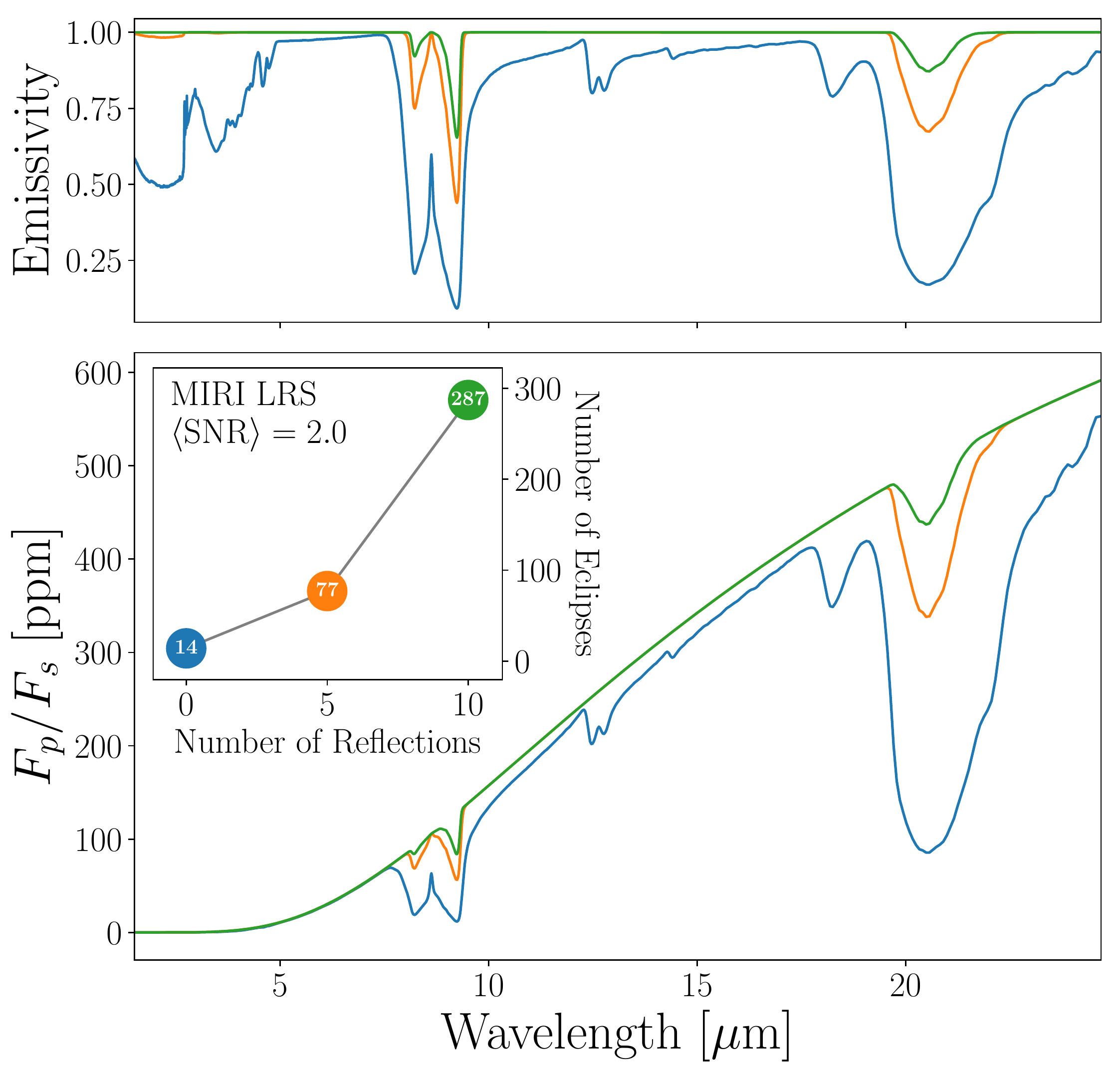}
\caption{Estimated detectability of pure quartz surface emissivity features in the secondary eclipse spectrum of TRAPPIST-1b with JWST's MIRI LRS instrument. The top panel shows the MIR effective emissivity of a quartz surface for three different assumed values for the number of surface reflections (see Equation \ref{eqn:epsilon_e}): 0 (blue), 5 (orange), and 10 (green). The bottom panel shows theoretical secondary eclipse spectra for TRAPPIST-1b assuming the effective emissivities from the top panel. The inset axis shows the number of secondary eclipses that would be required to observe with MIRI LRS to detect features in the emission spectrum with a $\left < \mathrm{SNR} \right > = 2.0$, for each of the three assumed effective emissivities. Only an extremely polished pure quartz surface could possibly be detected with JWST.}
\label{fig:quartz_emission_detectability}
\end{figure}

If TRAPPIST-1b does not possess an atmosphere, emissivity features from minerals on the surface of the planet are not likely to be detectable in secondary eclipse thermal emission spectra. Figure \ref{fig:quartz_emission_detectability} shows the estimated thermal emission spectrum of TRAPPIST-1b if it is airless and possesses a pure quartz surface. Three different effective emissivities are shown in the top panel, corresponding to 0 (blue), 5 (orange), and 10 (green) surface reflections (see Equation \ref{eqn:epsilon_e}). The bottom panel shows the MIR secondary eclipse spectrum for each surface emissivity, assuming the surface is at the (highly optimistic) zero bond albedo equilibrium temperature for TRAPPIST-1b ($392$ K). The inset axis shows the number of secondary eclipses that MIRI LRS would have to observe to detect features in each emission spectrum at a weak $\left < \mathrm{SNR} \right > = 2.0$. Increasing the number of surface reflections results in surface emissivity features that are not detectable. 

\subsection{Distinguishing Specific Molecules}
\label{sec:results:detect_molecules}

We now present the sensitivity of each JWST instrument to each gas in the \citet{Lincowski2018} TRAPPIST-1 model transmission spectra. Table \ref{tab:molec_detect} lists the molecules for which JWST could weakly detect ($\left < \mathrm{SNR} \right > = 3.0$) that molecule's contribution to the spectrum in 100 or fewer transits. For each molecule, the number of transits is listed in parentheses next to each molecular formula along with a footnote identifying which JWST instrument is used for that observation. While in some cases multiple instruments may be sensitive to the same molecule, we list only the instrument that can detect each gas in the minimum number of transits.

\begin{deluxetable}{lll}
\tablefontsize{\scriptsize}
\tablewidth{\columnwidth}
\tablecaption{\label{tab:molec_detect} Detectable molecules with transmission spectroscopy for different plausible TRAPPIST-1 planet atmospheres}
\tablehead{ \colhead{Planet}  &\colhead{Model}  &\colhead{Molecules}\\ 
\colhead{}  &\colhead{}  &\colhead{(Number of transits to $\left < \mathrm{SNR} \right > = 3.0$)}
}
\startdata
T-1b&  10 bar CO$_2$&  CO$_{2}$(1\tablenotemark{k}), CO(47\tablenotemark{k}), H$_{2}$O(44\tablenotemark{f}), SO$_{2}$(40\tablenotemark{f})\\
&  92 bar CO$_2$&  CO$_{2}$(1\tablenotemark{k}), CO(47\tablenotemark{k}), H$_{2}$O(41\tablenotemark{f}), SO$_{2}$(38\tablenotemark{f})\\
&  100 bar O$_2$
outgassing&  CO$_{2}$(1\tablenotemark{k}), H$_{2}$O(2\tablenotemark{k}), O$_{3}$(11\tablenotemark{e}), O$_{2}$(3\tablenotemark{c})\\
&  10 bar O$_2$
outgassing&  CO$_{2}$(1\tablenotemark{k}), H$_{2}$O(1\tablenotemark{k}), O$_{2}$(4\tablenotemark{c})\\
&  100 bar O$_2$
desiccated&  CO$_{2}$(1\tablenotemark{k}), O$_{2}$(3\tablenotemark{c})\\
&  10 bar O$_2$
desiccated&  CO(13\tablenotemark{d}), CO$_{2}$(1\tablenotemark{k}), O$_{3}$(81\tablenotemark{f}), O$_{2}$(3\tablenotemark{c})\\
T-1c&  10 bar CO$_2$&  CO$_{2}$(1\tablenotemark{k}), H$_{2}$O(45\tablenotemark{f}), SO$_{2}$(51\tablenotemark{f})\\
&  92 bar CO$_2$&  CO$_{2}$(1\tablenotemark{k}), H$_{2}$O(43\tablenotemark{f}), SO$_{2}$(52\tablenotemark{f})\\
&  100 bar O$_2$
outgassing&  CO$_{2}$(3\tablenotemark{k}), H$_{2}$O(4\tablenotemark{k}), O$_{3}$(5\tablenotemark{e}), O$_{2}$(7\tablenotemark{c})\\
&  10 bar O$_2$
outgassing&  CO$_{2}$(2\tablenotemark{k}), H$_{2}$O(2\tablenotemark{k}), O$_{3}$(86\tablenotemark{f}), O$_{2}$(5\tablenotemark{c})\\
&  100 bar O$_2$
desiccated&  CO$_{2}$(2\tablenotemark{k}), O$_{3}$(71\tablenotemark{f}), O$_{2}$(7\tablenotemark{c})\\
&  10 bar O$_2$
desiccated&  CO(61\tablenotemark{k}), CO$_{2}$(1\tablenotemark{k}), O$_{3}$(64\tablenotemark{f}), O$_{2}$(6\tablenotemark{c})\\
&  10 bar Venus&  CO$_{2}$(3\tablenotemark{k})\\
&  92 bar Venus&  CO$_{2}$(6\tablenotemark{k})\\
T-1d&  10 bar CO$_2$&  CO$_{2}$(1\tablenotemark{k}), CO(48\tablenotemark{k}), H$_{2}$O(18\tablenotemark{f}), SO$_{2}$(27\tablenotemark{f})\\
&  92 bar CO$_2$&  CO$_{2}$(1\tablenotemark{k}), CO(47\tablenotemark{k}), H$_{2}$O(17\tablenotemark{f}), SO$_{2}$(27\tablenotemark{f})\\
&  100 bar O$_2$
outgassing&  CO$_{2}$(2\tablenotemark{k}), H$_{2}$O(2\tablenotemark{k}), O$_{3}$(2\tablenotemark{e}), O$_{2}$(3\tablenotemark{c})\\
&  10 bar O$_2$
outgassing&  CO$_{2}$(1\tablenotemark{k}), H$_{2}$O(2\tablenotemark{k}), O$_{3}$(32\tablenotemark{f}), O$_{2}$(3\tablenotemark{c})\\
&  100 bar O$_2$
desiccated&  CO$_{2}$(1\tablenotemark{k}), O$_{3}$(26\tablenotemark{f}), O$_{2}$(3\tablenotemark{c})\\
&  10 bar O$_2$
desiccated&  CO(49\tablenotemark{k}), CO$_{2}$(1\tablenotemark{k}), O$_{3}$(30\tablenotemark{f}), O$_{2}$(3\tablenotemark{c})\\
&  10 bar Venus&  CO$_{2}$(6\tablenotemark{k})\\
&  92 bar Venus&  CO$_{2}$(7\tablenotemark{k})\\
T-1e&  10 bar CO$_2$&  CO$_{2}$(2\tablenotemark{k}), SO$_{2}$(86\tablenotemark{f})\\
&  92 bar CO$_2$&  CO$_{2}$(2\tablenotemark{k}), SO$_{2}$(88\tablenotemark{f})\\
&  100 bar O$_2$
outgassing&  CO$_{2}$(9\tablenotemark{k}), H$_{2}$O(64\tablenotemark{k}), O$_{3}$(4\tablenotemark{k}), O$_{2}$(22\tablenotemark{c})\\
&  10 bar O$_2$
outgassing&  CO$_{2}$(4\tablenotemark{k}), H$_{2}$O(78\tablenotemark{k}), O$_{3}$(72\tablenotemark{e}), O$_{2}$(24\tablenotemark{c})\\
&  100 bar O$_2$
desiccated&  CO$_{2}$(4\tablenotemark{k}), O$_{3}$(61\tablenotemark{e}), O$_{2}$(18\tablenotemark{c})\\
&  10 bar O$_2$
desiccated&  CO$_{2}$(3\tablenotemark{k}), O$_{3}$(40\tablenotemark{e}), O$_{2}$(19\tablenotemark{c})\\
&  10 bar Venus&  CO$_{2}$(9\tablenotemark{k})\\
&  92 bar Venus&  CO$_{2}$(9\tablenotemark{k})\\
&  1 bar H$_2$O&  CO$_{2}$(9\tablenotemark{k}), H$_{2}$O(15\tablenotemark{k}), N$_{2}$(68\tablenotemark{e})\\
&  1 bar H$_2$O
cloudy&  CO$_{2}$(13\tablenotemark{k})\\
T-1f&  10 bar CO$_2$&  CO$_{2}$(2\tablenotemark{k}), SO$_{2}$(76\tablenotemark{f})\\
&  92 bar CO$_2$&  CO$_{2}$(2\tablenotemark{k}), SO$_{2}$(79\tablenotemark{f})\\
&  100 bar O$_2$
outgassing&  CO$_{2}$(7\tablenotemark{k}), O$_{3}$(2\tablenotemark{k}), O$_{2}$(23\tablenotemark{c})\\
&  10 bar O$_2$
outgassing&  CO$_{2}$(4\tablenotemark{k}), O$_{3}$(39\tablenotemark{e}), O$_{2}$(30\tablenotemark{c})\\
&  100 bar O$_2$
desiccated&  CO$_{2}$(4\tablenotemark{k}), O$_{3}$(10\tablenotemark{e}), O$_{2}$(20\tablenotemark{c})\\
&  10 bar O$_2$
desiccated&  CO$_{2}$(3\tablenotemark{k}), O$_{3}$(7\tablenotemark{e}), O$_{2}$(20\tablenotemark{c})\\
&  10 bar Venus&  CO$_{2}$(3\tablenotemark{k})\\
&  92 bar Venus&  CO$_{2}$(3\tablenotemark{k})\\
T-1g&  10 bar CO$_2$&  CO$_{2}$(2\tablenotemark{k}), SO$_{2}$(82\tablenotemark{f})\\
&  92 bar CO$_2$&  CO$_{2}$(2\tablenotemark{k}), SO$_{2}$(89\tablenotemark{f})\\
&  100 bar O$_2$
outgassing&  CO$_{2}$(9\tablenotemark{k}), O$_{3}$(2\tablenotemark{k}), O$_{2}$(36\tablenotemark{c})\\
&  10 bar O$_2$
outgassing&  CO$_{2}$(4\tablenotemark{k}), O$_{3}$(29\tablenotemark{e}), O$_{2}$(43\tablenotemark{c})\\
&  100 bar O$_2$
desiccated&  CO$_{2}$(5\tablenotemark{k}), O$_{3}$(6\tablenotemark{e}), O$_{2}$(28\tablenotemark{c})\\
&  10 bar O$_2$
desiccated&  CO$_{2}$(3\tablenotemark{k}), O$_{3}$(5\tablenotemark{k}), O$_{2}$(28\tablenotemark{c})\\
&  10 bar Venus&  CO$_{2}$(3\tablenotemark{k})\\
&  92 bar Venus&  CO$_{2}$(3\tablenotemark{k})\\
T-1h&  10 bar CO$_2$&  CO$_{2}$(2\tablenotemark{k}), SO$_{2}$(85\tablenotemark{f})\\
&  92 bar CO$_2$&  CO$_{2}$(2\tablenotemark{k}), SO$_{2}$(86\tablenotemark{f})\\
&  100 bar O$_2$
outgassing&  CO$_{2}$(9\tablenotemark{k}), O$_{3}$(1\tablenotemark{k}), O$_{2}$(32\tablenotemark{c})\\
&  10 bar O$_2$
outgassing&  CO$_{2}$(4\tablenotemark{k}), SO$_{2}$(54\tablenotemark{f}), O$_{3}$(23\tablenotemark{e}), O$_{2}$(31\tablenotemark{c})\\
&  100 bar O$_2$
desiccated&  CO$_{2}$(4\tablenotemark{k}), O$_{3}$(3\tablenotemark{k}), O$_{2}$(22\tablenotemark{c})\\
&  10 bar O$_2$
desiccated&  CO$_{2}$(3\tablenotemark{k}), O$_{3}$(4\tablenotemark{k}), O$_{2}$(22\tablenotemark{c})\\
&  10 bar Venus&  CO$_{2}$(2\tablenotemark{k})\\
&  92 bar Venus&  CO$_{2}$(2\tablenotemark{k})
\enddata

\tablenotetext{c}{NIRSpec G140H}
\tablenotetext{d}{NIRSpec G235H}
\tablenotetext{e}{NIRSpec G395H}
\tablenotetext{f}{MIRI LRS}
\tablenotetext{k}{NIRSpec Prism sub512 ngroup6}

\end{deluxetable}

The presence of CO$_2$ dominates the detectability of all the atmospheres simulated in \citet{Lincowski2018} with JWST. Even a relatively small amount of \ce{CO2} (e.g. ${\sim} 290$ ppm for the 1 bar \ce{H2O} TRAPPIST-1 e) can saturate the strong 2.7, 4.3, 15 $\mu$m \ce{CO2} absorption features and lead to the detection of both the atmosphere and \ce{CO2}. As a result, the number of transits necessary to detect spectral features in a transmission spectrum, given in Section \ref{sec:results:ntran_featureless}, are close to the number of transits necessary to detect CO$_2$. In some cases, the number of transits to detect CO$_2$ is fewer than that needed to simply detect the atmosphere. This is because the spectral model without CO$_2$ deviates more significantly from the true spectrum than the best-fitting featureless spectrum. In these cases it is important to defer to the number of transits required to rule out a featureless spectrum, because it provides a more realistic fit to the spectrum. 

The inner TRAPPIST-1 planets may have several detectable molecules that can be used to distinguish between different evolutionary scenarios. \ce{H2O} and \ce{SO2} may be marginally detectable with MIRI LRS transmission spectra of the inner TRAPPIST-1 planets---b, c, and d--- if they possess clear sky \ce{CO2} atmospheres. However, if these planets possess \ce{H2SO4} clouds then the \ce{H2O} and \ce{SO2} may be undetectable. Alternatively, if TRAPPIST-1 b, c, and d have oxygen-dominated atmospheres then \ce{O2} should be distinguishable via O$_2$-O$_2$ (O$_4$) CIA features at 1.06 and 1.27 $\mu$m, which are slightly more detectable with NIRSpec G140M/H than the NIRSpec Prism*. An oxygen-dominated planet with outgassing may be distinguished from a completely desiccated world by detecting \ce{H2O}, which is only readily detectable in our models of \ce{O2} planets with outgassing. 

TRAPPIST-1 e offers an opportunity to characterize a planet in the habitable zone with JWST. Like the inner TRAPPIST-1 planets, \ce{O2} may be detectable in the spectrum of an oxygen-dominated atmosphere via the O$_2$-O$_2$ CIA features. However, modern Earth levels of \ce{O2} and \ce{O3}, and Earth geologic levels of \ce{CH4} in a 1 bar \ce{N2}-dominated atmosphere are not likely to be detectable. If TRAPPIST-1 e possesses such a habitable environment without clouds, tropospheric \ce{H2O} may be detectable using the NIRSpec Prism*. However, with full cloud coverage the detectability of \ce{H2O} strongly diminishes and becomes unobservable. The broad 4.3 $\mu$m N$_2$-N$_2$ CIA feature may be marginally detectable in a 1 bar habitable \ce{N2}-dominated atmosphere with NIRSpec G395M/H. The 9.6 $\mu$m \ce{O3} feature would require just over 100 transits with MIRI LRS to detect at $\left < \mathrm{SNR} \right > = 3.0$ in the clear sky 1 bar \ce{N2}-dominated atmosphere; approximately twice the number of transits would be required if the planet possesses 100\% water cloud coverage. 

For the outer planets---TRAPPIST-1 f, g, and h---modest atmospheric characterization with transmission spectroscopy may be possible. In particular, oxygen-dominated atmospheres may be distinguished not only by their prominent O$_2$-O$_2$ CIA features, but also by their \ce{O3} features, which become more detectable with the NIRSpec Prism* than the O$_2$-O$_2$ features for the coolest TRAPPIST-1 planets. 

\begin{figure*}[t!]
\centering
\includegraphics[width=0.8\textwidth]{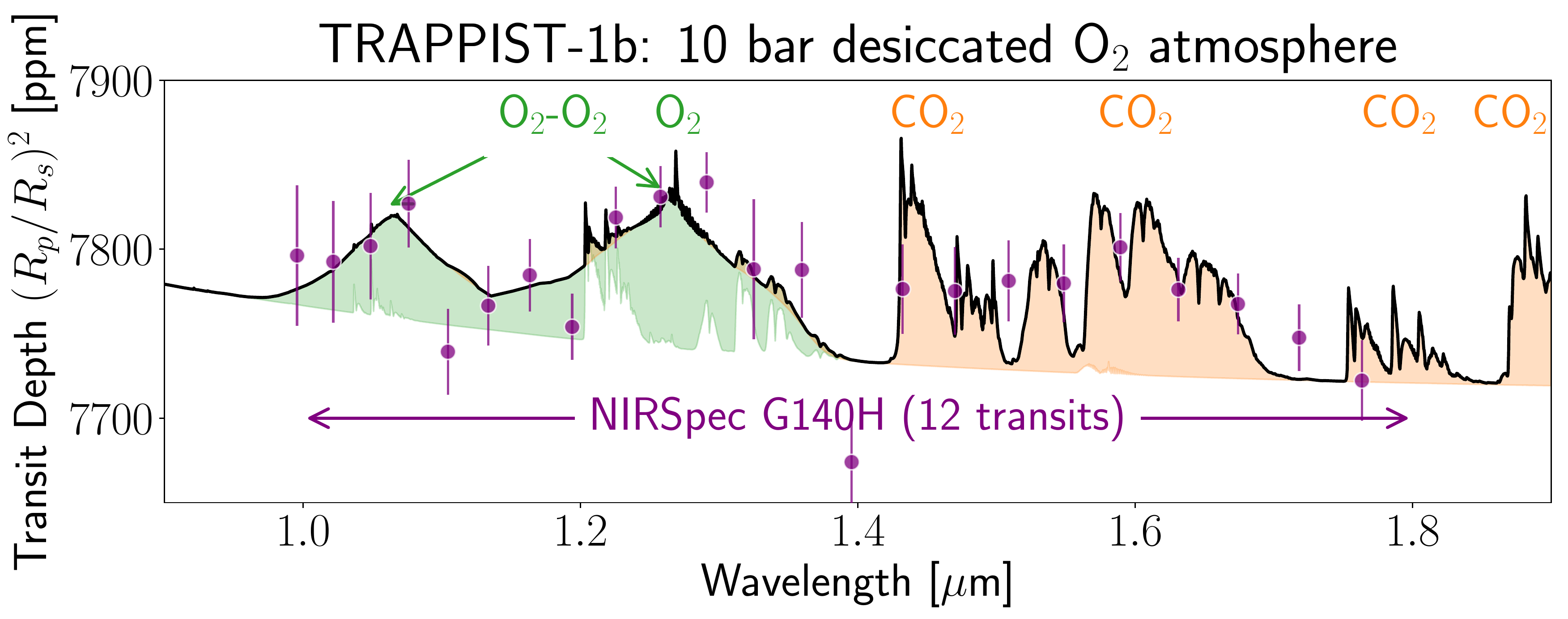}
\includegraphics[width=0.8\textwidth]{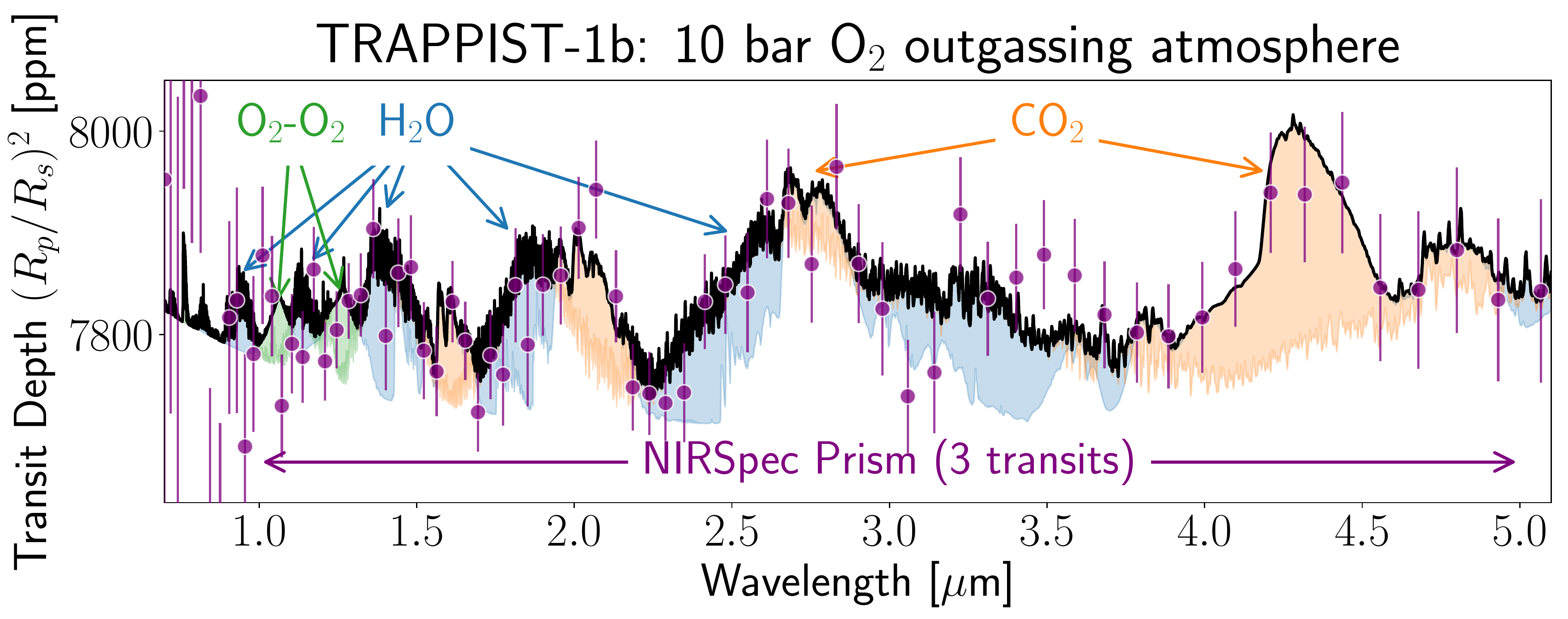}
\includegraphics[width=0.8\textwidth]{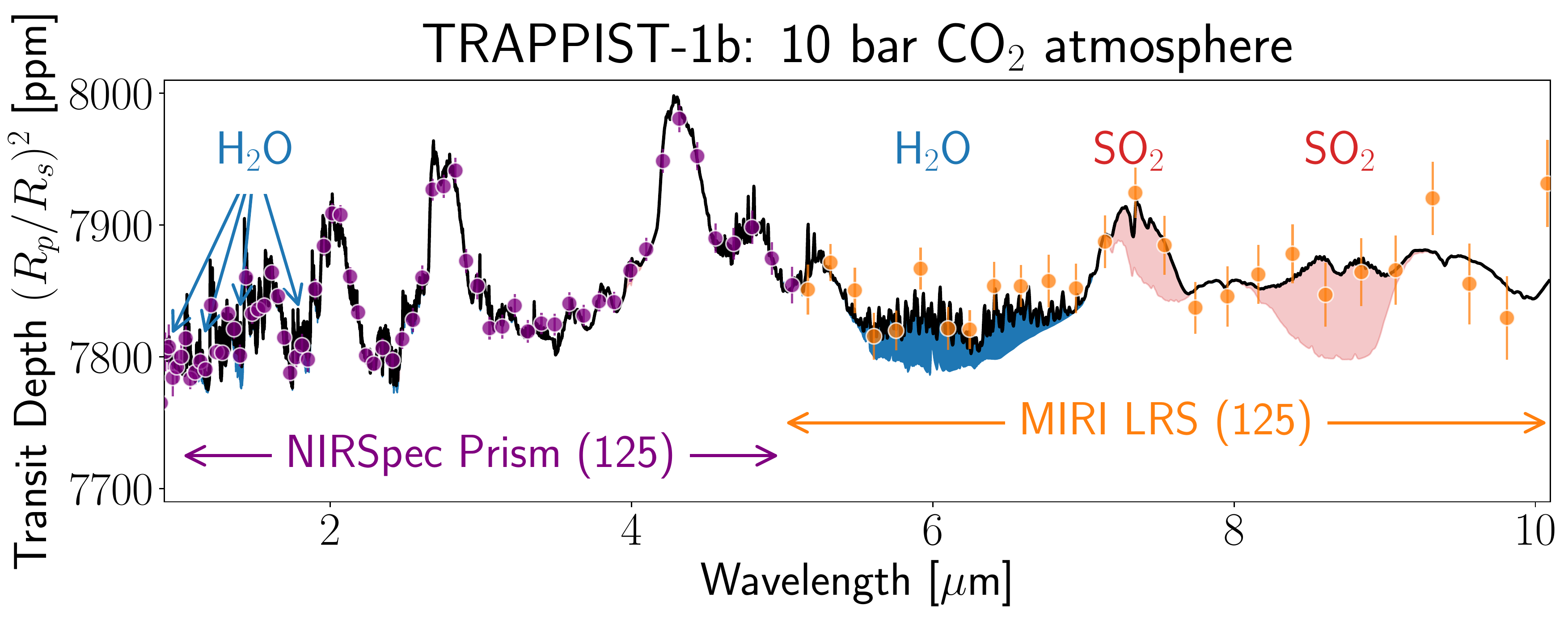}
\caption{Theoretical transmission spectra of TRAPPIST-1 b assuming three different atmospheric compositions with modeled noise for JWST observations. 
    \textit{Top:} Transmission spectrum of a 10 bar desiccated \ce{O2} atmosphere shown with error bars calculated for 12 transits with NIRSpec G140H---sufficient for $\left < \mathrm{SNR} \right > \sim 5$ on the \ce{O4} features. 
    \textit{Middle:} Transmission spectrum of a 10 bar outgassing \ce{O2} atmosphere shown with error bars calculated for 3 transits with NIRSpec Prism*---sufficient for $\left < \mathrm{SNR} \right > \sim 5$ on the \ce{H2O} features. 
    \textit{Bottom:} Transmission spectrum of a 10 bar \ce{CO2} atmosphere shown with error bars calculated for 125 transits with NIRSpec Prism*---sufficient for $\left < \mathrm{SNR} \right > \sim 5$ on the NIR \ce{H2O} features---and 125 transits with MIRI LRS---sufficient for $\left < \mathrm{SNR} \right > \sim 5$ on both the 6 $\mu$m \ce{H2O} feature and the 7.3 and 8.7 $\mu$m \ce{SO2} features. 
    }
\label{fig:sensitivity_3plot}
\end{figure*}

The observational difficulty with which individual molecules may be detected in a transmission spectrum---bulk atmospheric constituents or trace gases---varies substantially as a function of atmospheric composition. Consequently, the gases that are both relatively easy to detect and unique to a specific atmosphere make optimal testable hypotheses for that atmospheric composition. 
Figure \ref{fig:sensitivity_3plot} demonstrates a potential approach for distinguishing between three different atmospheric states for TRAPPIST-1 b using JWST transmission spectroscopy. The models are shown with calculated error bars that correspond to the amount of JWST observing time that would be required to detect specific molecules in the given spectrum. 

In the top panel of Fig. \ref{fig:sensitivity_3plot}, the spectrum of a 10 bar desiccated oxygen atmosphere displays prominent \ce{O4} features that may be detected with $\left < \mathrm{SNR} \right > \sim 5$ in 12 transits with NIRSpec G140H. This is a strong discriminant between an \ce{O2}-dominated atmosphere and a \ce{CO2}-dominated atmosphere, both of which have strong and detectable \ce{CO2} features.  

The spectrum of a 10 bar oxygen atmosphere with outgassing is shown in the middle panel of Fig. \ref{fig:sensitivity_3plot}. This spectrum has similarly detectable \ce{O4} features to the desiccated atmosphere, but also substantial \ce{H2O} features (due to Earth levels of geological fluxes) that may be detected with $\left < \mathrm{SNR} \right > \sim 5$ in just 3 transits with NIRSpec Prism*. Such strong water absorption in an oxygen-dominated atmosphere would indicate incomplete desiccation. 

The spectrum of a 10 bar \ce{CO2} atmosphere is shown in the bottom panel of Fig. \ref{fig:sensitivity_3plot}. This spectrum contrasts with the outgassing oxygen atmosphere in the detectability of \ce{H2O}. Water in a Venus-like atmosphere is scarce, particularly in the upper atmosphere, which would require ${\sim} 125$ transits with either NIRSpec Prism* or MIRI LRS to detect with $\left < \mathrm{SNR} \right > \sim 5$. The added benefit of making this costly observation with MIRI LRS is that \ce{SO2} could also be detected with $\left < \mathrm{SNR} \right > \sim 5$ from the 7.3 and 8.7 $\mu$m features. Nonetheless, detecting any gases other than \ce{CO2} in a \ce{CO2}-dominated atmosphere will be very difficult. 

\section{Discussion}
\label{sec:discussion}

Our results indicate that JWST observations may be able to place strong constraints on the presence of high mean molecular weight terrestrial atmospheres for all of the TRAPPIST-1 planets and, in some cases, detect individual touchstone molecules that may be used to distinguish between different evolutionary histories that the planets may have undergone. 

\subsection{Do the TRAPPIST-1 planets have atmospheres?} \label{sec:disc:atmospheres}

For detecting the presence of atmospheres on the TRAPPIST-1 planets, we find that NIRSpec Prism observations are optimal. Transmission spectroscopy with the NIRSpec Prism could lead to a $ \left < \mathrm{SNR} \right > = 5$ detection of atmospheric spectral features in as few as 2-11 transits for TRAPPIST-1 b out to TRAPPIST-1 h if the planets lack high altitude aerosols. However, if the TRAPPIST-1 planets possess Venus-like \ce{H2SO4} aerosols, reaching the same constraints on the presence of atmospheres may require up to 12 times more transits.  
\ce{CO2} possesses numerous strong absorption bands from the near- through the mid-IR, such as the 2.0, 2.7, 4.2, and 15 $\mu$m bands, which significantly contribute to the ability for JWST to detect the terrestrial atmospheres considered in this work.

Our self-consistent Venus-like \ce{H2SO4} aerosol modeling reveals trends in the detectability of such atmospheres with semi-major axis. \citet{Lincowski2018} found that TRAPPIST-1 b was too hot for Venus-like aerosols to form in the atmosphere, but that \ce{H2SO4} aerosols could form in all six exterior planets. These aerosols form at high altitudes in the atmospheres of TRAPPIST-1 c, d, and e and lead to muted \ce{CO2} features that will require about 4-12 times more transits to be detected with JWST. However, TRAPPIST-1 f, g, and h are cool enough for the \ce{H2SO4} aerosols to form at lower altitudes \citep{Lincowski2018}, and therefore contribute less to their observable transmission spectra such that their atmospheres may be detected in fewer than 2 times the number of transits compared to the clear sky \ce{CO2} cases for the same planets. Cloudy Venus-like atmospheres for TRAPPIST-1 f, g, and h require fewer transits to detect than cloudy Venus-like atmospheres for TRAPPIST-1 c, d, and e.  

Secondary eclipse spectroscopy may require a significantly greater JWST time commitment than transmission spectroscopy to achieve comparable constraints on the detection of the TRAPPIST-1 planet atmospheres. Because of its access to longer wavelengths, MIRI LRS is the only JWST instrument capable of observing eclipse spectra of the TRAPPIST-1 planets with high enough SNR to detect absorption features in the spectra that we considered. However, our estimates for the  JWST observing time required to detect \textit{emission} spectrum features with MIRI LRS dwarfed the time required for spectral features to be detected in transmission spectra with NIRSpec, NIRISS, and NIRCam (see Figs. \ref{fig:miri_imaging_nocc}, \ref{fig:colortable_featureless_T-1b_transit}-\ref{fig:colortable_featureless_T-1h_transit}). Furthermore, the disparity between transmission and emission spectroscopy only broadens with semi-major access, making precise MIRI LRS emission spectroscopy beyond TRAPPIST-1c infeasible with JWST. Our self-consistent planet models reveal that atmospheres with high altitude aerosols---that may appear featureless in a transmission spectrum---may also appear featureless in an emission spectrum as thermal flux is emitted and/or scattered from near the top of the cloud deck. 

Initial photometric assessments of the TRAPPIST-1 system with MIRI filter photometry are unlikely to provide more efficient preliminary results than transmission spectroscopy in the NIR with NIRSpec, NIRISS, or NIRCam. Our MIRI transit and eclipse photometry modeling showed that detecting the presence of atmospheres would require approximately an order of magnitude more JWST time than NIRSpec Prism transmission spectroscopy. This is due to (1) the higher SNR on transits afforded to NIRSpec near the peak of the stellar SED in the 1-3 $\mu$m range, (2) the usefulness of spectral resolution for atmospheric detections via deviations from a featureless spectrum, and (3) the need to observe transits in each MIRI filter separately to gain any meaningful wavelength resolution. 

However, targeted MIRI photometric observations may still provide useful atmospheric constraints. Detecting the presence of atmospheric greenhouse heating may be done by inferring brightness temperatures that exceed the zero bond albedo equilibrium temperature. However, this will depend on the accuracy of MIRI's absolute flux measurements. Specific wavelength bands where the observed atmosphere is optically thin, and therefore emits from hotter depths, provide the key observable. However, these observations may only be feasible for TRAPPIST-1 b and c due to their higher expected thermal emission.  The 11.3 $\mu$m MIRI filter (F1130W) may be optimal for such measurements because it is sufficiently separated from the strongly absorbing 15 $\mu$m CO$_2$ band. The 5.6 $\mu$m MIRI filter (F560W) may also be optimal for detecting high thermal emission from TRAPPIST-1 b and c, if saturation on the star can be avoided in this bandpass. 
Transit and/or eclipse photometry targeting the 15 $\mu$m \ce{CO2} band with the F1500W MIRI filter, and neighboring filters, could be used to detect the atmosphere, but our work shows that targeting the strong 4.2 $\mu$m \ce{CO2} feature with NIRSpec Prism transmission spectroscopy will be a much more efficient approach for detecting \ce{CO2} in an atmosphere. 

Detecting wavelength-dependent surface emissivity features in the emission spectrum of a planet without an atmosphere will be highly unlikely with JWST. Even for the optimistic case---TRAPPIST-1 b with a pure and smooth quartz surface---we find it would require ${\sim}90$ secondary eclipses with MIRI LRS to detect the mid-IR silicate absorption (see Fig. \ref{fig:quartz_emission_detectability}). Consequently, detecting surface features on an airless world would be extremely challenging. This validates our assumption that an airless planet will likely appear featureless in emission, and that detecting spectral features due to surface emissivity variations is an inefficient means to confirm that a planet is airless. 
Furthermore, this strengthens the case for transmission spectroscopy, which can detect the high mean molecular weight atmosphere of a Venus-like planet enshrouded in \ce{H2SO4} clouds in just ${\sim}$20-30 transits (for TRAPPIST-1 c, d, and e). 
However, actually \textit{confirming} that a TRAPPIST-1 planet does not possess an atmosphere---even if featureless spectra lend favor to that hypothesis---will be a very difficult task, that may require thermal phase curve \citep{Selsis2011, Maurin2012, Kreidberg2016, Meadows2018} or planet-planet occultation observations \citep{Luger2017b} to probe the day-night temperature contrast.

\paragraph{Comparison with Previous Works}

Our results are in agreement with previous investigations on optimal ways to detect  terrestrial exoplanet atmospheres of different compositions.   
Although emission spectroscopy has been suggested as a means of detecting terrestrial atmospheres \citep[e.g.][]{Belu2011}, we agree with previous works that transmission spectroscopy is more viable \citep[e.g.][]{Hedelt2013, Barstow2016, Barstow2016b, Greene2016}, particularly for temperate and cool planets which emit considerably less thermal flux. 
We agree with \citet{Betremieux2018} that the JWST integration times needed to detect molecules strongly depend on the composition of the atmosphere---particularly the presence of high altitude aerosols. 

Although we considered more atmospheric compositions, our predicted exposure times for  the detection of planetary atmospheres agree for the subset of similar atmospheres modeled by \citet{Morley2017}, with many of the discrepancies attributable to differences in the assumed planetary masses used in the transmission spectrum models. 
Our work and that of \citet{Morley2017} has a common focus of distinguishing clear sky Venus-like atmospheres from a featureless spectrum with the NIRSpec/G235 instrument, which offers a basis for comparison. 
The detectability of these atmospheres is dominated by \ce{CO2} features, which are largely unaffected by photochemistry, which was not included in \citet{Morley2017}.
We find that a 10 bar CO$_2$ atmosphere would require 5, 8, 4, 17, 15, 16, and 15 transits to distinguish TRAPPIST-1 b, c, d, e, f, g, and h, respectively, from a featureless spectrum, and \citet{Morley2017} find that a 1 bar aerosol-free Venus-like atmosphere would require 6, 36, 13, 4, 17, 10, and 4 transits. 

The cases of largest disagreement (e.g. c, e, h) appear consistent with the different masses used in each study. Planet mass affects the atmospheric scale height (via the surface gravity) and therefore the size and detectability of molecular features in a transmission spectrum. \citet{Morley2017} used masses from \citet{Wang2017}, with the exception of TRAPPIST-1 f, for which the mass from \citet{Gillon2017} was used. In this paper, and in \citet{Lincowski2018}, we use the TRAPPIST-1 planet masses from \citet{Grimm2018}. 
For the case of TRAPPIST-1 c where we use a significantly smaller planet mass than \citet{Morley2017} we find that fewer transits are required to detect spectral features in transit. For the cases of TRAPPIST-1 e and h where we use a significantly larger planet mass than \citet{Morley2017} we find that more transits are required to detect spectral features. This scaling with mass is consistent with the findings of \citet{Morley2017} when they repeated calculations with masses derived from the \citet{Weiss2014} mass-radius relationship, and further underscores the need for accurate masses for spectral modeling and fitting.

Our focus on thicker atmospheres (1-100 bar) than \citet{Morley2017} (0.01-1 bar) both explains discrepancies between the studies and further demonstrates that such thick and aerosol-free atmospheres may be more easily detected than thinner atmospheres. Our results for TRAPPIST-1 d, where our masses agree best, show 9 fewer transits required to detect spectral features than \citet{Morley2017}, which is consistent with different planet surface pressures used in each study. \citet{Morley2017} demonstrated a trend of increasing atmospheric detectability with increasing surface pressure in simulations of 0.01, 0.1, and 1 bar CO$_2$ atmospheres. Our results continue this nearly linear trend in log-pressure out to 10 bars. However, our 92 bar CO$_2$ atmospheres are consistent with the detectability of our 10 bar atmospheres, indicating that this trend in pressure saturates for thick atmospheres as the planet surface drops below the atmospheric regions that are sensed with transmission spectroscopy.

The \citet{Krissansen-Totton2018} work on the detectability of biogenic gases in an anoxic atmosphere for TRAPPIST-1 e with JWST suggests that relatively few transits are required to constrain atmospheric abundances. \citet{Krissansen-Totton2018} showed that a retrieval using ${\sim}10$ transits of TRAPPIST-1 e with the NIRSpec Prism was sufficient to begin to constrain the abundances of CO, CO$_2$, and CH$_4$. 
Consequently, the 13 transits of TRAPPIST-1 e that we show may be required to detect a clear 1 bar H$_2$O atmosphere to a $\left < \mathrm{SNR} \right > = 5$, may also provide spectra that are sufficient to begin to constrain molecular abundances in an atmospheric retrieval framework. 

HST transmission spectroscopy of hot gaseous exoplanets has demonstrated how the presence of clouds and hazes can diminish sensitivity to molecular absorption features \citep[e.g.][]{Berta2012, Ehrenreich2014, Knutson2014, Kreidberg2014, Nikolov2015, Sing2016}, but we know \textit{a priori} that planets with such large radii must have atmospheres. However, for terrestrials we will rely on absorption features to test for the presence of atmospheres, and clouds and hazes, if present, will make this more difficult. \citet{Batalha2018} concluded that 10 transits should be sufficient to detect the dominant molecular absorber in the transmission spectrum of TRAPPIST-1f when observed using a partial saturation strategy with the NIRSpec Prism, and that additional observations are unlikely to reveal more information. While our results agree with \citet{Batalha2018} for the case of clear atmospheres, cloudy and/or hazy terrestrial atmospheres may require significantly more observations. Venus-like planets with \ce{H2SO4} aerosols tend to require more than 10 transits (and up to ${\sim} 30$ for TRAPPIST-1e) to reach the same confidence in the detection of the atmosphere. We recommend testing the hypothesis that a TRAPPIST-1 (or similar) planet has a \textit{clear} atmosphere in ${\le} 10$ transits, and then evaluating the scientific value of additional transits to test the hypothesis that the planet has aerosols.

\subsection{What is the nature of the TRAPPIST-1 planet atmospheres?} \label{sec:disc:nature}

Detecting specific molecules in JWST transmission spectra of the TRAPPIST-1 planets may be possible, and allow for the discrimination between different climate/composition states and evolutionary histories. 
As previously stated, \ce{CO2} should be the easiest molecule for JWST to detect in the atmospheres of the TRAPPIST-1 planets.  
Due to its prevalence in these simulated atmospheres \textit{regardless of evolutionary history}, \ce{CO2} makes for a strong indicator of a terrestrial atmosphere, but a weak discriminant of specific atmospheric state. 
Furthermore, \ce{CO2} produces nearly the same strength features regardless of atmospheric abundance, particularly the 4.3 and 15 $\mu$m bands, which are saturated in a Venus-like atmosphere, an \ce{O2} outgassing atmosphere, and an Earth-like atmosphere, even though the abundance ranges from 90 bars down to 360 ppm \citep{Lincowski2018}. 
Other molecules, such as \ce{O2}, \ce{O3}, \ce{H2O}, and \ce{SO2}, may be detectable with JWST and may help to distinguish between the suite of atmospheres that we considered. 

Although oxygen as a biosignature may not be detectable for the potentially habitable TRAPPIST-1 planets, oxygen as a remnant of pre-main-sequence water loss may be easily detected or ruled out. 
We find that biogenic \ce{O2} in the atmosphere of TRAPPIST-1e may be too difficult to detect with JWST, but the 9.6 $\mu$m \ce{O3} feature may be weakly detectable at $\left < \mathrm{SNR} \right > = 3$ in over 100 transits with MIRI LRS, which is in general agreement with the findings of \citet{Wunderlich2019}. 
However, the 1.06 and 1.27 $\mu$m O$_2$-O$_2$ CIA features are key discriminants of a planet that has an oxygen abundance greatly exceeding biogenic oxygen production on Earth and may therefore indicate a planet that has undergone vigorous water photolysis and subsequent loss during the protracted super-luminous pre-main-sequence phase faced by late M dwarfs \citep{Luger2015, Schwieterman2016}. We find that NIRSpec G140M/H is the optimal JWST instrument for detecting these O$_2$-O$_2$ features and could lead to their detection in as few as 7-9, 15, 8, 49-67, 55-82, 79-100, and 62-89 transits of TRAPPIST-1b, c, d, e, f, g, and h, respectively, should they possess such an atmosphere. These quoted number of transits may be sufficient to rule out the existence of oxygen-dominated atmospheres in the TRAPPIST-1 system. Additional evidence of ocean loss could be provided by detection of isotope fractionation, which may also be possible in as few as 11 transits with JWST, for strong isotopologue bands such as HDO \citep{Lincowski2019}. 

Detecting ozone absorption may be another strong indicator of a post-runaway, oxygen-dominated atmosphere.
For the 10 bar desiccated \ce{O2} atmospheres, ozone absorption features become more detectable for the outer planets. 
Targeting \ce{O3} at 9.6 $\mu$m with MIRI LRS is optimal for the inner planets (b, c, d, e), although it may require upwards of 100 transits to detect at $\left < \mathrm{SNR} \right > = 5$. However, targeting the weaker \ce{O3} bands between 3 -- 5 $\mu$m with NIRSpec G395M/H or NIRSpec Prism is optimal for the outer planets (f, g, h), due to their larger ozone column abundances. The \ce{O2}-\ce{O2} CIA bands for these cooler, desiccated planets are more difficult to detect, which may make \ce{O3} a more efficient observational discriminant for such a planet. 

The photochemistry underlaying the detectable ozone buildup in the cooler planets has implications beyond the TRAPPIST-1 system. \citet{Lincowski2018} noted that the competing effects of the Chapman cycle (with declining photolysis rates with distance from the star), was primarily responsible for the ozone accumulation in the atmospheres of the outer planets, an effect previously noticed by \citet{Grenfell2007}. While this is a driving factor, the differences among the planets can more specifically be attributed to catalytic cycles of nitrogen oxides (primarily, \ce{N2O}, NO, and \ce{NO2}), which drive the destruction of \ce{O3}, as in the stratosphere of Earth \citep{Seinfeld2006}.
Because these atmospheres contain \ce{N2}, \ce{O^1D} produced from photolysis of oxygen-bearing molecules can react with \ce{N2} to generate nitrogen oxides. The availability of \ce{O^1D} declines with distance from the star due to lower UV fluxes. Beginning with planet e, the production of nitrogen oxides declines substantially, removing them as a mechanism for the destruction of \ce{O3}, so that \ce{O3} accumulates and becomes well-mixed, generating a large column density. In the atmosphere of an \ce{O2}-dominated atmosphere without \ce{N2}, \ce{O3} levels would likely be higher due to the lack of the nitrogen oxide catalysts. 

Detecting water in the atmospheres of the TRAPPIST-1 planets may also help to constrain evolutionary scenarios. The presence of water may be readily detectable for TRAPPIST-1 b, c, and d with the NIRSpec Prism* if they possess high O$_2$ atmospheres that have not been completely desiccated. High \ce{O2} atmospheres for planets that exited the pre-main-sequence with their atmospheres and interiors completely desiccated, however, will have no water to detect, making water in an oxygen-dominated atmosphere a potentially detectable discriminant of incomplete desiccation or outgassing from the interior. In CO$_2$-dominated atmospheres, H$_2$O may be prohibitively difficult to detect due to its scarcity in the atmospheres of Venus-like worlds, before even considering obscuration by \ce{H2SO4} clouds. 

Detecting water in the atmosphere of one of the potentially habitable TRAPPIST-1 planets could indirectly hint at surface habitability \citep{Robinson2018}, but this will be challenging for JWST, even with an ideal system like TRAPPIST-1.   
Tropospheric \ce{H2O} may be detectable in the transmission spectrum of TRAPPIST-1 e in ${\sim} 35$ transits with NIRSpec Prism* should the planet have a clear-sky 1-bar N$_2$/O$_2$ atmosphere with H$_2$O. However, the cold trap that keeps water vapor concentrated in the lower atmosphere makes habitability difficult to infer with transmission spectroscopy, which cannot readily probe surface environments. 
Additionally, we find that 100\% cloud coverage strongly increases the required JWST time to detect H$_2$O in such a habitable atmosphere. 
Given these difficulties for detecting water in the atmosphere of a HZ planet with JWST, robust habitability assessments may ultimately require a future direct imaging telescope that can readily probe rocky planet surfaces to search for more direct evidence of surface liquid water \citep[e.g.][]{Cowan2009, Robinson2010, Lustig-Yaeger2018}.

\subsection{Further Considerations}

The results in this paper lean optimistic, and therefore represent lower limits on the amount of observing time needed to detect and characterize evolved terrestrial atmospheres in the TRAPPIST-1 system. 
As a result, TRAPPIST-1 observing plans that include fewer transits/eclipses than reported here may require additional observations to make robust inferences on the existence and nature of atmospheres. 
One source of optimism is that our $\left < \mathrm{SNR} \right >$ approach (see Equation \ref{eqn:SNR_delta_chi2}) to atmospheric and molecular detection, rather than a full atmospheric retrieval, implicitly assumes that any retrieval will have converged on the true underlying spectrum. As a result, our signal-to-noise metric is not equal to a detection of that specific atmosphere to a given significance (e.g. $3 \sigma$). In practice, many different atmospheric compositions will likely be capable of fitting JWST spectra of TRAPPIST-1 planets, potentially leaving large regions of atmospheric parameter space unconstrained. Thus, the confidence of any ``one'' composition representing the true state of the planet may remain low despite having high confidence in the presence of features in the spectrum that would indicate the presence an atmosphere. Our approach is simply a wavelength range and resolution agnostic method to quantify the detectability of spectral signals emanating from self-consistent atmospheres above the expected noise of JWST. 

Our calculations also lean optimistic because we have neglected astrophysical and systematic sources of noise that may make precise time-series exoplanet observations with JWST more difficult than our estimates suggest. 
Stellar effects such as limb-darkening \citep[e.g.][]{Csizmadia2013} and heterogeneous photospheres \citep[e.g.][]{Rackham2016, Rackham2018, Zhang2018} are major concerns for precision exoplanet transmission spectroscopy. These effects may be particularly perilous for observations of planets orbiting low mass stars, as the stars may contain large, cool spots and absorbing molecules in their photospheres. 
We also did not assume a systematic noise floor for any JWST instruments \citep[e.g.][]{Greene2016}, but this is likely optimistic in terms of the currently unknown on-target performance. 
Astrophysical and systematic effects must be well understood to detect molecular features at the ${\sim}10$ ppm level, where many terrestrial features reside. However, many of the major absorption features in our TRAPPIST-1 planet spectral models are much larger and more detectable, spanning 50-200 ppm. 

During the preparation of this paper, a new version of \texttt{Pandeia} was released (version 1.3) that increased JWST background noise predictions by 10-40 ppm. For bright targets like TRAPPIST-1 where the noise budget is dominated by stellar photon noise, such an increase in the background has a negligible effect on our estimated number of transits and eclipses to detect and characterize exoplanet atmospheres in the near- through the mid-IR.

We considered a limited number of atmospheres that may not resemble the true state of the TRAPPIST-1 planets, especially since the star's UV spectrum is poorly constrained. 
Here, as in \citet{Lincowski2018}, we used a UV spectrum scaled from measurements of the nearby mid M dwarf Proxima Centauri, but these stars have slightly different spectral types and activity levels. Recently, three synthetic SEDs have been produced by \citet{Peacock2019} for TRAPPIST-1, which are constrained by the Lyman alpha flux and upper limits of GALEX UV photometry. These synthetic spectra differ by having higher NUV and a different distribution of FUV flux than \citet{Lincowski2018}, which could slightly change the photochemistry of these planetary atmospheres and the observable molecular absorption in their upper atmospheres. 
However, only observations---including those outlined in this paper---will ultimately reveal what planetary processes dominate in sculpting the observable atmospheric signatures of terrestrial exoplanets, particularly those orbiting M dwarfs. 
In the mean time, further study is warranted on the range of possible atmospheric conditions for the TRAPPIST-1 planets, particularly photochemically and climatically self-consistent reducing atmospheres \citep[e.g.][]{Arney2017}, and UV characterizations of late M dwarfs to improve photochemical predictions and interpretation of observations. 

Although the atmospheric states considered in this paper represent only a selection of possible states, many of their characteristics and molecular features may exist in other atmospheres, making our results apply more broadly. For example, the strength of the \ce{H2O}, \ce{CO2}, and \ce{O3} bands in the \ce{O2}-dominated outgassing atmospheres may be similar in other clear atmospheres that may contain more inert gasses, like \ce{N2}. Our JWST detectability calculations for these gases may be useful beyond that particular atmospheric case. The \ce{O2}-\ce{O2} features, however, are specific evidence of high \ce{O2} content. Since all of the atmospheres considered in this work were high mean molecular weight, our results help to elucidate the detectability of such atmospheres and the molecules within, even if the true compositions differ. 

Constraining the atmospheres of Earth-size planets transiting even the smallest stars like TRAPPIST-1 will require pushing the limits of JWST. As we demonstrated with the NIRSpec Prism partial saturation, alternate JWST modes that can improve observations of transiting exoplanets can enhance the science return, and dramatically decrease the time investment required to detect and characterize terrestrial atmospheres. Although modes such as NIRCam's Dispersed Hartmann Sensor \citep[DHS;][]{Schlawin2017} and high efficiency readout patterns for NIRSpec Prism \citep{Batalha2018} have not been officially approved, they represent promising avenues towards improved JWST observations that may ultimately make the difference in enabling JWST to constrain terrestrial exoplanet atmospheric compositions.

\section{Conclusion}
\label{sec:conclusion}

We investigated the potential to detect and characterize the atmospheres of all seven known TRAPPIST-1 exoplanets with JWST. Although the planets are small and likely possess high mean molecular weight atmospheres with relatively low scale heights, we found that many molecular absorption features may be detectable with JWST in ${\sim}$2-15 transits. 
These observations may be used to diagnose the presence of atmospheres and, in some cases, discriminate between different plausible atmospheric compositions. 

However, we find that an initial photometric assessment of the TRAPPIST-1 planets with MIRI is, perhaps non-intuitively, not as efficient as spectroscopic atmosphere detection in the NIR. To achieve comparable constraints on the detection of atmospheres approximately an order of magnitude more transits or eclipses will need to be observed with MIRI, compared to transits observed with spectrometers in the NIR. 

Transmission spectroscopy with NIRSpec Prism may be the most efficient path to detect the presence of atmospheres for the TRAPPIST-1 planets, via detecting \ce{CO2} bands between 1-5 $\mu$m. Venus-like atmospheres with high altitude \ce{H2SO4} aerosols will be more difficult to detect in transmission and emission spectra, however, these aerosols will form at lower altitudes for the temperate and cooler planets such that they obscure less molecular absorption in transmission spectra. Furthermore, post-runaway oxygen-dominated worlds may be identified in transmission using (1) \ce{O2}-\ce{O2} CIA observed with NIRSpec G140M/H, NIRSpec Prism, or NIRISS SOSS, or (2) \ce{O3} absorption observed with MIRI LRS (for b, c, d, and e) or NIRSpec G395M/H (for f, g, and h). If TRAPPIST-1 e is habitable and cloud-free, water could be detected in the troposphere with NIRSpec Prism in about 35 transits, but the presence of water clouds could completely obscure the water vapor absorption features.   

We outlined a particular path for characterising the TRAPPIST-1 planets with JWST that narrows down the possible evolutionary histories that the planets may have trod to exist in their current observable states. We recommend using transmission spectroscopy to 
\begin{easylist}[enumerate]
    & detect the planet atmospheres via \ce{CO2} absorption; 
    & detect or rule out a post-runaway oxygen-dominated atmospheres via \ce{O2}-\ce{O2} CIA or \ce{O3} absorption; 
    & constrain the extent of (atmosphere and interior) desiccation, and potentially the habitability, via the \ce{H2O} abundance.
\end{easylist}
\noindent 
However, our results may be used to construct countless additional observing strategies that best augment existing projects and proposals with our testable hypotheses on the nature of TRAPPIST-1 system of Earth-sized exoplanets. 

\acknowledgments

\

We thank Eric Agol for insights into our photometric signial-to-noise calculations, Joshua Bandfield and Elena Amador for crucial discussions on the emissivity signatures of rocks, and Rodrigo Luger for countless useful discussions that contributed to this work. 
We also thank the anonymous referee whose helpful comments contributed to the quality and clarity of this paper. 
This work was supported by NASA's Virtual Planetary Laboratory under NASA Astrobiology Institute Cooperative Agreement Number NNA13AA93A, and Grant Number 80NSSC18K0829. This work benefited from participation in the NASA Nexus for Exoplanet Systems Science research coordination network. This work made use of the advanced computational, storage, and networking infrastructure provided by the Hyak supercomputer system at the University of Washington. 

\software{Astropy \citep{Astropy2013, Astropy2018}, Matplotlib \citep{Hunter2007}, Numpy \citep{Walt2011}, \texttt{SMART} \citep{Meadows1996}, \texttt{Pandeia} \citep{Pontoppidan2016}, \texttt{PandExo} \citep{Batalha2017b, Pandexo2018}, pysynphot \citep{STScI2013}}

\bibliography{ms}

\begin{thebibliography}{}
\expandafter\ifx\csname natexlab\endcsname\relax\def\natexlab#1{#1}\fi

\bibitem[{{Airapetian} {et~al.}(2017){Airapetian}, {Glocer}, {Khazanov},
  {Loyd}, {France}, {Sojka}, {Danchi}, \& {Liemohn}}]{Airapetian2017}
{Airapetian}, V.~S., {Glocer}, A., {Khazanov}, G.~V., {et~al.} 2017, \apjl,
  836, L3

\bibitem[{{Anglada-Escud{\'e}} {et~al.}(2016){Anglada-Escud{\'e}}, {Amado},
  {Barnes}, {Berdi{\~n}as}, {Butler}, {Coleman}, {de La Cueva}, {Dreizler},
  {Endl}, {Giesers}, {Jeffers}, {Jenkins}, {Jones}, {Kiraga}, {K{\"u}rster},
  {L{\'o}pez-Gonz{\'a}lez}, {Marvin}, {Morales}, {Morin}, {Nelson}, {Ortiz},
  {Ofir}, {Paardekooper}, {Reiners}, {Rodr{\'{\i}}guez},
  {Rodr{\'{\i}}guez-L{\'o}pez}, {Sarmiento}, {Strachan}, {Tsapras}, {Tuomi}, \&
  {Zechmeister}}]{Anglada-Escude2016}
{Anglada-Escud{\'e}}, G., {Amado}, P.~J., {Barnes}, J., {et~al.} 2016, Nature,
  536, 437

\bibitem[{{Arney} {et~al.}(2018){Arney}, {Domagal-Goldman}, \&
  {Meadows}}]{Arney2018}
{Arney}, G., {Domagal-Goldman}, S.~D., \& {Meadows}, V.~S. 2018, Astrobiology,
  18, 311

\bibitem[{{Arney} {et~al.}(2017){Arney}, {Meadows}, {Domagal-Goldman},
  {Deming}, {Robinson}, {Tovar}, {Wolf}, \& {Schwieterman}}]{Arney2017}
{Arney}, G.~N., {Meadows}, V.~S., {Domagal-Goldman}, S.~D., {et~al.} 2017,
  \apj, 836, 49

\bibitem[{{Astropy Collaboration} {et~al.}(2013){Astropy Collaboration},
  {Robitaille}, {Tollerud}, {Greenfield}, {Droettboom}, {Bray}, {Aldcroft},
  {Davis}, {Ginsburg}, {Price-Whelan}, {Kerzendorf}, {Conley}, {Crighton},
  {Barbary}, {Muna}, {Ferguson}, {Grollier}, {Parikh}, {Nair}, {Unther},
  {Deil}, {Woillez}, {Conseil}, {Kramer}, {Turner}, {Singer}, {Fox}, {Weaver},
  {Zabalza}, {Edwards}, {Azalee Bostroem}, {Burke}, {Casey}, {Crawford},
  {Dencheva}, {Ely}, {Jenness}, {Labrie}, {Lim}, {Pierfederici}, {Pontzen},
  {Ptak}, {Refsdal}, {Servillat}, \& {Streicher}}]{Astropy2013}
{Astropy Collaboration}, {Robitaille}, T.~P., {Tollerud}, E.~J., {et~al.} 2013,
  \aap, 558, A33

\bibitem[{{Bagnasco} {et~al.}(2007){Bagnasco}, {Kolm}, {Ferruit}, {Honnen},
  {Koehler}, {Lemke}, {Maschmann}, {Melf}, {Noyer}, {Rumler}, {Salvignol},
  {Strada}, \& {Te Plate}}]{Bagnasco2007}
{Bagnasco}, G., {Kolm}, M., {Ferruit}, P., {et~al.} 2007, in \procspie, Vol.
  6692, Cryogenic Optical Systems and Instruments XII, 66920M

\bibitem[{{Baraffe} {et~al.}(2015){Baraffe}, {Homeier}, {Allard}, \&
  {Chabrier}}]{Baraffe2015}
{Baraffe}, I., {Homeier}, D., {Allard}, F., \& {Chabrier}, G. 2015, \aap, 577,
  A42

\bibitem[{{Barstow} {et~al.}(2016){Barstow}, {Aigrain}, {Irwin}, {Kendrew}, \&
  {Fletcher}}]{Barstow2016}
{Barstow}, J.~K., {Aigrain}, S., {Irwin}, P.~G.~J., {Kendrew}, S., \&
  {Fletcher}, L.~N. 2016, Monthly Notices of the Royal Astronomical Society,
  458, 2657

\bibitem[{{Barstow} \& {Irwin}(2016)}]{Barstow2016b}
{Barstow}, J.~K., \& {Irwin}, P.~G.~J. 2016, Monthly Notices of the Royal
  Astronomical Society, 461, L92

\bibitem[{Batalha {et~al.}(2018)Batalha, Stevenson, Hill, Fraine, eas342, \&
  Cubillos}]{Pandexo2018}
Batalha, N., Stevenson, K., Hill, M., {et~al.} 2018, natashabatalha/PandExo:
  Starting PandExo Releases, doi:10.5281/zenodo.1256955

\bibitem[{{Batalha} {et~al.}(2018){Batalha}, {Lewis}, {Line}, {Valenti}, \&
  {Stevenson}}]{Batalha2018}
{Batalha}, N.~E., {Lewis}, N.~K., {Line}, M.~R., {Valenti}, J., \& {Stevenson},
  K. 2018, \apjl, 856, L34

\bibitem[{{Batalha} {et~al.}(2017){Batalha}, {Mandell}, {Pontoppidan},
  {Stevenson}, {Lewis}, {Kalirai}, {Earl}, {Greene}, {Albert}, \&
  {Nielsen}}]{Batalha2017b}
{Batalha}, N.~E., {Mandell}, A., {Pontoppidan}, K., {et~al.} 2017, Publications
  of the Astronomical Society of the Pacific, 129, 064501

\bibitem[{{Belu} {et~al.}(2011){Belu}, {Selsis}, {Morales}, {Ribas}, {Cossou},
  \& {Rauer}}]{Belu2011}
{Belu}, A.~R., {Selsis}, F., {Morales}, J.-C., {et~al.} 2011, Astronomy \&
  Astrophysics, 525, A83

\bibitem[{{Berta} {et~al.}(2012){Berta}, {Charbonneau}, {D{\'e}sert},
  {Miller-Ricci Kempton}, {McCullough}, {Burke}, {Fortney}, {Irwin}, {Nutzman},
  \& {Homeier}}]{Berta2012}
{Berta}, Z.~K., {Charbonneau}, D., {D{\'e}sert}, J.-M., {et~al.} 2012, The
  Astrophysical Journal, 747, 35

\bibitem[{{Berta-Thompson} {et~al.}(2015){Berta-Thompson}, {Irwin},
  {Charbonneau}, {Newton}, {Dittmann}, {Astudillo-Defru}, {Bonfils}, {Gillon},
  {Jehin}, {Stark}, {Stalder}, {Bouchy}, {Delfosse}, {Forveille}, {Lovis},
  {Mayor}, {Neves}, {Pepe}, {Santos}, {Udry}, \&
  {W{\"u}nsche}}]{Berta-Thompson2015}
{Berta-Thompson}, Z.~K., {Irwin}, J., {Charbonneau}, D., {et~al.} 2015, Nature,
  527, 204

\bibitem[{{B{\'e}tr{\'e}mieux} \& {Kaltenegger}(2014)}]{Betremieux2014}
{B{\'e}tr{\'e}mieux}, Y., \& {Kaltenegger}, L. 2014, The Astrophysical Journal,
  791, 7

\bibitem[{{B{\'e}tr{\'e}mieux} \& {Swain}(2018)}]{Betremieux2018}
{B{\'e}tr{\'e}mieux}, Y., \& {Swain}, M.~R. 2018, ArXiv e-prints,
  arXiv:1801.00738

\bibitem[{{Bolmont} {et~al.}(2017){Bolmont}, {Selsis}, {Owen}, {Ribas},
  {Raymond}, {Leconte}, \& {Gillon}}]{Bolmont2017}
{Bolmont}, E., {Selsis}, F., {Owen}, J.~E., {et~al.} 2017, \mnras, 464, 3728

\bibitem[{{Bouchet} {et~al.}(2015){Bouchet}, {Garc{\'{\i}}a-Mar{\'{\i}}n},
  {Lagage}, {Amiaux}, {Augu{\'e}res}, {Bauwens}, {Blommaert}, {Chen}, {Detre},
  {Dicken}, {Dubreuil}, {Galdemard}, {Gastaud}, {Glasse}, {Gordon}, {Gougnaud},
  {Guillard}, {Justtanont}, {Krause}, {Leboeuf}, {Longval}, {Martin}, {Mazy},
  {Moreau}, {Olofsson}, {Ray}, {Rees}, {Renotte}, {Ressler}, {Ronayette},
  {Salasca}, {Scheithauer}, {Sykes}, {Thelen}, {Wells}, {Wright}, \&
  {Wright}}]{Bouchet2015}
{Bouchet}, P., {Garc{\'{\i}}a-Mar{\'{\i}}n}, M., {Lagage}, P.-O., {et~al.}
  2015, \pasp, 127, 612

\bibitem[{{Cowan} {et~al.}(2009){Cowan}, {Agol}, {Meadows}, {Robinson},
  {Livengood}, {Deming}, {Lisse}, {A'Hearn}, {Wellnitz}, {Seager},
  {Charbonneau}, \& {EPOXI Team}}]{Cowan2009}
{Cowan}, N.~B., {Agol}, E., {Meadows}, V.~S., {et~al.} 2009, \apj, 700, 915

\bibitem[{{Cowan} {et~al.}(2015){Cowan}, {Greene}, {Angerhausen}, {Batalha},
  {Clampin}, {Col{\'o}n}, {Crossfield}, {Fortney}, {Gaudi}, {Harrington},
  {Iro}, {Lillie}, {Linsky}, {Lopez-Morales}, {Mandell}, \&
  {Stevenson}}]{Cowan2015}
{Cowan}, N.~B., {Greene}, T., {Angerhausen}, D., {et~al.} 2015, \pasp, 127, 311

\bibitem[{{Csizmadia} {et~al.}(2013){Csizmadia}, {Pasternacki}, {Dreyer},
  {Cabrera}, {Erikson}, \& {Rauer}}]{Csizmadia2013}
{Csizmadia}, S., {Pasternacki}, T., {Dreyer}, C., {et~al.} 2013, \aap, 549, A9

\bibitem[{{de Wit} {et~al.}(2016){de Wit}, {Wakeford}, {Gillon}, {Lewis},
  {Valenti}, {Demory}, {Burgasser}, {Burdanov}, {Delrez}, {Jehin}, {Lederer},
  {Queloz}, {Triaud}, \& {Van Grootel}}]{deWit2016}
{de Wit}, J., {Wakeford}, H.~R., {Gillon}, M., {et~al.} 2016, Nature, 537, 69

\bibitem[{{de Wit} {et~al.}(2018){de Wit}, {Wakeford}, {Lewis}, {Delrez},
  {Gillon}, {Selsis}, {Leconte}, {Demory}, {Bolmont}, {Bourrier}, {Burgasser},
  {Grimm}, {Jehin}, {Lederer}, {Owen}, {Stamenkovi{\'c}}, \&
  {Triaud}}]{deWit2018}
{de Wit}, J., {Wakeford}, H.~R., {Lewis}, N.~K., {et~al.} 2018, Nature
  Astronomy, 2, 214

\bibitem[{{Delrez} {et~al.}(2018){Delrez}, {Gillon}, {Triaud}, {Demory}, {de
  Wit}, {Ingalls}, {Agol}, {Bolmont}, {Burdanov}, {Burgasser}, {Carey},
  {Jehin}, {Leconte}, {Lederer}, {Queloz}, {Selsis}, \& {Van
  Grootel}}]{Delrez2018}
{Delrez}, L., {Gillon}, M., {Triaud}, A.~H.~M.~J., {et~al.} 2018, \mnras, 475,
  3577

\bibitem[{{Dittmann} {et~al.}(2017){Dittmann}, {Irwin}, {Charbonneau},
  {Bonfils}, {Astudillo-Defru}, {Haywood}, {Berta-Thompson}, {Newton},
  {Rodriguez}, {Winters}, {Tan}, {Almenara}, {Bouchy}, {Delfosse}, {Forveille},
  {Lovis}, {Murgas}, {Pepe}, {Santos}, {Udry}, {W{\"u}nsche}, {Esquerdo},
  {Latham}, \& {Dressing}}]{Dittmann2017}
{Dittmann}, J.~A., {Irwin}, J.~M., {Charbonneau}, D., {et~al.} 2017, \nat, 544,
  333

\bibitem[{{Dong} {et~al.}(2018){Dong}, {Jin}, {Lingam}, {Airapetian}, {Ma}, \&
  {van der Holst}}]{Dong2018}
{Dong}, C., {Jin}, M., {Lingam}, M., {et~al.} 2018, Proceedings of the National
  Academy of Science, 115, 260

\bibitem[{{Doyon} {et~al.}(2012){Doyon}, {Hutchings}, {Beaulieu}, {Albert},
  {Lafreni{\`e}re}, {Willott}, {Touahri}, {Rowlands}, {Maszkiewicz},
  {Fullerton}, {Volk}, {Martel}, {Chayer}, {Sivaramakrishnan}, {Abraham},
  {Ferrarese}, {Jayawardhana}, {Johnstone}, {Meyer}, {Pipher}, \&
  {Sawicki}}]{Doyon2012}
{Doyon}, R., {Hutchings}, J.~B., {Beaulieu}, M., {et~al.} 2012, in \procspie,
  Vol. 8442, Space Telescopes and Instrumentation 2012: Optical, Infrared, and
  Millimeter Wave, 84422R

\bibitem[{{Ehrenreich} {et~al.}(2014){Ehrenreich}, {Bonfils}, {Lovis},
  {Delfosse}, {Forveille}, {Mayor}, {Neves}, {Santos}, {Udry}, \&
  {S{\'e}gransan}}]{Ehrenreich2014}
{Ehrenreich}, D., {Bonfils}, X., {Lovis}, C., {et~al.} 2014, Astronomy \&
  Astrophysics, 570, A89

\bibitem[{{Feng} {et~al.}(2018){Feng}, {Robinson}, {Fortney}, {Lupu}, {Marley},
  {Lewis}, {Macintosh}, \& {Line}}]{Feng2018}
{Feng}, Y.~K., {Robinson}, T.~D., {Fortney}, J.~J., {et~al.} 2018, \aj, 155,
  200

\bibitem[{{Ferruit} {et~al.}(2014){Ferruit}, {Birkmann}, {B{\"o}ker},
  {Sirianni}, {Giardino}, {de Marchi}, {Alves de Oliveira}, \&
  {Dorner}}]{Ferruit2014}
{Ferruit}, P., {Birkmann}, S., {B{\"o}ker}, T., {et~al.} 2014, in \procspie,
  Vol. 9143, Space Telescopes and Instrumentation 2014: Optical, Infrared, and
  Millimeter Wave, 91430A

\bibitem[{{Garcia-Sage} {et~al.}(2017){Garcia-Sage}, {Glocer}, {Drake},
  {Gronoff}, \& {Cohen}}]{Garcia-Sage2017}
{Garcia-Sage}, K., {Glocer}, A., {Drake}, J.~J., {Gronoff}, G., \& {Cohen}, O.
  2017, \apjl, 844, L13

\bibitem[{{Gillon} {et~al.}(2016){Gillon}, {Jehin}, {Lederer}, {Delrez}, {de
  Wit}, {Burdanov}, {Van Grootel}, {Burgasser}, {Triaud}, {Opitom}, {Demory},
  {Sahu}, {Bardalez Gagliuffi}, {Magain}, \& {Queloz}}]{Gillon2016}
{Gillon}, M., {Jehin}, E., {Lederer}, S.~M., {et~al.} 2016, Nature, 533, 221

\bibitem[{{Gillon} {et~al.}(2017){Gillon}, {Triaud}, {Demory}, {Jehin}, {Agol},
  {Deck}, {Lederer}, {de Wit}, {Burdanov}, {Ingalls}, {Bolmont}, {Leconte},
  {Raymond}, {Selsis}, {Turbet}, {Barkaoui}, {Burgasser}, {Burleigh}, {Carey},
  {Chaushev}, {Copperwheat}, {Delrez}, {Fernandes}, {Holdsworth}, {Kotze}, {Van
  Grootel}, {Almleaky}, {Benkhaldoun}, {Magain}, \& {Queloz}}]{Gillon2017}
{Gillon}, M., {Triaud}, A.~H.~M.~J., {Demory}, B.-O., {et~al.} 2017, Nature,
  542, 456

\bibitem[{{Glasse} {et~al.}(2015){Glasse}, {Rieke}, {Bauwens},
  {Garc{\'{\i}}a-Mar{\'{\i}}n}, {Ressler}, {Rost}, {Tikkanen}, {Vandenbussche},
  \& {Wright}}]{Glasse2015}
{Glasse}, A., {Rieke}, G.~H., {Bauwens}, E., {et~al.} 2015, \pasp, 127, 686

\bibitem[{{Greene} {et~al.}(2007){Greene}, {Beichman}, {Eisenstein}, {Horner},
  {Kelly}, {Mao}, {Meyer}, {Rieke}, \& {Shi}}]{Greene2007}
{Greene}, T., {Beichman}, C., {Eisenstein}, D., {et~al.} 2007, in \procspie,
  Vol. 6693, Techniques and Instrumentation for Detection of Exoplanets III,
  66930G

\bibitem[{{Greene} {et~al.}(2016){Greene}, {Line}, {Montero}, {Fortney},
  {Lustig-Yaeger}, \& {Luther}}]{Greene2016}
{Greene}, T.~P., {Line}, M.~R., {Montero}, C., {et~al.} 2016, The Astrophysical
  Journal, 817, 17

\bibitem[{{Greene} {et~al.}(2017){Greene}, {Kelly}, {Stansberry}, {Leisenring},
  {Egami}, {Schlawin}, {Chu}, {Hodapp}, \& {Rieke}}]{Greene2017}
{Greene}, T.~P., {Kelly}, D.~M., {Stansberry}, J., {et~al.} 2017, Journal of
  Astronomical Telescopes, Instruments, and Systems, 3, 035001

\bibitem[{{Grenfell} {et~al.}(2007){Grenfell}, {Grie{\ss}meier}, {Patzer},
  {Rauer}, {Segura}, {Stadelmann}, {Stracke}, {Titz}, \& {Von
  Paris}}]{Grenfell2007}
{Grenfell}, J.~L., {Grie{\ss}meier}, J.-M., {Patzer}, B., {et~al.} 2007,
  Astrobiology, 7, 208

\bibitem[{{Grimm} {et~al.}(2018){Grimm}, {Demory}, {Gillon}, {Dorn}, {Agol},
  {Burdanov}, {Delrez}, {Sestovic}, {Triaud}, {Turbet}, {Bolmont}, {Caldas},
  {Wit}, {Jehin}, {Leconte}, {Raymond}, {Grootel}, {Burgasser}, {Carey},
  {Fabrycky}, {Heng}, {Hernandez}, {Ingalls}, {Lederer}, {Selsis}, \&
  {Queloz}}]{Grimm2018}
{Grimm}, S.~L., {Demory}, B.-O., {Gillon}, M., {et~al.} 2018, \aap, 613, A68

\bibitem[{{Hedelt} {et~al.}(2013){Hedelt}, {von Paris}, {Godolt}, {Gebauer},
  {Grenfell}, {Rauer}, {Schreier}, {Selsis}, \& {Trautmann}}]{Hedelt2013}
{Hedelt}, P., {von Paris}, P., {Godolt}, M., {et~al.} 2013, Astronomy \&
  Astrophysics, 553, A9

\bibitem[{{Hu} {et~al.}(2012){Hu}, {Ehlmann}, \& {Seager}}]{Hu2012}
{Hu}, R., {Ehlmann}, B.~L., \& {Seager}, S. 2012, \apj, 752, 7

\bibitem[{Hunter(2007)}]{Hunter2007}
Hunter, J.~D. 2007, Computing In Science \& Engineering, 9, 90

\bibitem[{{Husser} {et~al.}(2013){Husser}, {Wende-von Berg}, {Dreizler},
  {Homeier}, {Reiners}, {Barman}, \& {Hauschildt}}]{Husser2013}
{Husser}, T.~O., {Wende-von Berg}, S., {Dreizler}, S., {et~al.} 2013, \aap,
  553, A6

\bibitem[{{Kalirai}(2018)}]{Kalirai2018}
{Kalirai}, J. 2018, Contemporary Physics, 59, 251

\bibitem[{{Kendrew} {et~al.}(2015){Kendrew}, {Scheithauer}, {Bouchet},
  {Amiaux}, {Azzollini}, {Bouwman}, {Chen}, {Dubreuil}, {Fischer}, {Glasse},
  {Greene}, {Lagage}, {Lahuis}, {Ronayette}, {Wright}, \&
  {Wright}}]{Kendrew2015}
{Kendrew}, S., {Scheithauer}, S., {Bouchet}, P., {et~al.} 2015, \pasp, 127, 623

\bibitem[{{Kirkland} {et~al.}(2003){Kirkland}, {Herr}, \&
  {Adams}}]{Kirkland2003}
{Kirkland}, L.~E., {Herr}, K.~C., \& {Adams}, P.~M. 2003, Journal of
  Geophysical Research (Planets), 108, 5137

\bibitem[{{Knutson} {et~al.}(2014){Knutson}, {Benneke}, {Deming}, \&
  {Homeier}}]{Knutson2014}
{Knutson}, H.~A., {Benneke}, B., {Deming}, D., \& {Homeier}, D. 2014, Nature,
  505, 66

\bibitem[{{Kreidberg} \& {Loeb}(2016)}]{Kreidberg2016}
{Kreidberg}, L., \& {Loeb}, A. 2016, \apjl, 832, L12

\bibitem[{{Kreidberg} {et~al.}(2014){Kreidberg}, {Bean}, {D{\'e}sert},
  {Benneke}, {Deming}, {Stevenson}, {Seager}, {Berta-Thompson}, {Seifahrt}, \&
  {Homeier}}]{Kreidberg2014}
{Kreidberg}, L., {Bean}, J.~L., {D{\'e}sert}, J.-M., {et~al.} 2014, Nature,
  505, 69

\bibitem[{{Krissansen-Totton} {et~al.}(2018){Krissansen-Totton}, {Garland},
  {Irwin}, \& {Catling}}]{Krissansen-Totton2018}
{Krissansen-Totton}, J., {Garland}, R., {Irwin}, P., \& {Catling}, D.~C. 2018,
  \aj, 156, 114

\bibitem[{{Liebert} \& {Gizis}(2006)}]{Liebert2006}
{Liebert}, J., \& {Gizis}, J.~E. 2006, \pasp, 118, 659

\bibitem[{{Lincowski} {et~al.}(2019){Lincowski}, {Lustig-Yaeger}, \&
  {Meadows}}]{Lincowski2019}
{Lincowski}, A., {Lustig-Yaeger}, J., \& {Meadows}, V. 2019, in American
  Astronomical Society Meeting Abstracts, Vol. 233, American Astronomical
  Society Meeting Abstracts 233, 226.05

\bibitem[{{Lincowski} {et~al.}(2018){Lincowski}, {Meadows}, {Crisp},
  {Robinson}, {Luger}, {Lustig-Yaeger}, \& {Arney}}]{Lincowski2018}
{Lincowski}, A.~P., {Meadows}, V.~S., {Crisp}, D., {et~al.} 2018, \apj, 867, 76

\bibitem[{{Luger} \& {Barnes}(2015)}]{Luger2015}
{Luger}, R., \& {Barnes}, R. 2015, Astrobiology, 15, 119

\bibitem[{{Luger} {et~al.}(2015){Luger}, {Barnes}, {Lopez}, {Fortney},
  {Jackson}, \& {Meadows}}]{Luger2015b}
{Luger}, R., {Barnes}, R., {Lopez}, E., {et~al.} 2015, Astrobiology, 15, 57

\bibitem[{{Luger} {et~al.}(2017{\natexlab{a}}){Luger}, {Lustig-Yaeger}, \&
  {Agol}}]{Luger2017b}
{Luger}, R., {Lustig-Yaeger}, J., \& {Agol}, E. 2017{\natexlab{a}}, \apj, 851,
  94

\bibitem[{{Luger} {et~al.}(2017{\natexlab{b}}){Luger}, {Sestovic}, {Kruse},
  {Grimm}, {Demory}, {Agol}, {Bolmont}, {Fabrycky}, {Fernandes}, {Van Grootel},
  {Burgasser}, {Gillon}, {Ingalls}, {Jehin}, {Raymond}, {Selsis}, {Triaud},
  {Barclay}, {Barentsen}, {Howell}, {Delrez}, {de Wit}, {Foreman-Mackey},
  {Holdsworth}, {Leconte}, {Lederer}, {Turbet}, {Almleaky}, {Benkhaldoun},
  {Magain}, {Morris}, {Heng}, \& {Queloz}}]{Luger2017}
{Luger}, R., {Sestovic}, M., {Kruse}, E., {et~al.} 2017{\natexlab{b}}, Nature
  Astronomy, 1, 0129

\bibitem[{{Lustig-Yaeger} {et~al.}(2018){Lustig-Yaeger}, {Meadows}, {Tovar
  Mendoza}, {Schwieterman}, {Fujii}, {Luger}, \&
  {Robinson}}]{Lustig-Yaeger2018}
{Lustig-Yaeger}, J., {Meadows}, V.~S., {Tovar Mendoza}, G., {et~al.} 2018, \aj,
  156, 301

\bibitem[{{Maurin} {et~al.}(2012){Maurin}, {Selsis}, {Hersant}, \&
  {Belu}}]{Maurin2012}
{Maurin}, A.~S., {Selsis}, F., {Hersant}, F., \& {Belu}, A. 2012, \aap, 538,
  A95

\bibitem[{{Meadows} \& {Crisp}(1996)}]{Meadows1996}
{Meadows}, V.~S., \& {Crisp}, D. 1996, \jgr, 101, 4595

\bibitem[{{Meadows} {et~al.}(2018){Meadows}, {Arney}, {Schwieterman},
  {Lustig-Yaeger}, {Lincowski}, {Robinson}, {Domagal-Goldman}, {Deitrick},
  {Barnes}, {Fleming}, {Luger}, {Driscoll}, {Quinn}, \& {Crisp}}]{Meadows2018}
{Meadows}, V.~S., {Arney}, G.~N., {Schwieterman}, E.~W., {et~al.} 2018,
  Astrobiology, 18, 133

\bibitem[{{Miller-Ricci} {et~al.}(2009){Miller-Ricci}, {Seager}, \&
  {Sasselov}}]{Miller-Ricci2009}
{Miller-Ricci}, E., {Seager}, S., \& {Sasselov}, D. 2009, The Astrophysical
  Journal, 690, 1056

\bibitem[{{Misra} {et~al.}(2014){Misra}, {Meadows}, \& {Crisp}}]{Misra2014a}
{Misra}, A., {Meadows}, V., \& {Crisp}, D. 2014, The Astrophysical Journal,
  792, 61

\bibitem[{{Moran} {et~al.}(2018){Moran}, {H{\"o}rst}, {Batalha}, {Lewis}, \&
  {Wakeford}}]{Moran2018}
{Moran}, S.~E., {H{\"o}rst}, S.~M., {Batalha}, N.~E., {Lewis}, N.~K., \&
  {Wakeford}, H.~R. 2018, ArXiv e-prints, arXiv:1810.05210

\bibitem[{{Morley} {et~al.}(2017){Morley}, {Kreidberg}, {Rustamkulov},
  {Robinson}, \& {Fortney}}]{Morley2017}
{Morley}, C.~V., {Kreidberg}, L., {Rustamkulov}, Z., {Robinson}, T., \&
  {Fortney}, J.~J. 2017, \apj, 850, 121

\bibitem[{{Morris} {et~al.}(2018){Morris}, {Agol}, {Davenport}, \&
  {Hawley}}]{Morris2018}
{Morris}, B.~M., {Agol}, E., {Davenport}, J.~R.~A., \& {Hawley}, S.~L. 2018,
  \apj, 857, 39

\bibitem[{{Nicodemus}(1965)}]{Nicodemus1965}
{Nicodemus}, F.~E. 1965, \ao, 4, 767

\bibitem[{{Nikolov} {et~al.}(2015){Nikolov}, {Sing}, {Burrows}, {Fortney},
  {Henry}, {Pont}, {Ballester}, {Aigrain}, {Wilson}, {Huitson}, {Gibson},
  {D{\'e}sert}, {Lecavelier Des Etangs}, {Showman}, {Vidal-Madjar}, {Wakeford},
  \& {Zahnle}}]{Nikolov2015}
{Nikolov}, N., {Sing}, D.~K., {Burrows}, A.~S., {et~al.} 2015, Monthly Notices
  of the Royal Astronomical Society, 447, 463

\bibitem[{{Peacock} {et~al.}(2019){Peacock}, {Barman}, {Shkolnik},
  {Hauschildt}, \& {Baron}}]{Peacock2019}
{Peacock}, S., {Barman}, T., {Shkolnik}, E.~L., {Hauschildt}, P.~H., \&
  {Baron}, E. 2019, \apj, 871, 235

\bibitem[{{Pontoppidan} {et~al.}(2016){Pontoppidan}, {Pickering}, {Laidler},
  {Gilbert}, {Sontag}, {Slocum}, {Sienkiewicz}, {Hanley}, {Earl}, {Pueyo},
  {Ravindranath}, {Karakla}, {Robberto}, {Noriega-Crespo}, \&
  {Barker}}]{Pontoppidan2016}
{Pontoppidan}, K.~M., {Pickering}, T.~E., {Laidler}, V.~G., {et~al.} 2016, in
  \procspie, Vol. 9910, Observatory Operations: Strategies, Processes, and
  Systems VI, 991016

\bibitem[{{Price-Whelan} {et~al.}(2018){Price-Whelan}, {Sip{\H{o}}cz},
  {G{\"u}nther}, {Lim}, {Crawford}, {Conseil}, {Shupe}, {Craig}, {Dencheva},
  {Ginsburg}, {VanderPlas}, {Bradley}, {P{\'e}rez-Su{\'a}rez}, {de Val-Borro},
  {Paper Contributors}, {Aldcroft}, {Cruz}, {Robitaille}, {Tollerud},
  {Coordination Committee}, {Ardelean}, {Babej}, {Bach}, {Bachetti}, {Bakanov},
  {Bamford}, {Barentsen}, {Barmby}, {Baumbach}, {Berry}, {Biscani}, {Boquien},
  {Bostroem}, {Bouma}, {Brammer}, {Bray}, {Breytenbach}, {Buddelmeijer},
  {Burke}, {Calderone}, {Cano Rodr{\'\i}guez}, {Cara}, {Cardoso}, {Cheedella},
  {Copin}, {Corrales}, {Crichton}, {D{\textquoteright}Avella}, {Deil},
  {Depagne}, {Dietrich}, {Donath}, {Droettboom}, {Earl}, {Erben}, {Fabbro},
  {Ferreira}, {Finethy}, {Fox}, {Garrison}, {Gibbons}, {Goldstein}, {Gommers},
  {Greco}, {Greenfield}, {Groener}, {Grollier}, {Hagen}, {Hirst}, {Homeier},
  {Horton}, {Hosseinzadeh}, {Hu}, {Hunkeler}, {Ivezi{\'c}}, {Jain}, {Jenness},
  {Kanarek}, {Kendrew}, {Kern}, {Kerzendorf}, {Khvalko}, {King}, {Kirkby},
  {Kulkarni}, {Kumar}, {Lee}, {Lenz}, {Littlefair}, {Ma}, {Macleod},
  {Mastropietro}, {McCully}, {Montagnac}, {Morris}, {Mueller}, {Mumford},
  {Muna}, {Murphy}, {Nelson}, {Nguyen}, {Ninan}, {N{\"o}the}, {Ogaz}, {Oh},
  {Parejko}, {Parley}, {Pascual}, {Patil}, {Patil}, {Plunkett}, {Prochaska},
  {Rastogi}, {Reddy Janga}, {Sabater}, {Sakurikar}, {Seifert}, {Sherbert},
  {Sherwood-Taylor}, {Shih}, {Sick}, {Silbiger}, {Singanamalla}, {Singer},
  {Sladen}, {Sooley}, {Sornarajah}, {Streicher}, {Teuben}, {Thomas},
  {Tremblay}, {Turner}, {Terr{\'o}n}, {van Kerkwijk}, {de la Vega}, {Watkins},
  {Weaver}, {Whitmore}, {Woillez}, {Zabalza}, \& {Contributors}}]{Astropy2018}
{Price-Whelan}, A.~M., {Sip{\H{o}}cz}, B.~M., {G{\"u}nther}, H.~M., {et~al.}
  2018, \aj, 156, 123

\bibitem[{Rackham {et~al.}(2016)Rackham, Espinoza, Apai, L{\'o}pez-Morales,
  Jord{\'a}n, Osip, Lewis, Rodler, Fraine, Morley, {et~al.}}]{Rackham2016}
Rackham, B., Espinoza, N., Apai, D., {et~al.} 2016, arXiv preprint
  arXiv:1612.00228

\bibitem[{{Rackham} {et~al.}(2018){Rackham}, {Apai}, \&
  {Giampapa}}]{Rackham2018}
{Rackham}, B.~V., {Apai}, D., \& {Giampapa}, M.~S. 2018, \apj, 853, 122

\bibitem[{{Robinson}(2017)}]{Robinson2017a}
{Robinson}, T.~D. 2017, \apj, 836, 236

\bibitem[{{Robinson}(2018)}]{Robinson2018}
---. 2018, {Characterizing Exoplanet Habitability}, 67

\bibitem[{{Robinson} \& {Crisp}(2018)}]{Robinson2018b}
{Robinson}, T.~D., \& {Crisp}, D. 2018, \jqsrt, 211, 78

\bibitem[{{Robinson} {et~al.}(2010){Robinson}, {Meadows}, \&
  {Crisp}}]{Robinson2010}
{Robinson}, T.~D., {Meadows}, V.~S., \& {Crisp}, D. 2010, \apjl, 721, L67

\bibitem[{{Roettenbacher} \& {Kane}(2017)}]{Roettenbacher2017}
{Roettenbacher}, R.~M., \& {Kane}, S.~R. 2017, \apj, 851, 77

\bibitem[{{Rugheimer} {et~al.}(2015){Rugheimer}, {Kaltenegger}, {Segura},
  {Linsky}, \& {Mohanty}}]{Rugheimer2015}
{Rugheimer}, S., {Kaltenegger}, L., {Segura}, A., {Linsky}, J., \& {Mohanty},
  S. 2015, \apj, 809, 57

\bibitem[{{Schaefer} {et~al.}(2016){Schaefer}, {Wordsworth}, {Berta-Thompson},
  \& {Sasselov}}]{Schaefer2016}
{Schaefer}, L., {Wordsworth}, R.~D., {Berta-Thompson}, Z., \& {Sasselov}, D.
  2016, \apj, 829, 63

\bibitem[{{Schlawin} {et~al.}(2017){Schlawin}, {Rieke}, {Leisenring}, {Walker},
  {Fraine}, {Kelly}, {Misselt}, {Greene}, {Line}, {Lewis}, \&
  {Stansberry}}]{Schlawin2017}
{Schlawin}, E., {Rieke}, M., {Leisenring}, J., {et~al.} 2017, \pasp, 129,
  015001

\bibitem[{{Schwieterman} {et~al.}(2016){Schwieterman}, {Meadows},
  {Domagal-Goldman}, {Deming}, {Arney}, {Luger}, {Harman}, {Misra}, \&
  {Barnes}}]{Schwieterman2016}
{Schwieterman}, E.~W., {Meadows}, V.~S., {Domagal-Goldman}, S.~D., {et~al.}
  2016, The Astrophysical Journal Letters, 819, L13

\bibitem[{{Schwieterman} {et~al.}(2018){Schwieterman}, {Kiang}, {Parenteau},
  {Harman}, {DasSarma}, {Fisher}, {Arney}, {Hartnett}, {Reinhard}, {Olson},
  {Meadows}, {Cockell}, {Walker}, {Grenfell}, {Hegde}, {Rugheimer}, {Hu}, \&
  {Lyons}}]{Schwieterman2018}
{Schwieterman}, E.~W., {Kiang}, N.~Y., {Parenteau}, M.~N., {et~al.} 2018,
  Astrobiology, 18, 663

\bibitem[{{Segura} {et~al.}(2005){Segura}, {Kasting}, {Meadows}, {Cohen},
  {Scalo}, {Crisp}, {Butler}, \& {Tinetti}}]{Segura2005}
{Segura}, A., {Kasting}, J.~F., {Meadows}, V., {et~al.} 2005, Astrobiology, 5,
  706

\bibitem[{Seinfeld \& Pandis(2006)}]{Seinfeld2006}
Seinfeld, J., \& Pandis, S. 2006, Atmospheric Chemistry and Physics

\bibitem[{{Selsis} {et~al.}(2011){Selsis}, {Wordsworth}, \&
  {Forget}}]{Selsis2011}
{Selsis}, F., {Wordsworth}, R.~D., \& {Forget}, F. 2011, \aap, 532, A1

\bibitem[{{Sing} {et~al.}(2016){Sing}, {Fortney}, {Nikolov}, {Wakeford},
  {Kataria}, {Evans}, {Aigrain}, {Ballester}, {Burrows}, {Deming},
  {D{\'e}sert}, {Gibson}, {Henry}, {Huitson}, {Knutson}, {Lecavelier Des
  Etangs}, {Pont}, {Showman}, {Vidal-Madjar}, {Williamson}, \&
  {Wilson}}]{Sing2016}
{Sing}, D.~K., {Fortney}, J.~J., {Nikolov}, N., {et~al.} 2016, Nature, 529, 59

\bibitem[{{STScI Development Team}(2013)}]{STScI2013}
{STScI Development Team}. 2013, {pysynphot: Synthetic photometry software
  package}, Astrophysics Source Code Library, ascl:1303.023

\bibitem[{{van der Walt} {et~al.}(2011){van der Walt}, {Colbert}, \&
  {Varoquaux}}]{Walt2011}
{van der Walt}, S., {Colbert}, S.~C., \& {Varoquaux}, G. 2011, Computing in
  Science and Engineering, 13, 22

\bibitem[{{Van Grootel} {et~al.}(2018){Van Grootel}, {Fernandes}, {Gillon},
  {Jehin}, {Manfroid}, {Scuflaire}, {Burgasser}, {Barkaoui}, {Benkhaldoun},
  {Burdanov}, {Delrez}, {Demory}, {de Wit}, {Queloz}, \&
  {Triaud}}]{VanGrootel2018}
{Van Grootel}, V., {Fernandes}, C.~S., {Gillon}, M., {et~al.} 2018, \apj, 853,
  30

\bibitem[{{Vida} {et~al.}(2017){Vida}, {K{\H o}v{\'a}ri}, {P{\'a}l},
  {Ol{\'a}h}, \& {Kriskovics}}]{Vida2017}
{Vida}, K., {K{\H o}v{\'a}ri}, Z., {P{\'a}l}, A., {Ol{\'a}h}, K., \&
  {Kriskovics}, L. 2017, \apj, 841, 124

\bibitem[{{Wang} {et~al.}(2017){Wang}, {Wu}, {Barclay}, \&
  {Laughlin}}]{Wang2017}
{Wang}, S., {Wu}, D.-H., {Barclay}, T., \& {Laughlin}, G.~P. 2017, ArXiv
  e-prints, arXiv:1704.04290

\bibitem[{{Weiss} \& {Marcy}(2014)}]{Weiss2014}
{Weiss}, L.~M., \& {Marcy}, G.~W. 2014, \apjl, 783, L6

\bibitem[{{Wordsworth} {et~al.}(2018){Wordsworth}, {Schaefer}, \&
  {Fischer}}]{Wordsworth2018}
{Wordsworth}, R.~D., {Schaefer}, L.~K., \& {Fischer}, R.~A. 2018, \aj, 155, 195

\bibitem[{Wunderlich {et~al.}(2019)Wunderlich, Godolt, Grenfell, St{\"a}dt,
  Smith, Gebauer, Schreier, Hedelt, \& Rauer}]{Wunderlich2019}
Wunderlich, F., Godolt, M., Grenfell, J.~L., {et~al.} 2019

\bibitem[{{Zhang} {et~al.}(2018){Zhang}, {Zhou}, {Rackham}, \&
  {Apai}}]{Zhang2018}
{Zhang}, Z., {Zhou}, Y., {Rackham}, B.~V., \& {Apai}, D. 2018, \aj, 156, 178

\end{thebibliography}

\appendix

\section{Signal-to-Noise Ratio on Transit and Eclipse Depths}
\label{sec:appendix:snr}

Transit and eclipse depth measurements are both relative to the out-of-occultation measurement, and thus require accounting for photon fluence in- and out-of the the event of interest. Below we derive the signal-to-noise on measurements of transit and eclipse depths by considering the signal as the occultation depth, which must be calculated from the difference between in- and out-of-event observations, and the noise by propagating individual measurement errors through the signal calculation. 

For primary transit, the signal of interest is the number of photons missing when the star is occulted by the planet. Neglecting photons emitted and reflected by the planet prior to and during transit, the out-of-transit photon fluence is measured as 
\begin{equation}
    N_{\text{out}} = (N_s+N_{bg})n_{\text{out}}, 
\end{equation}
where $n_{\text{out}}$ is the time of measurement outside of transit in units of the transit duration, and $N_s$ and $N_{bg}$ are the total number of photons counted over a transit duration. The in-transit photon fluence measured is
\begin{equation}
    N_{tr} = N_s \left ( 1 - \left (\frac{R_p}{R_s} \right )^2 \right ) + N_{bg}.
\end{equation}
The number of stellar photons blocked by the planet can be estimated from
\begin{equation}
    N_{sp} = N_{\text{out}} / n_{\text{out}} - N_{tr}, 
\end{equation}
this is the signal we seek to measure. The noise on $N_{sp}$ can be calculated by considering the variance given standard uncorrelated error propagation:
\begin{align}
    \sigma^2 &= \left ( \frac{\partial N_{sp}}{\partial N_{\text{out}}}\right )^2 \sigma N_{\text{out}}^2 + \left (\frac{\partial N_{sp}}{\partial N_{tr}} \right )^2 \sigma N_{tr}^2 \\
    &= \left ( \frac{1}{ n_{\text{out}}} \right )^2 N_{\text{out}} + ( -1 )^2 N_{tr} \\
    &= \frac{(N_s+N_{bg})}{n_{\text{out}}} + N_s \left ( 1 - \left (\frac{R_p}{R_s} \right )^2 \right ) + N_{bg}
\end{align}
Finally, the signal-to-noise ratio on the transit depth can be constructed by dividing the blocked photons $N_{sp}$ by the standard deviation on that estimate:
\begin{equation}
    \text{SNR}_T = \frac{N_s \left ( R_p/R_s \right )^2}{\sqrt{(N_s+N_{bg})/n_{\text{out}} + N_s \left ( 1 - \left ( R_p/R_s \right )^2 \right ) + N_{bg}}}.
\end{equation}

For secondary eclipse, the signal of interest is the number of photons missing when the planet is occulted by the star, which is measured assuming that the star is not varying. In this case, the out-of-eclipse photon fluence is measured as 
\begin{equation}
    N_{\text{out}} = (N_p+N_s+N_{bg})n_{\text{out}}, 
\end{equation}
where $n_{\text{out}}$ is the time of measurement outside of eclipse in units of the eclipse duration, $N_p$ is the total number of planet photons over an eclipse duration, and the sum of $N_p$, $N_s$ and $N_{bg}$ is the total number of photons counted over an eclipse duration. The in-eclipse photon fluence measured is 
\begin{equation}
    N_{ec} = N_s + N_{bg},
\end{equation}  
which allows the planet photon counts to be estimated as 
\begin{equation}
    N_p = N_{\text{out}}/n_{\text{out}} - N_{ec}.
\end{equation}
Again, analogous to the transit calculation, the noise term can be calculated by considering the variance on $N_p$ given standard error propagation:  
\begin{align}
    \sigma^2 &= \left ( \frac{\partial N_p}{\partial N_{\text{out}}}\right )^2 \sigma N_{\text{out}}^2 + \left (\frac{\partial N_p}{\partial N_{ec}} \right )^2 \sigma N_{ec}^2 \\
    &= \left ( \frac{1}{ n_{\text{out}}} \right )^2 N_{\text{out}} + ( -1 )^2 N_{ec}
\end{align}
where we have used the fact that $N_{\text{out}}$ has a variance of $N_{\text{out}}$.
Substituting in for $N_{\text{out}}$ and $N_{ec}$ gives:
\begin{equation}
    \sigma^2 = \frac{(N_p + N_s + N_{bg})}{n_{\text{out}}} + N_s + N_{bg}. 
\end{equation}
Finally, the signal-to-noise ratio on the eclipse depth can be constructed by dividing the estimated planet photons by the standard deviation on that estimate:
\begin{equation}
    \text{SNR}_E = \frac{N_p}{\sqrt{(N_p + N_s + N_{bg})/n_{\text{out}} + N_s + N_{bg}}}.
\end{equation}

\section{Atmospheric Detectability by Instrument}
\label{app:atmos_detect}

\begin{figure*}[h!]
\centering
\includegraphics[width=0.97\textwidth]{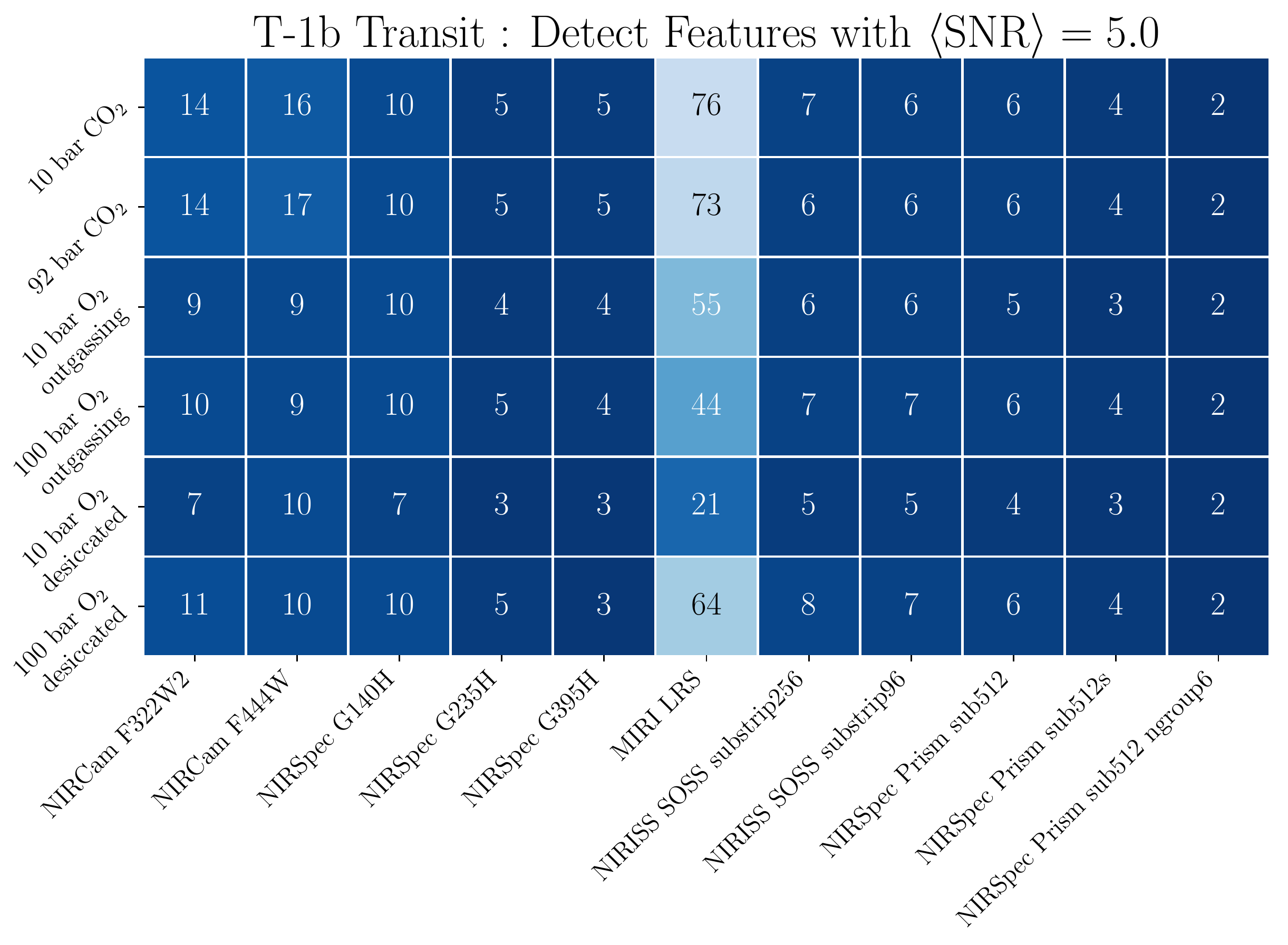}
\caption{Number of TRAPPIST-1b transits necessary to rule out a featureless spectrum with $\left < \mathrm{SNR} \right > = 5$ for different atmospheric compositions and using different JWST instruments and modes.}
\label{fig:colortable_featureless_T-1b_transit}
\end{figure*}

\begin{figure*}[h!]
\centering
\includegraphics[width=0.97\textwidth]{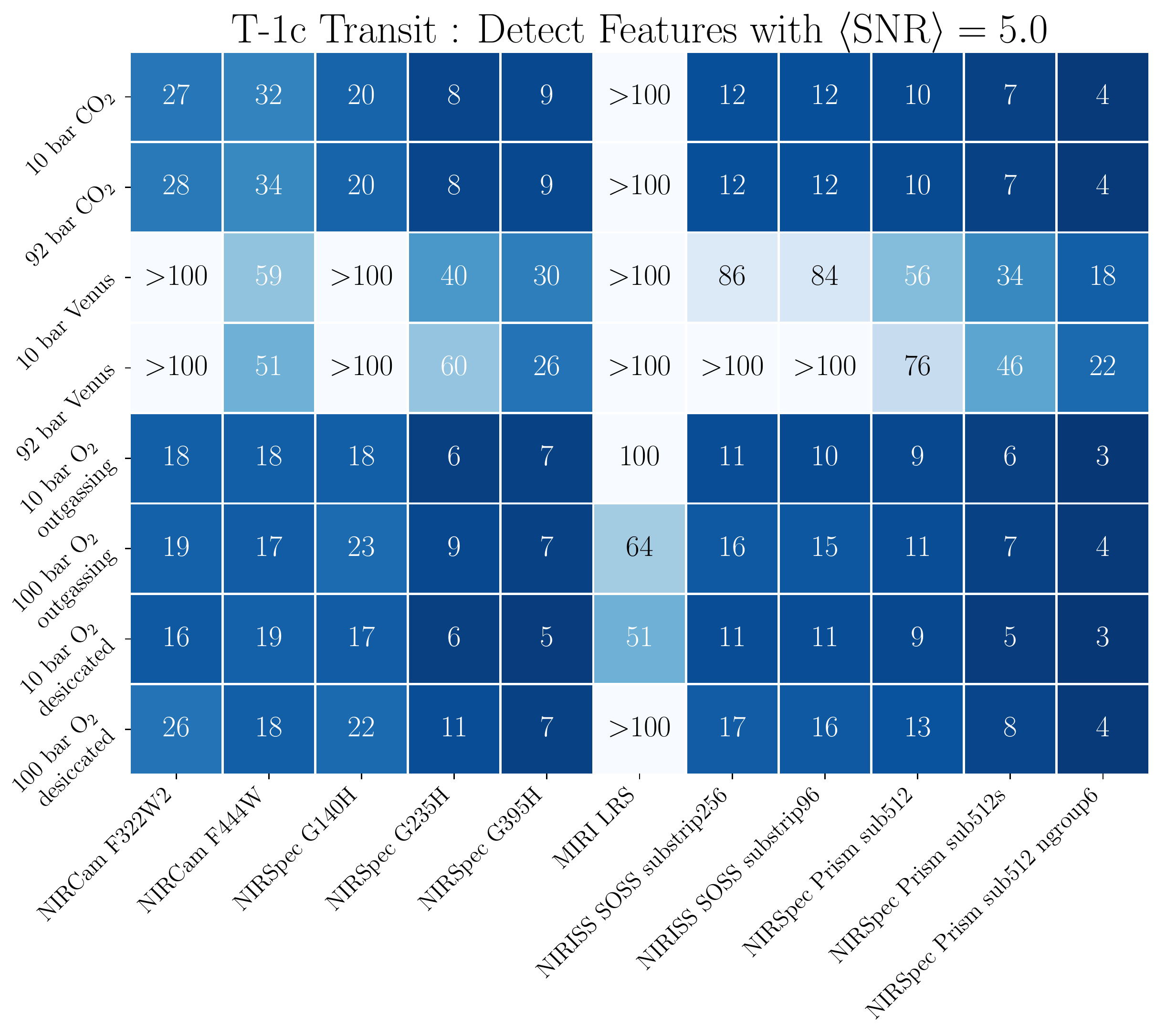}
\caption{Number of TRAPPIST-1c transits necessary to rule out a featureless spectrum with $\left < \mathrm{SNR} \right > = 5$ for different atmospheric compositions and using different JWST instruments and modes.}
\label{fig:colortable_featureless_T-1c_transit}
\end{figure*}

\begin{figure*}[h!]
\centering
\includegraphics[width=0.97\textwidth]{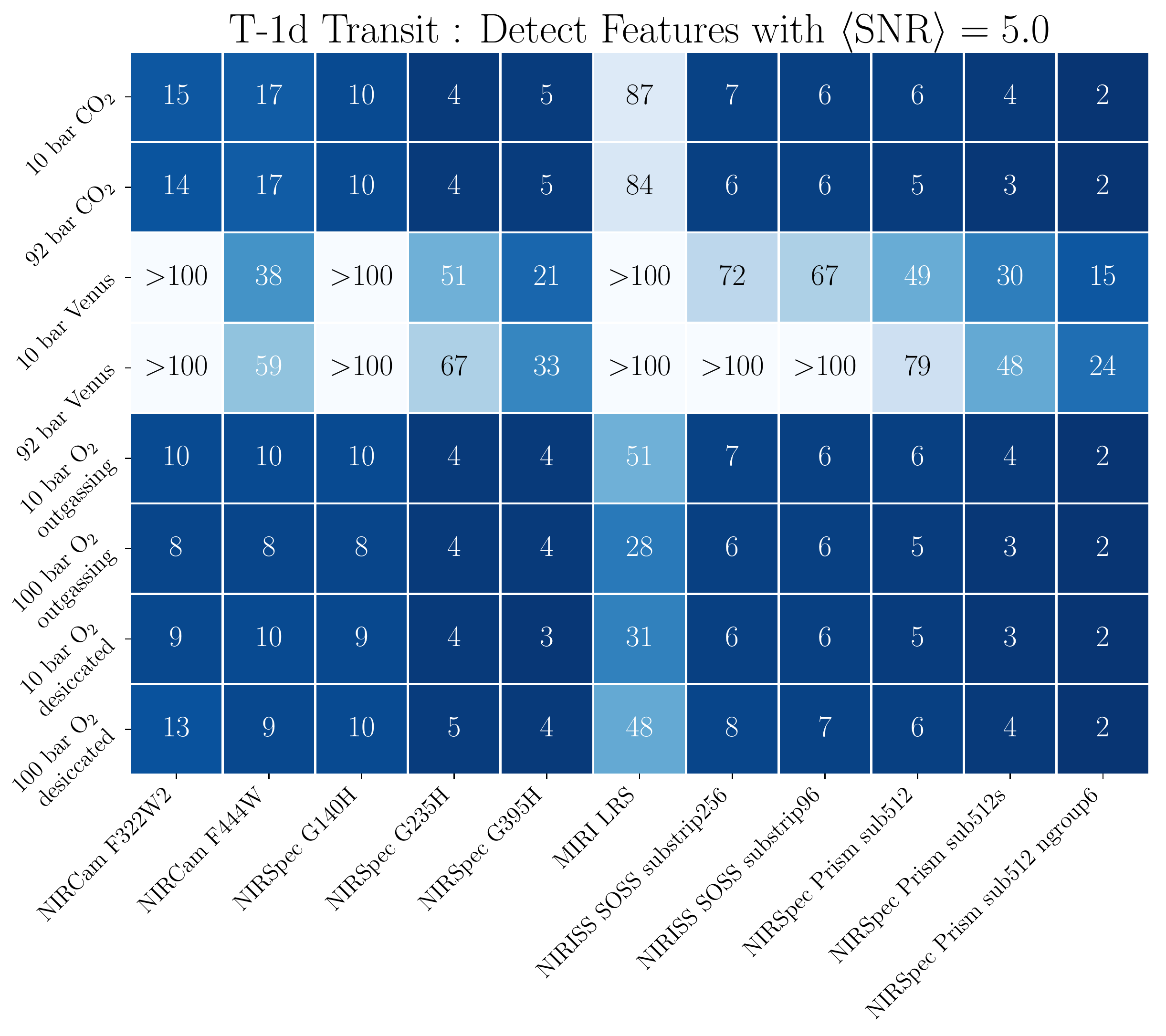}
\caption{Number of TRAPPIST-1d transits necessary to rule out a featureless spectrum with $\left < \mathrm{SNR} \right > = 5$ for different atmospheric compositions and using different JWST instruments and modes.}
\label{fig:colortable_featureless_T-1d_transit}
\end{figure*}

\begin{figure*}[h!]
\centering
\includegraphics[width=0.97\textwidth]{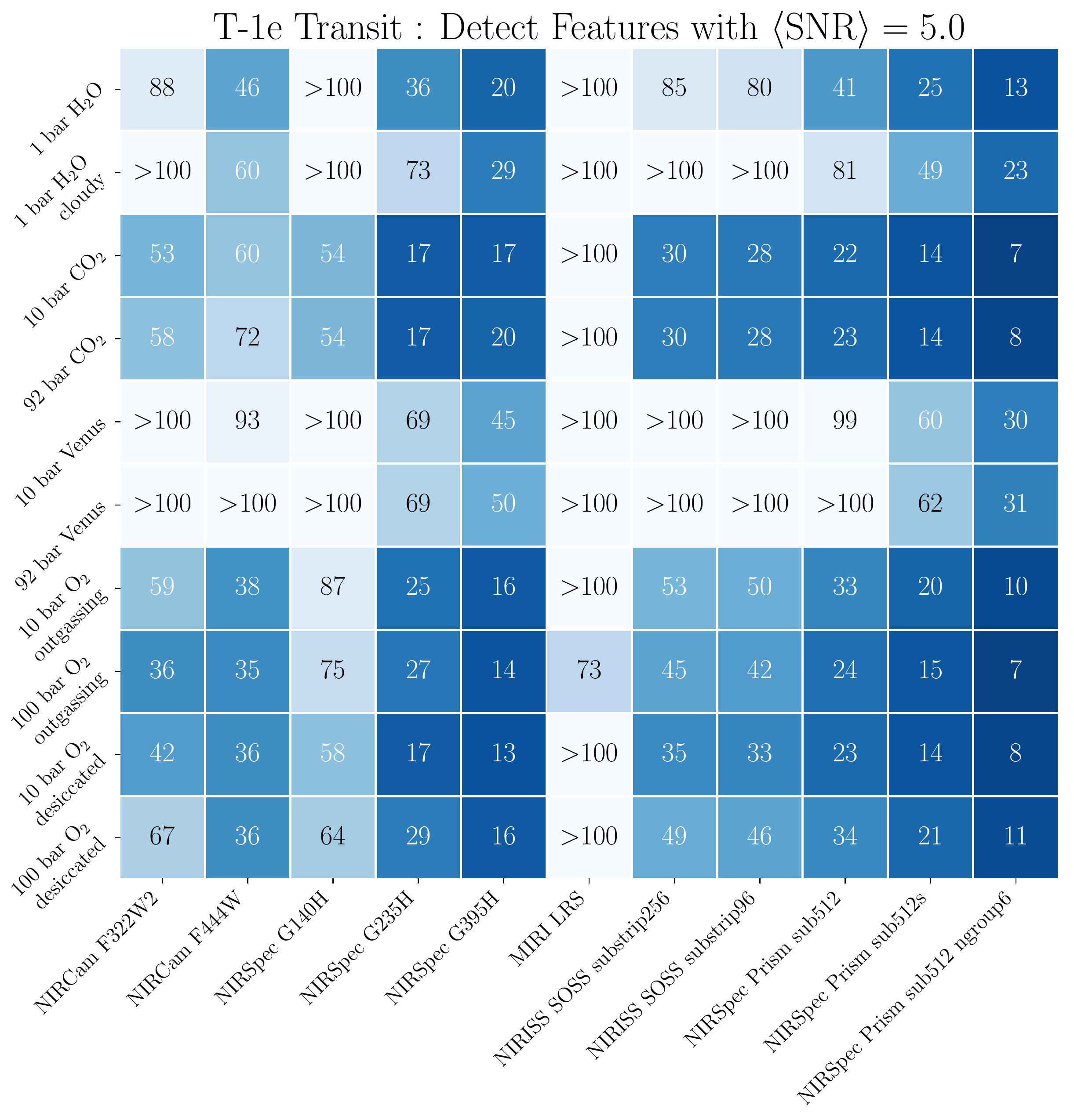}
\caption{Number of TRAPPIST-1e transits necessary to rule out a featureless spectrum with $\left < \mathrm{SNR} \right > = 5$ for different atmospheric compositions and using different JWST instruments and modes.}
\label{fig:colortable_featureless_T-1e_transit}
\end{figure*}

\begin{figure*}[h!]
\centering
\includegraphics[width=0.97\textwidth]{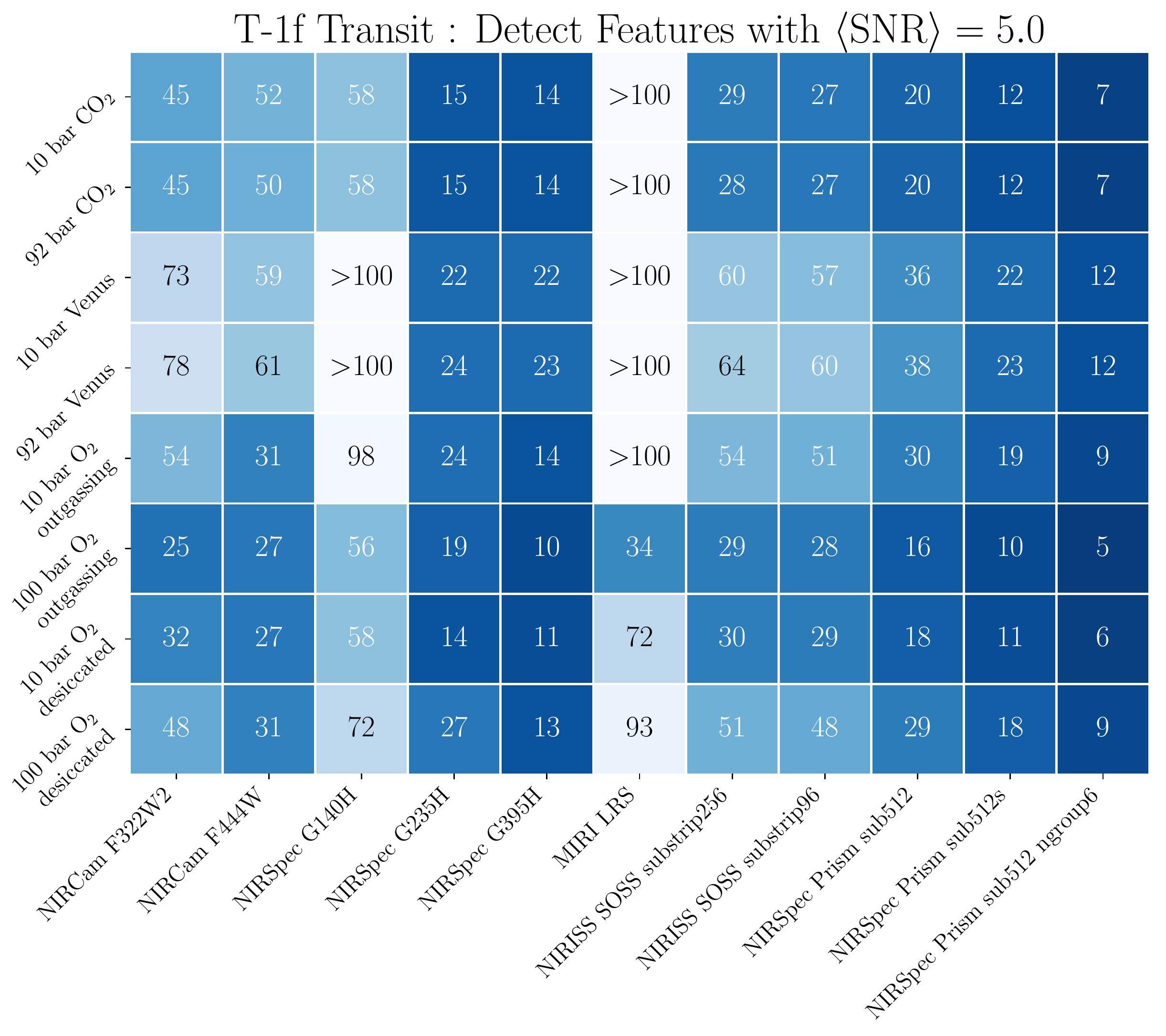}
\caption{Number of TRAPPIST-1f transits necessary to rule out a featureless spectrum with $\left < \mathrm{SNR} \right > = 5$ for different atmospheric compositions and using different JWST instruments and modes.}
\label{fig:colortable_featureless_T-1f_transit}
\end{figure*}

\begin{figure*}[h!]
\centering
\includegraphics[width=0.97\textwidth]{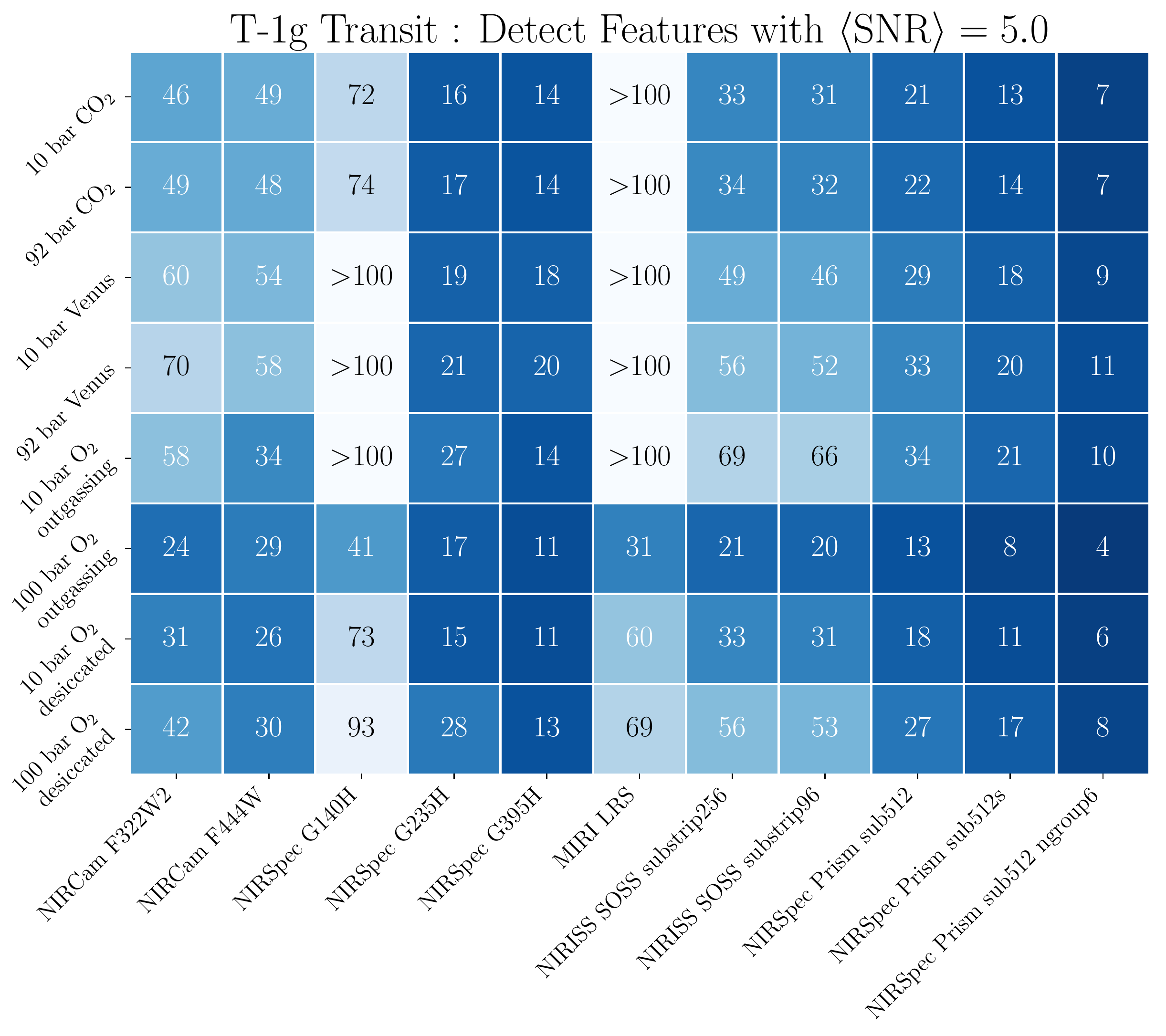}
\caption{Number of TRAPPIST-1g transits necessary to rule out a featureless spectrum with $\left < \mathrm{SNR} \right > = 5$ for different atmospheric compositions and using different JWST instruments and modes.}
\label{fig:colortable_featureless_T-1g_transit}
\end{figure*}

\begin{figure*}[h!]
\centering
\includegraphics[width=0.97\textwidth]{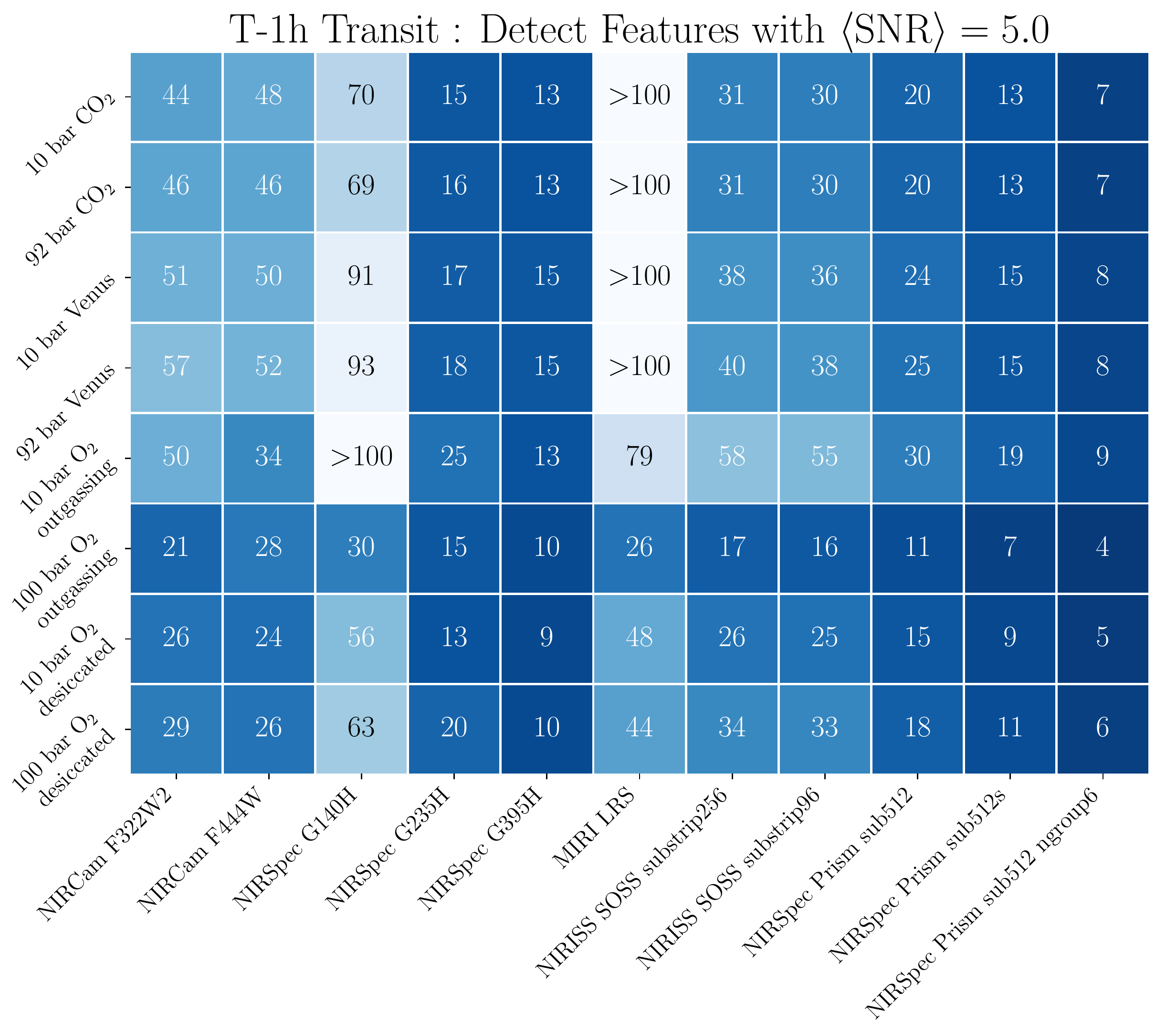}
\caption{Number of TRAPPIST-1h transits necessary to rule out a featureless spectrum with $\left < \mathrm{SNR} \right > = 5$ for different atmospheric compositions and using different JWST instruments and modes.}
\label{fig:colortable_featureless_T-1h_transit}
\end{figure*}

Figures \ref{fig:colortable_featureless_T-1b_transit},  \ref{fig:colortable_featureless_T-1c_transit},  \ref{fig:colortable_featureless_T-1d_transit},  \ref{fig:colortable_featureless_T-1e_transit},  \ref{fig:colortable_featureless_T-1f_transit},  \ref{fig:colortable_featureless_T-1g_transit}, and  \ref{fig:colortable_featureless_T-1h_transit}, show the number of transits needed to detect the atmospheres of TRAPPIST-1 b, c, d, e, f, g, and h, respectively, as a function of both atmospheric compositions and JWST instrument/mode. The detection of an atmosphere requires $\left < \mathrm{SNR} \right > = 5$ on spectral features in the transmission spectrum. 

\section{Brightness Temperatures}
\label{app:bright_temp}

Table \ref{tab:bright_temps} shows the brightness temperature of each \citet{Lincowski2018} TRAPPIST-1 atmospheric model in each of the JWST/MIRI photometric filters and the two warm \textit{Spitzer} bands. 

\begin{longrotatetable}

\begin{deluxetable*}{llccccccccccc}
\tablewidth{\linewidth}
\tablecaption{\label{tab:bright_temps} Brightness temperatures in different photometric filters for different models. }
\tablehead{ \colhead{Planet}  &\colhead{Model}  &\colhead{IRAC1}  &\colhead{IRAC2}  &\colhead{F560W}  &\colhead{F770W}  &\colhead{F1000W}  &\colhead{F1130W}  &\colhead{F1280W}  &\colhead{F1500W}  &\colhead{F1800W}  &\colhead{F2100W}  &\colhead{F2550W}   \\
\colhead{}  &\colhead{}  &\colhead{(K)}  &\colhead{(K)}  &\colhead{(K)}  &\colhead{(K)}  &\colhead{(K)}  &\colhead{(K)}  &\colhead{(K)}  &\colhead{(K)}  &\colhead{(K)}  &\colhead{(K)}  &\colhead{(K)}   }
\startdata
T-1b&  10 bar CO$_2$&  498&  328&  432&  384&  340&  356&  330&  291&  315&  362&  430\\
&  92 bar CO$_2$&  495&  328&  432&  383&  340&  357&  330&  289&  315&  362&  430\\
&  100 bar O$_2$
outgassing&  500&  388&  372&  396&  379&  432&  385&  305&  355&  381&  346\\
&  10 bar O$_2$
outgassing&  514&  391&  357&  394&  394&  425&  377&  300&  345&  371&  329\\
&  100 bar O$_2$
desiccated&  425&  375&  390&  387&  389&  404&  376&  306&  368&  408&  364\\
&  10 bar O$_2$
desiccated&  436&  357&  401&  405&  369&  397&  360&  292&  346&  408&  380\\
T-1c&  10 bar CO$_2$&  445&  292&  385&  328&  304&  326&  295&  250&  275&  315&  386\\
&  92 bar CO$_2$&  437&  290&  380&  324&  303&  326&  293&  248&  273&  313&  382\\
&  100 bar O$_2$
outgassing&  434&  337&  325&  346&  312&  390&  343&  266&  325&  345&  308\\
&  10 bar O$_2$
outgassing&  424&  334&  313&  350&  337&  378&  330&  252&  304&  332&  294\\
&  100 bar O$_2$
desiccated&  387&  325&  330&  328&  333&  339&  327&  272&  325&  342&  310\\
&  10 bar O$_2$
desiccated&  395&  313&  339&  341&  322&  345&  318&  257&  310&  347&  322\\
&  10 bar Venus&  427&  292&  373&  318&  300&  320&  296&  272&  282&  313&  376\\
&  92 bar Venus&  435&  287&  382&  317&  297&  324&  291&  250&  275&  303&  383\\
T-1d&  10 bar CO$_2$&  384&  256&  337&  274&  269&  295&  260&  219&  239&  272&  340\\
&  92 bar CO$_2$&  380&  254&  333&  271&  268&  295&  259&  215&  237&  269&  337\\
&  100 bar O$_2$
outgassing&  370&  290&  280&  301&  264&  352&  295&  227&  290&  305&  270\\
&  10 bar O$_2$
outgassing&  361&  283&  270&  299&  287&  324&  285&  215&  267&  291&  260\\
&  100 bar O$_2$
desiccated&  356&  281&  278&  274&  280&  279&  280&  239&  282&  284&  260\\
&  10 bar O$_2$
desiccated&  362&  275&  286&  286&  279&  293&  277&  225&  275&  292&  269\\
&  10 bar Venus&  326&  258&  240&  230&  230&  233&  233&  227&  231&  239&  243\\
&  92 bar Venus&  375&  251&  332&  264&  263&  288&  256&  223&  245&  271&  335\\
T-1e&  10 bar CO$_2$&  339&  230&  297&  235&  247&  272&  232&  193&  208&  237&  301\\
&  92 bar CO$_2$&  335&  226&  292&  229&  243&  267&  228&  189&  206&  233&  294\\
&  100 bar O$_2$
outgassing&  305&  236&  244&  259&  231&  301&  245&  203&  253&  265&  245\\
&  10 bar O$_2$
outgassing&  346&  243&  243&  256&  243&  267&  244&  191&  239&  259&  238\\
&  100 bar O$_2$
desiccated&  341&  250&  244&  238&  242&  239&  246&  214&  251&  243&  226\\
&  10 bar O$_2$
desiccated&  346&  246&  250&  246&  242&  249&  244&  200&  247&  251&  232\\
&  10 bar Venus&  303&  217&  213&  203&  204&  204&  202&  192&  204&  206&  211\\
&  92 bar Venus&  311&  223&  287&  228&  235&  254&  225&  195&  220&  242&  285\\
&  1 bar H$_2$O&  296&  267&  243&  259&  272&  277&  267&  213&  258&  260&  236\\
&  1 bar H$_2$O
cloudy&  287&  253&  241&  247&  250&  253&  248&  212&  247&  249&  243\\
T-1f&  10 bar CO$_2$&  277&  201&  260&  200&  221&  235&  200&  172&  185&  206&  263\\
&  92 bar CO$_2$&  271&  196&  262&  195&  216&  225&  194&  171&  183&  202&  260\\
&  100 bar O$_2$
outgassing&  229&  209&  218&  219&  202&  234&  208&  192&  221&  224&  222\\
&  10 bar O$_2$
outgassing&  326&  212&  218&  216&  206&  223&  210&  178&  209&  218&  213\\
&  100 bar O$_2$
desiccated&  268&  225&  228&  216&  210&  209&  219&  192&  227&  213&  194\\
&  10 bar O$_2$
desiccated&  298&  222&  229&  223&  210&  216&  218&  184&  227&  219&  200\\
&  10 bar Venus&  272&  189&  210&  188&  193&  193&  188&  171&  184&  189&  199\\
&  92 bar Venus&  247&  197&  239&  199&  211&  218&  196&  177&  195&  210&  237\\
T-1g&  10 bar CO$_2$&  231&  181&  236&  180&  203&  205&  179&  160&  170&  185&  232\\
&  92 bar CO$_2$&  229&  177&  240&  175&  200&  197&  174&  158&  167&  181&  231\\
&  100 bar O$_2$
outgassing&  209&  191&  199&  199&  188&  202&  192&  180&  205&  204&  194\\
&  10 bar O$_2$
outgassing&  312&  195&  202&  194&  187&  199&  191&  166&  188&  192&  191\\
&  100 bar O$_2$
desiccated&  224&  207&  210&  200&  192&  187&  201&  180&  208&  189&  176\\
&  10 bar O$_2$
desiccated&  276&  205&  211&  205&  192&  192&  200&  174&  209&  195&  178\\
&  10 bar Venus&  242&  173&  210&  173&  183&  181&  173&  158&  168&  174&  190\\
&  92 bar Venus&  226&  182&  222&  181&  200&  199&  177&  165&  176&  191&  217\\
T-1h&  10 bar CO$_2$&  174&  157&  217&  155&  176&  167&  154&  145&  150&  157&  186\\
&  92 bar CO$_2$&  175&  154&  206&  151&  174&  162&  151&  143&  147&  153&  182\\
&  100 bar O$_2$
outgassing&  195&  168&  171&  170&  166&  169&  167&  163&  174&  172&  164\\
&  10 bar O$_2$
outgassing&  296&  181&  184&  170&  165&  167&  170&  150&  165&  164&  160\\
&  100 bar O$_2$
desiccated&  199&  183&  185&  179&  170&  161&  177&  165&  181&  162&  149\\
&  10 bar O$_2$
desiccated&  253&  184&  188&  183&  171&  164&  179&  160&  186&  167&  150\\
&  10 bar Venus&  209&  155&  185&  153&  162&  161&  153&  145&  149&  154&  162\\
&  92 bar Venus&  189&  159&  195&  156&  185&  169&  146&  145&  144&  161&  187
\enddata
\tablecomments{Table note 1}
\end{deluxetable*}

\end{longrotatetable}


\end{document}